\documentclass[prd,aps,twocolumn,nofootinbib,preprintnumbers,superscriptaddress,balancelastpage,longbibliography]{revtex4-2}
\usepackage[utf8]{inputenc}
\usepackage{graphicx}
\graphicspath{ {./figures/} }
\usepackage{booktabs}
\usepackage{caption}
\DeclareCaptionLabelFormat{cont}{#1 #2 (cont.)}
\usepackage{hyperref}
\usepackage{slashed}
\usepackage{aas_macros}
\usepackage{amsmath}
\usepackage{array}
\usepackage{tabularx}
\usepackage{multirow}
\usepackage[dvipsnames]{xcolor}
\usepackage{listings}
\usepackage{xspace}
\usepackage[normalem]{ulem}
\hypersetup{
     colorlinks   = true,
     citecolor    = colorforlinks,
     urlcolor     = colorforlinks,
     linkcolor    = colorforlinks
}

\definecolor{colorforlinks}{HTML}{e69138} 

\newcommand{\preMAX}{\texttt{preMAX}\xspace}
\newcommand{\MAX}{\texttt{MAX}\xspace}
\newcommand{\postMAX}{\texttt{postMAX}\xspace}
\newcommand{\conv}{\texttt{conv}\xspace}
\newcommand{\reacc}{\texttt{reacc}\xspace}
\newcommand{\FF}{\texttt{FF}\xspace}

\newcommand{\Ext}{\texttt{Ext}\xspace}

\begin{document}

\title{Addressing the Impact of Solar Modulation Systematic Uncertainties \\ on Cosmic-Ray Propagation Models}

\author{Isabelle John}
\thanks{E-mail: \href{mailto:isabelle.john@unito.it}{isabelle.john@unito.it}, ORCID: \href{https://orcid.org/0000-0003-2550-7038}{0000-0003-2550-7038}}
\affiliation{Dipartimento di Fisica, Universit\`a degli Studi di Torino, \\ via P.\ Giuria, 1 10125 Torino, Italy}
\affiliation{INFN -- Istituto Nazionale di Fisica Nucleare, Sezione di Torino, \\ via P.\ Giuria 1, 10125 Torino, Italy}

\author{Alessandro Cuoco}
\thanks{E-mail: \href{mailto:alessandro.cuoco@unito.it}{alessandro.cuoco@unito.it}, ORCID: \href{https://orcid.org/0000-0003-1504-894X}{0000-0003-1504-894X}}
\affiliation{Dipartimento di Fisica, Universit\`a degli Studi di Torino, \\ via P.\ Giuria, 1 10125 Torino, Italy}
\affiliation{INFN -- Istituto Nazionale di Fisica Nucleare, Sezione di Torino, \\ via P.\ Giuria 1, 10125 Torino, Italy}

\author{Mattia di Mauro}
\thanks{E-mail: \href{mailto:mattia.dimauro@to.infn.it}{mattia.dimauro@to.infn.it}, ORCID: \href{https://orcid.org/0000-0003-2759-5625}{0000-0003-2759-5625}}
\affiliation{INFN -- Istituto Nazionale di Fisica Nucleare, Sezione di Torino, \\ via P.\ Giuria 1, 10125 Torino, Italy}


\begin{abstract}
\noindent
We perform a comprehensive analysis of cosmic-ray propagation using the time-dependent AMS-02 flux measurements covering a full solar cycle, with particular emphasis on the role of solar modulation. We fit two representative Galactic propagation scenarios, convection- and re-acceleration–dominated models, in combination with three solar modulation prescriptions: the standard force-field approximation, an extended force-field model with a rigidity break, and the heliospheric propagation code \texttt{HelMod}. The inclusion of time-resolved antiproton data provides a unique probe of charge-sign–dependent modulation effects and low-energy systematics. We find that the force-field approximation can describe positively charged nuclei reasonably well outside the solar maximum in convection-dominated models, but fails during periods of high solar activity and for antiprotons at all times. In re-acceleration scenarios, strong degeneracies between solar modulation and low-energy propagation lead to unphysical results when simple modulation models are employed. Across all models, we identify systematic uncertainties of order 10–15\% in the reconstructed local interstellar spectra and propagation parameters, driven by limitations in current solar modulation modelling. Compared to the percent level error of current measurements, these uncertainties significantly limit the precision of cosmic-ray studies. Future time-dependent measurements spanning a full 22-year solar cycle will be crucial to reduce these uncertainties.
\end{abstract}

\maketitle

\section{Introduction}
\label{sec: introduction}

The Sun plays a crucial role in shaping the local cosmic-ray fluxes. Due to a combination of solar winds and the heliospheric magnetic field, the cosmic-ray fluxes below rigidities of $\mathcal{O}(30-40)$~GV are strongly suppressed, a process referred to as solar modulation~\cite{2013LRSP...10....3P, Rankin:2022poh, Parker1958, Parker1965, Jokipii1971}. The exact mechanisms are complex and still subject to sizeable uncertainties, which, in turn, introduces systematic uncertainties in cosmic-ray propagation models ~\cite{Usoskin:2017cli, Wang:2019xtu, Corti:2015bqi, Kuhlen:2019hqb, Tomassetti:2017hbe, Tomassetti:2025nna, Tomassetti:2018tnl, Zhang:2026ohu, Boschini:2017fxq, Boschini:2018baj, Silver:2024ero}.

However, detailed and dependable knowledge of the cosmic-ray fluxes is crucial to understand the origin of cosmic rays and the processes that take place during their acceleration and propagation throughout the Galaxy and within the solar system. Additionally, precise and reliable cosmic-ray models are necessary in the search for unexpected or exotic components in the cosmic-ray fluxes, for example from dark matter~\cite{Cuoco:2016eej, Calore:2022stf, Bergstrom:2013jra, John:2021ugy, Nguyen:2024kwy}.

The effects of solar modulation were spectacularly confirmed in 2012 and 2018, when the Voyager I and Voyager II space probes, respectively, left the heliosphere and entered interstellar space. As expected, the cosmic-ray fluxes increased at lower energies~\cite{Stone:2013zlg, Corti:2015bqi}. To this date, the Voyager probes provide the only measurements of the unmodulated cosmic-ray fluxes, \textit{i.e.} the local interstellar spectra (LIS), offering valuable insights into the Sun's impact on cosmic-ray propagation. 

\begin{figure*}[tbp]
    \centering
    \includegraphics[width=\linewidth]{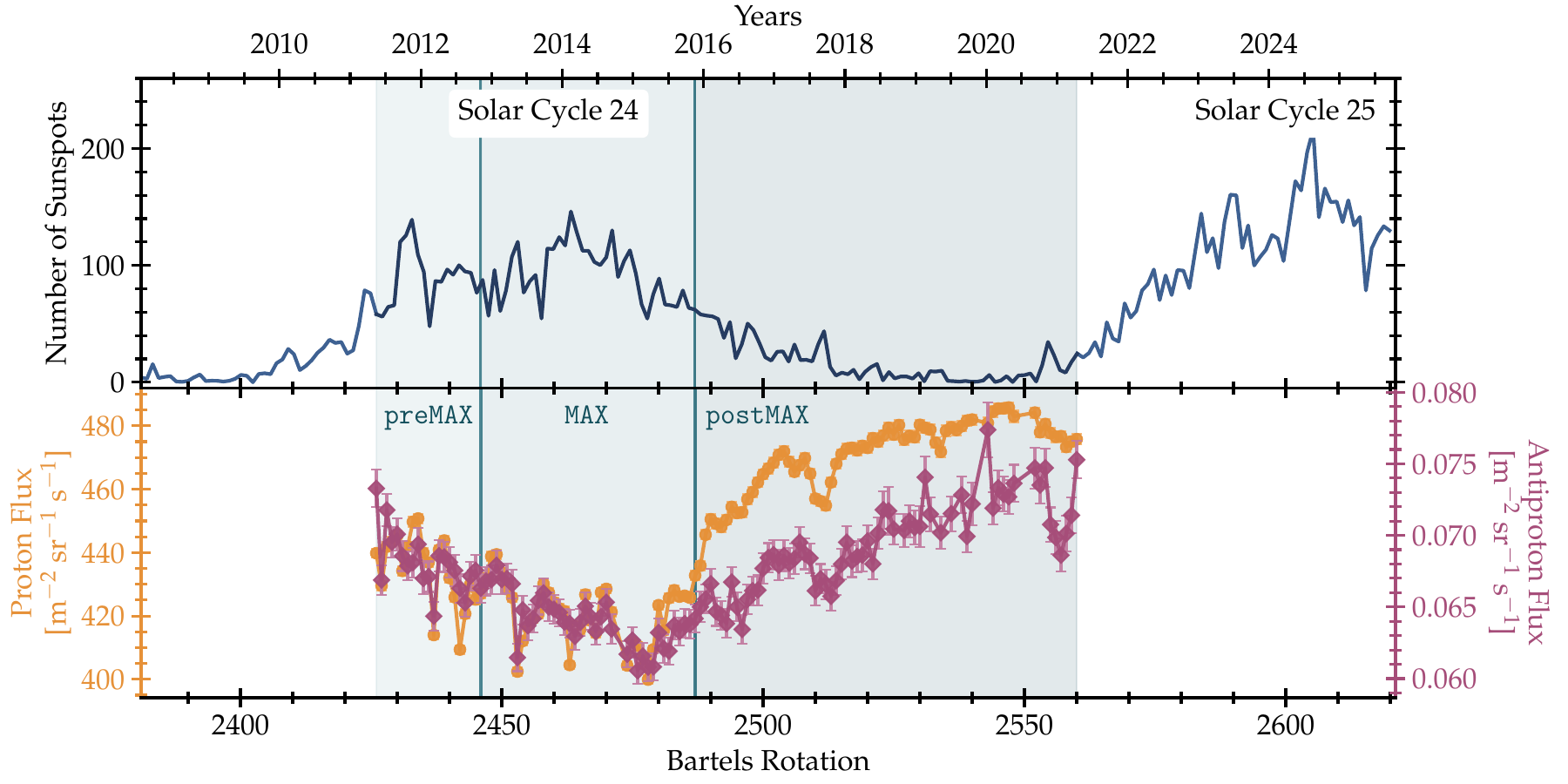}
    \caption{The monthly-averaged number of sunspots~\cite{SILSO_Sunspot_Number} (top panel) and the cosmic-ray proton (orange circle markers) and antiproton (burgundy diamond markers) fluxes as measured by AMS-02~\cite{AMS:2025npj, AMS:2025pgu} (bottom panel). The cosmic-ray fluxes are shown for rigidities between 5~GV and 42~GV for each Bartels rotation from May 2011 to January 2021. The \preMAX, \MAX and \postMAX periods indicate the separate fitting periods in our analysis. A larger number of sunspots indicates stronger solar activity, which is anti-correlated with the intensity of the cosmic-ray fluxes. Due to the time-lag of the propagation of the solar winds and magnetic field, the imprint of solar activity on the cosmic-ray fluxes is typically delayed by a few months to a year.  }
    \label{fig: ssn}
\end{figure*}

Moreover, the Sun's activity changes over an 11-year cycle, which adds a temporal dependence to the modulation of the cosmic-ray fluxes~\cite{2013LRSP...10....3P, Rankin:2022poh}. During solar maximum, solar activity is strongest, and the heliospheric magnetic field is the most turbulent, as it is in the process of reversing its polarity. During solar minimum, the Sun is in a calmer state and its magnetic field more regular. This time-variation in solar activity is imprinted on the cosmic-ray fluxes by suppressing them more during periods of higher solar activity and less during periods of lower activity. This is illustrated in Figure~\ref{fig: ssn}: The top panel shows the number of sunspots, averaged over each month from the beginning of 2008 to the end of 2025~\cite{SILSO_Sunspot_Number}, \textit{i.e.} from the beginning of solar cycle 24 until the most recent data of solar cycle 25. The number of sunspots strongly correlates with solar activity, and therefore anti-correlates with the modulation of the cosmic-ray fluxes, shown in the bottom panel for protons (orange circle markers) and antiprotons (burgundy diamond markers) as measured by AMS-02~\cite{AMS:2025pgu, AMS:2025npj}, spanning nearly a full solar cycle. Fluxes are shown at each Bartels rotation (\textit{i.e.} the 27-day period that it takes until the same location on the Sun's equator is again visible from Earth), and are integrated over the rigidity range from 5~to 41~GV. Temporal changes in the intensity of the fluxes of $\sim 20$\% are observed. The fluxes are most suppressed in 2014, which corresponds to the maximum of solar cycle 24 (spanning from 2008 to 2019). Currently, the Sun is undergoing solar cycle 25, which started in 2020 and recently reached its peak in $\sim 2024$. The anti-correlation between solar activity and cosmic-ray fluxes was also shown by PAMELA, through the measurement of time-dependent proton~\cite{AdrianiEtAl2013PAMELAProtonTime}, electron~\cite{Adriani:2015kxa} and positron fluxes \cite{Adriani:2016uhu} across solar cycles 23 and 24.

Temporal changes correlating with the modulation of the cosmic-ray fluxes have also been detected in the solar $\gamma$-ray emission of the solar disk and halo, associated with interactions of cosmic rays with both photospheric gas and sunlight~\cite{Ng:2015gya, Linden:2020lvz, Linden:2025xom, Acharyya:2025xya}, further highlighting the extensive impact of solar activity on the solar system.

Cosmic-ray solar modulation can be modelled in terms of propagation in the heliosphere through the Parker equation, which includes convection, diffusion, drifts, and adiabatic energy losses~\cite{Parker1958, Parker1965, Jokipii1971}. To this purpose, various codes have been developed, such as \texttt{Helioprop}~\cite{Maccione:2012cu}, \texttt{SolarProp}~\cite{Kappl:2015hxv} and \texttt{HelMod}~\cite{Boschini:2022jwz}, of which the latter is the most advanced, since it considers time-dependent effects in the heliosphere, and is tuned against the latest data from space probes such as PAMELA, AMS-02 and Ulysses~\cite{2015ApJ...812..140S}. Nonetheless, even within the framework of Parker propagation, numerous assumptions and ingredients (for example on the geometry of the diffusion tensor or on the overall level of diffusion) are required to derive the modulated flux of cosmic rays at Earth, which are subject to further uncertainties~\cite{Boschini:2022jwz}. For this reason, many approximate-phenomenological models have been developed, which are easier to handle at the price of neglecting some of the complexity of the true process. Among these, the most commonly employed is the force-field approximation~\cite{1968ApJ...154.1011G, CaballeroLopezMoraal2004, GieselerHeberHerbst2017} which compresses the whole solar modulation process into a single parameter, \textit{i.e.}, the force-field potential.
A number of extended and modified versions of the force-field potential also exists, which improve the modelling of the time-, charge-, and rigidity dependence of solar modulation~(\textit{e.g.}~\cite{Cholis:2015gna, Kuhlen:2019hqb, Cholis:2020tpi, Li:2022zio, Long:2024nty, Zhu:2024ega, Zhu:2025lrc, Zhu:2025qqx, Zhang:2026ohu}). In the following we will use both force-field and extended force-field modulation models, as well as \texttt{HelMod} modulation in order to explore the systematic uncertainties related to solar modulation, and quantify their effect on the reconstruction of the cosmic-ray LIS and propagation in the Galaxy.

Recently, the AMS-02 experiment has released data of the fluxes of many cosmic-ray nuclei over a full solar cycle~\cite{AMS:2025npj, AMS:2025pgu}, ranging from May 2011 until June 2022. Importantly, AMS-02 has also measured the fluxes of antiprotons, which, as negatively charged cosmic rays, are impacted differently compared to protons and other positively charged nuclei, due to charge-sign effects involved in solar modulation. This can be seen in the bottom panel of Fig.~\ref{fig: ssn}, where, indeed, the time behaviour of the proton and antiproton fluxes are very different. The high-statistical precision of the AMS-02 data over a full solar cycle allows for a detailed study of the impact of solar activity on the locally measured cosmic rays. Studying the temporal effects makes it possible to investigate solar modulation during different configurations of solar activity, which can be governed by a variety of uncertainties, and in turn results in different modelling requirements. 

In this work, we utilise the time-dependent AMS-02 data to investigate solar modulation, and the underlying systematic uncertainties that affect our ability to precisely model cosmic-ray propagation. In particular, we conduct detailed parameter fits of cosmic-ray Galactic propagation models during three periods of different levels of solar activity, highlighted in Fig.~\ref{fig: ssn}. We combine this with various solar modulation models to gain an understanding of their impact on cosmic-ray propagation models and their systematic limitations. Ideally, if both the Galactic propagation and solar modulation models are correct, the results should be independent of the analysed time-period. We will thus investigate the consistency of the LIS fluxes and of the solar modulation potentials derived in the three differing periods to examine if a coherent picture emerges.

The remainder of this article is structured as follows: In Section~\ref{sec: cosmic-ray data}, we list the cosmic-ray datasets used throughout our analysis. In Section~\ref{sec: cosmic-ray models}, we describe our cosmic-ray propagation models, and in Section~\ref{sec: solar modulation models} the solar modulation models investigated in our analysis. We present and discuss our results in Section~\ref{sec: results and discussion}, and provide conclusions in Section~\ref{sec: summary and conclusions}.

\section{Cosmic-Ray Data Selection}
\label{sec: cosmic-ray data}

AMS-02 has recently released time-dependent data for many cosmic-ray species with high statistical precision. These datasets span a time period corresponding to a full solar cycle\footnote{Note the liberal use of the term ``solar cycle". While a solar cycle is defined from solar minimum to solar minimum, we also adopt the term to represent any $\sim 11$-year period of solar activity, such as the period covered by the time-dependent AMS-02 data.}, from May 2011 to November 2022 (mostly solar cycle 24), allowing us to track changes in the local cosmic-ray flux over a full cycle of solar activity.  To study these effects, we conduct both time-dependent and time-integrated fits to the data. 

For reference, we perform fits using the \textit{7-year time-integrated} AMS-02 dataset~\cite{AMS:2021nhj}. This dataset spans from May 2011 to May 2018 and includes protons, Helium, Carbon, Oxygen, and Nitrogen, as well as the antiproton-to-proton ratio, Boron-to-Carbon ratio and Helium-3-to-Helium-4 ratio. We also include the proton data from CALET~\cite{CALET:2022vro} up to $\sim 10$~TeV/nuc, the Helium data from DAMPE~\cite{Alemanno:2021gpb} up to $\sim 10$~TeV/nuc, and the Boron-to-Carbon ratio from DAMPE~\cite{DAMPE:2022jgy}. Furthermore, we use the data from the Voyager probes, as they provide measurements of the unmodulated LIS fluxes for protons and Helium outside the heliosphere at rigidities below $\sim 1$~GV~\cite{Stone:2013zlg}.

We then perform further fits using the recently published \textit{time-dependent} AMS-02 datasets. Specifically, we include the time-dependent data for antiprotons and protons~\cite{AMS:2025npj}, and for the cosmic-ray nuclei Helium, Boron, Carbon, Nitrogen, and Oxygen~\cite{AMS:2025pgu}, as well as the Boron-to-Helium ratio~\cite{AMS:2025pgu} (as, unfortunately, the time-dependent Boron-to-Carbon ratio has not been provided by AMS-02), and the time-dependent Helium-3-to-Helium-4 ratio~\cite{AMS:2024idr}. Due to large uncertainties in their production cross sections, we do not fit to data for Lithium and Beryllium~\cite{Maurin:2025gsz}. The above time-dependent datasets, however, only reach up to ${ \sim 41 }$~GV for antiprotons and protons, and ${ \sim 60 }$~GV for the other nuclei. Above these rigidities, where solar modulation is sub-dominant, we thus complement the data with the 7-year time-integrated AMS-02 dataset~\cite{AMS:2021nhj}.

Furthermore, for consistency, we only include data above a rigidity of 2~GV, as data below this rigidity is only given for protons and antiprotons, but not the other nuclei. This does not apply to the Voyager data, which we fit directly to the LIS fluxes below 1~GV.

Since the Helium-3-to-Helium-4 time-dependent AMS-02 data is only provided until April 2021, we restrict our time range from May 2011 until April 2021 for all datasets, even if time-dependent data for other nuclei is given over the full solar cycle.

To study the different phases of solar activity over a solar cycle, we divide this data into three time-integrated periods (see Fig.~\ref{fig: ssn}). These correspond to 

\begin{itemize}
    \item \texttt{preMAX}: May 2011 to November 2012 (Bartels rotations 2426 to 2446), \textit{i.e.} a few months before the maximum of the solar activity, 
    \item \texttt{MAX}: December 2012 to November 2015 (Bartels rotations 2447 to 2487), \textit{i.e.} the period where solar activity is maximal, and
    \item \texttt{postMAX}: December 2015 to April 2021 (Bartels rotations 2488 to 2560), \textit{i.e.} the time after the maximum, where the polarity of the solar magnetic field is flipped compared to \texttt{preMAX}. This period also includes the solar minimum in around 2020.
\end{itemize}

Since the time-dependent data is given for each Bartels rotation (or every four Bartels rotation, in the case of the Helium-3-to-Helium-4 ratio), we create each time-integrated dataset by time-averaging the fluxes in each time period. We add statistical uncertainties in quadrature, divided by the square root of their total number. The combined systematic uncertainty is, instead, derived as a time-average of the individual uncertainties of each datapoint. We then compute the total uncertainty by adding statistical and systematic uncertainties in quadrature. We note that our treatment of the systematic uncertainties is somewhat approximate, as we do not have detailed information on the origin and interplay of the underlying systematics. For the future, besides the monthly datasets, we advocate the release of yearly or bi-yearly aggregated datasets from the AMS-02 collaboration in order to perform this kind of time-dependent analyses more easily. In principle, one could perform analyses using the monthly datasets directly. However, this is not optimal since it requires the use of many fit parameters (\textit{i.e.}, an independent solar modulation potential in each Bartels rotation), or a model of the time-dependent modulation (as \textit{e.g.} in~\cite{Cholis:2015gna, Cholis:2020tpi, Zhu:2025lrc, Zhu:2025qqx, Tomassetti:2017hbe, Kuhlen:2019hqb}), which, however, introduces further model dependency.

Fig.~\ref{fig: ssn} illustrates our choice of these periods. The number of sunspots strongly anti-correlates with the cosmic-ray fluxes, clearly indicating periods of lower solar activity before and after the maximum phase. However, the imprint of solar activity on the cosmic-ray fluxes is subject to a time-lag due to the propagation time of the heliospheric magnetic field and solar winds. This lag is estimated to be around 2 to 14 months, and strongly depends on the rigidity of the cosmic rays, such that the lag time decreases with increasing rigidity~\cite{Tomassetti:2017gkx, Tomassetti:2022fal}. It can also be observed that the time lag is stronger during higher solar activity, by comparing the peak of the sunspot number in around 2014 to the dip in the cosmic-ray fluxes in around 2015, spanning almost a year. Conversely, during low solar activity, the lag is much shorter, as can be seen in the small peak in late 2020 that corresponds to a dip in the cosmic-ray fluxes occurring only a few months afterward. These observations allow us to estimate the period where the cosmic-ray fluxes are maximally modulated to be around December 2012 to November 2015 (\MAX). The \preMAX phase includes all data available before this, and the \postMAX phase all data after this. However, since the available data before the \MAX phase only encompass about 1.5~years, which is several years after the previous solar minimum, the \texttt{preMAX} phase somewhat resembles the \texttt{MAX} phase, and, as can be observed in Fig.~\ref{fig: ssn}, also includes periods of higher solar activity, like the spike in late 2011. The 7-yr dataset, instead, spanning from May 2011 to May 2018 encompasses all the \preMAX and \MAX periods and the beginning of the \postMAX phase, \textit{i.e.} it mostly includes the phase of high solar activity.

\section{Cosmic-Ray Propagation Models}
\label{sec: cosmic-ray models}

Our analysis is based on the framework for cosmic-ray propagation models developed in previous works~\cite{Cuoco:2019kuu, Korsmeier:2021brc, Korsmeier:2021bkw, DiMauro:2023oqx, DiMauro:2023jgg}. We use the numerical cosmic-ray propagation code \texttt{GALPROP}~\cite{Strong:1998pw, Strong:2015zva} that takes into account a variety of processes such as cosmic-ray injection, diffusion, convection, energy losses and re-acceleration. In order to fit over the complex multi-dimensional parameter space, we use \texttt{MultiNest}~\cite{Feroz:2008xx}. Our free parameters include various parameters for cosmic-ray injection and propagation in the Galaxy. Very importantly, we also include parameters that take into account and parametrize the uncertainties in the production cross-sections of secondary cosmic rays \cite{diMauro:2014zea, Donato:2017ywo, Korsmeier:2018gcy, Maurin:2025gsz, Genolini:2018ekk, Genolini:2023kcj}. Depending on the solar modulation model (discussed in detail in Section~\ref{sec: solar modulation models}), a number of additional free parameters is introduced. Within our framework, we fit solar-modulation and cosmic-ray parameters simultaneously, which accounts for the possible presence of degeneracies and interplay among all the parameters. We break the degeneracy between the diffusion coefficient and the halo height by fixing the latter to the canonical value of $z = 4.2$~kpc~\cite{2010ApJ...722L..58S, DeLaTorreLuque:2024wtv}. For a full list of the fit parameters, see the figures and tables in Appendix~\ref{app: propagation models}, and the references above for more details.

We use two benchmark propagation models that differ in a few free parameters, focusing on cosmic-ray convection (\conv) and re-acceleration (\reacc), respectively:

\begin{itemize}
    \item \texttt{conv}: The \textit{convection} propagation model includes the convection velocity through the parameter $v_{0, c}$, which describes the bulk motion of cosmic rays due to Galactic winds, perpendicular to the Galactic plane. In this model, re-acceleration is absent, and primary species have a single power-law injection spectrum without any breaks. The slope of the spectrum is $\gamma_{1,p}$ for protons, $\gamma_{\rm 1,He}$ for Helium and $\gamma_{\rm 1,CNO}$ for all other primary nuclei. This model also features a break in the diffusion spectrum, $R_{D, 1}$, with a prior of $(2 - 6) \times 10^{3}$~MV and a corresponding smoothing factor of the spectrum, $s_{D, 1}$. Note that also a second diffusion break is included at higher rigidities of about $200 \times 10^{3}$~GV, with $R_{D, 2}$ and $s_{D, 2}$, as a standard free parameter for both our propagation models.
    
    \item \texttt{reacc}: In the \textit{re-acceleration} propagation model, the Alfv\'en velocity, $v_A$, that determines the magnitude of re-acceleration of low-energetic cosmic rays, is a free parameter with prior $0 - 35$~km/s. Convection is not included. Furthermore, the additional low-energy break in the diffusion spectrum is not present. Instead, we introduce a break in the injection spectrum, $R_{\rm inj}$, with prior $(3 - 15) \times 10^3$~MV, and a corresponding smoothing factor, $s_{\rm inj}$. The break is the same for all the species, while three more parameters corresponding to the three injection slopes below the break are introduced, $\gamma_{0,p}$ for protons, $\gamma_{\rm 0,He}$ for Helium and $\gamma_{\rm 0,CNO}$ for the other primary nuclei.
\end{itemize}

Our choice to distinguish between models with and without re-acceleration is due to the uncertainties related to this process. The role of re-acceleration has been debated in the recent literature, and it is unclear to what extent it contributes to (re-)accelerating cosmic rays at energies of a few to tens of GeV~\cite{Drury:2016ubm}.

\section{Solar modulation of Galactic Cosmic Rays}
\label{sec: solar modulation models}

\subsection{General Framework}

Galactic cosmic-rays entering the heliosphere experience significant,
time-dependent modifications to their spectra (and, to a lesser extent, composition)
at kinetic energies below $\mathcal{O}(30$--$40)\,\mathrm{GeV}$.
These effects arise from transport through the expanding solar wind and the
large-scale heliospheric magnetic field (HMF), superimposed with a turbulent
magnetic component that governs particle scattering.
Solar modulation is therefore controlled by solar-cycle conditions and depends on
rigidity $R\equiv pc/|q|$ (rather than energy alone), particle speed $\beta$,
and, crucially, the sign of the electric charge
\cite{Parker1958,Parker1965,Jokipii1971,2013LRSP...10....3P}.
As a result, cosmic-ray spectra measured near Earth differ from the LIS fluxes, and an explicit heliospheric transport model is required whenever
low-energy data are used for precision physics, including antimatter studies and
dark-matter searches.

The standard description is provided by the Parker transport equation, a
Fokker--Planck equation for the phase-space density $f(\vec{r},p,t)$
\cite{Parker1958,Parker1965}:

\begin{equation}
  \frac{\partial f}{\partial t}
  =
  - (\vec{V}_{\rm sw} + \vec{v}_{\rm d}) \cdot \nabla f
  + \nabla \cdot \left( \mathbf{K} \cdot \nabla f \right)
  + \frac{1}{3} (\nabla \cdot \vec{V}_{\rm sw})\,
  p\,\frac{\partial f}{\partial p}.
  \label{eq:parker}
\end{equation}

Here, $\vec{V}_{\rm sw}$ is the solar-wind velocity (typically outward and
approximately radial), $\mathbf{K}$ is the diffusion tensor (encoding scattering
on magnetic turbulence), and $\vec{v}_{\rm d}$ is the drift velocity associated
with the antisymmetric part of $\mathbf{K}$, accounting for gradient/curvature
drifts and current-sheet drifts.
The last term describes adiabatic energy changes due to the diverging solar wind,
which generally act as energy losses for cosmic rays propagating inward.
Although all terms are always present, their relative importance depends strongly
on rigidity and on whether the heliosphere is in a solar-minimum or solar-maximum
state \cite{Jokipii1971,2013LRSP...10....3P}.

\textit{Diffusion:}
Spatial diffusion, described by the symmetric part of $\mathbf{K}$, is controlled
by scattering on the turbulent HMF and is usually modelled with different parallel
and perpendicular components with respect to the large-scale field.
In practice, modulation at fixed rigidity increases when the diffusion
normalization decreases (stronger scattering), or when the HMF magnitude and
turbulence level increase.
Modulation is, thus, stronger during solar maximum when the
enhanced solar activity tends to increase magnetic irregularities, reducing the
effective mean free path and thereby lowering the diffusion coefficient
\cite{2013LRSP...10....3P, VosPotgieter2015}.
Conversely, modulation is weaker during solar minimum when the field is more ordered and diffusion is more
efficient.
Independently of the solar cycle, diffusion effects rapidly weaken above
$\sim 10 - 20$~GV, where heliospheric transport becomes subdominant and the
observed spectra approach the LIS.

\textit{Convection and adiabatic energy losses: }
Convection by the solar wind and adiabatic losses are largely charge independent
and together set a baseline modulation that is always present, most relevant at
rigidities below a few GV~\cite{Parker1965,1968ApJ...154.1011G}.
Increasing $V_{\rm sw}$ (or, more generally, the effective residence time in the
expanding wind) enhances adiabatic losses and steepens the low-energy suppression.
These terms are therefore particularly important when modelling the overall
softening of low-rigidity spectra, even in periods when charge-sign effects are
weak.

\textit{Drifts and charge-sign dependence:}
A distinctive feature of Eq.~\eqref{eq:parker} is particle drift, which introduces
a strong dependence on the sign of the charge $q$ \cite{JokipiiLevyHubbard1977}.
The drift pattern depends on the global HMF polarity, commonly denoted by $A$,
which reverses approximately every 11 years.
In an $A>0$ polarity epoch (defined as the HMF lines pointing outwards at the solar north pole, and inwards at the south
pole),
positively charged particles preferentially drift inward through the polar
regions, where they can reach the
inner heliosphere  efficiently, resulting in a weak
modulation.
On the other hand, negatively charged particles enter more efficiently through the equatorial plane along the
heliospheric current sheet (HCS), \textit{i.e.}, the thin border between positive and negative polarity of the
magnetic field. The HCS can be highly warped, and its geometry---often summarized by a
tilt angle that increases from solar minimum to maximum---strongly inhibits propagation and increases energy
losses for particles whose access to the inner heliosphere
is governed by current-sheet trajectories~\cite{2013LRSP...10....3P, VosPotgieter2015}, reducing the flux of these particles near Earth. Conversely, the situation is flipped during negative polarity ($A < 0$); positively charged particles will be more modulated, and negatively charged particles less.

\textit{Solar minimum vs.\ maximum:}
The solar-cycle dependence can be summarized as follows:
(i) during \emph{solar minimum}, the HMF is more ordered and the HCS is less
tilted, so drift effects are comparatively large and the modulation becomes
more charge-sign dependent; at the same time, diffusion is more efficient,
reducing the overall modulation strength at a given rigidity.
(ii) during \emph{solar maximum}, enhanced turbulence suppresses drifts and lowers
the effective diffusion coefficient, leading to stronger overall modulation but a
more charge-symmetric behavior~\cite{2013LRSP...10....3P}.
Accordingly, the most pronounced particle/antiparticle differences are expected
at low rigidities (typically $\lesssim 5-10$~GV) during solar minima, and
especially when comparing data taken in opposite polarity epochs.
These expectations are borne out by time-dependent measurements from PAMELA and
AMS-02, which show both strong temporal evolution of the absolute fluxes and
clear signatures of charge-sign dependent modulation in appropriate phases of
the solar cycle~\cite{AdrianiEtAl2013PAMELAProtonTime, AguilarEtAl2018AMSProtonHeliumTime, Corti:2015bqi}.

Putting together the above considerations, this means that during our defined \preMAX phase, when $A<0$, protons are more modulated than antiprotons. In the \postMAX phase, when $A>0$ the opposite is true, and the antiproton flux is more suppressed than the proton flux. This can be seen in Fig.~\ref{fig: ssn} where, lasting for about one year from 2015 to 2016, the proton flux rises much more quickly than the antiproton flux, highlighting the transition between the two regimes. The above considerations will thus guide our expectations for the level of modulation of protons and antiprotons derived from the results of the fits performed in the three periods.

\subsection{Solar Modulation Models}

Numerical solutions of Eq.~\eqref{eq:parker} with realistic heliospheric inputs, including the HMF
magnitude, HCS tilt evolution, and drift suppression by turbulence, are implemented
in dedicated frameworks \cite{2013LRSP...10....3P,Maccione:2012cu,Kappl:2015hxv,  Boschini:2017gic, Boschini:2019ubh}.
However, these approaches are typically time-consuming and numerically expensive.  For this reason, alternative simplified frameworks have been developed.
The most commonly employed approach is the force-field approximation, which compresses the combined effects of diffusion, convection, and adiabatic losses into a single modulation  potential. This neglects drift effects and assumes steady-state homogeneous and spherically symmetric propagation~\cite{1968ApJ...154.1011G}. However, since this approach does not include particle drifts, it also does not describe charge-sign dependence and polarity effects. Nonetheless, charge-sign effects can be re-introduced in an effective phenomenological way using separate potentials for positively and negatively charged particles. Further extensions of the force-field approach consider a rigidity-dependent potential~\cite{Cholis:2020tpi, Zhu:2024ega, Kuhlen:2019hqb}.

In the following, we will thus study the impact of solar modulation and the chosen solar-modulation models by carrying out our analysis for three different solar modulation models: (1) the simple and widely used force-field potential~\cite{1968ApJ...154.1011G}, (2) an extended force-field potential~\cite{Zhu:2024ega} that allows for a rigidity dependence of the solar modulation potential, and (3) by applying the \texttt{HelMod} model~\cite{Boschini:2017gic, Boschini:2019ubh, Boschini:2022jwz}. Below, we briefly describe each model.

\textit{The force-field potential} relates the locally measured flux, $\Phi_\text{loc}$, to the flux in the local interstellar medium (\textit{i.e.}, the LIS), $\Phi_\text{LIS}$, through~\cite{1968ApJ...154.1011G} 

\begin{equation}\label{eq: force field}
\begin{aligned}
\Phi_\text{loc}\left(E\right) &= \Phi_\text{LIS}\left(E + \phi Z/A \right) \\ & \times \frac{E \times \left(E + 2 m_p\right)}{\left(E + \phi Z/A\right) \times \left(E + \phi Z/A + 2m_p\right)},
\end{aligned}
\end{equation}

\noindent where $E$ is the kinetic energy per nucleon, $Z$ and $A$ are the charge and mass number of the nuclei, $m_p$ is the proton mass, and $\phi$ the solar modulation potential. In our fit, this corresponds to two free parameters, $\phi_p$ for positively charged cosmic rays, and $\phi_{\bar{p}}$ for negatively charged cosmic rays (\textit{i.e.} antiprotons, as we do not fit to electron (or positron) data).

\textit{The extended model} modifies the force-field potential $\phi$ in Eq.~\ref{eq: force field} by introducing a rigidity-dependence on the potential $\phi_\text{Ext}(R)$, defined by~\cite{Zhu:2024ega}

\begin{equation}\label{eq: zhu model}
\phi_\text{Ext}\left(R\right) = \phi_l + \frac{\phi_h - \phi_l}{1 + \exp{\left(-R + R_{br}\right)}},
\end{equation}

\noindent where $\phi_l$ and $\phi_h$ are the modulation potential at low and high rigidities, respectively, and $R_{br}$ is a break in rigidity separating the two potentials. In our fits, this results in 5 free parameters: a low and high potential for positively and negatively charged cosmic rays denoted by $\phi_{l, p}$, $\phi_{h, p}$, $\phi_{l, \bar{p}}$ and $\phi_{h, \bar{p}}$, respectively, as well as the break $R_{br}$, which is applied to both negatively and positively charged cosmic rays simultaneously. We note that, since this is a phenomenological model, it is not clear from a physical point of view if this break is expected to be the same or different for positively and negatively charged nuclei. We therefore apply a universal break, to keep the number of free parameters in our fits as modest as possible.

\textit{The heliospheric modulation model}, \texttt{HelMod}~\cite{Boschini:2017gic, Boschini:2019ubh, Boschini:2022jwz}, is a cosmic-ray propagation code based on the full solution of Eq.~\ref{eq:parker}. While a full simulation is complex, the \texttt{HelMod} team provides matrices that allow a simple calculation of solar modulation from a LIS, for specific fits to AMS-02 datasets. We make use of the \texttt{HelMod} matrices corresponding to our selected AMS-02 datasets in our analysis, but note that these matrices were developed for an assumed LIS \cite{Boschini:2017fxq,Boschini:2018baj,Boschini:2020jty}. This means that, since we fit our own propagation model, \texttt{HelMod} cannot provide fully self-consistent modulation in our analysis. Despite this, \texttt{HelMod} is the only solar modulation model, among the three we consider, that fully models cosmic-ray propagation in the heliosphere, and thus offers an important comparison to the other solar modulation models investigated in our analysis. Note that for the \texttt{HelMod} fits, there is no free parameter describing solar modulation.

\bigskip\noindent
We conduct time-integrated fits for the force-field, extended model and \texttt{HelMod}. We also perform time-dependent fits for the force-field and extended solar modulation models. Due to the limitations described above, however, we do not perform a time-dependent fit with \texttt{HelMod}. For these time-dependent fits, we separately and independently fit to each of the three time periods described in Section~\ref{sec: cosmic-ray data}, \textit{i.e.} \texttt{preMAX}, \texttt{MAX} and \texttt{postMAX}, which allows us to investigate each modulation model for different configurations of solar activity. Furthermore, this enables us to study in detail the effects of the chosen solar modulation model on the propagation parameters.

\section{Results and Discussion}
\label{sec: results and discussion}

\begin{figure*}[tbp]
    \begin{minipage}[t]{0.48\linewidth}
        \centering
        \includegraphics[width=1\linewidth]{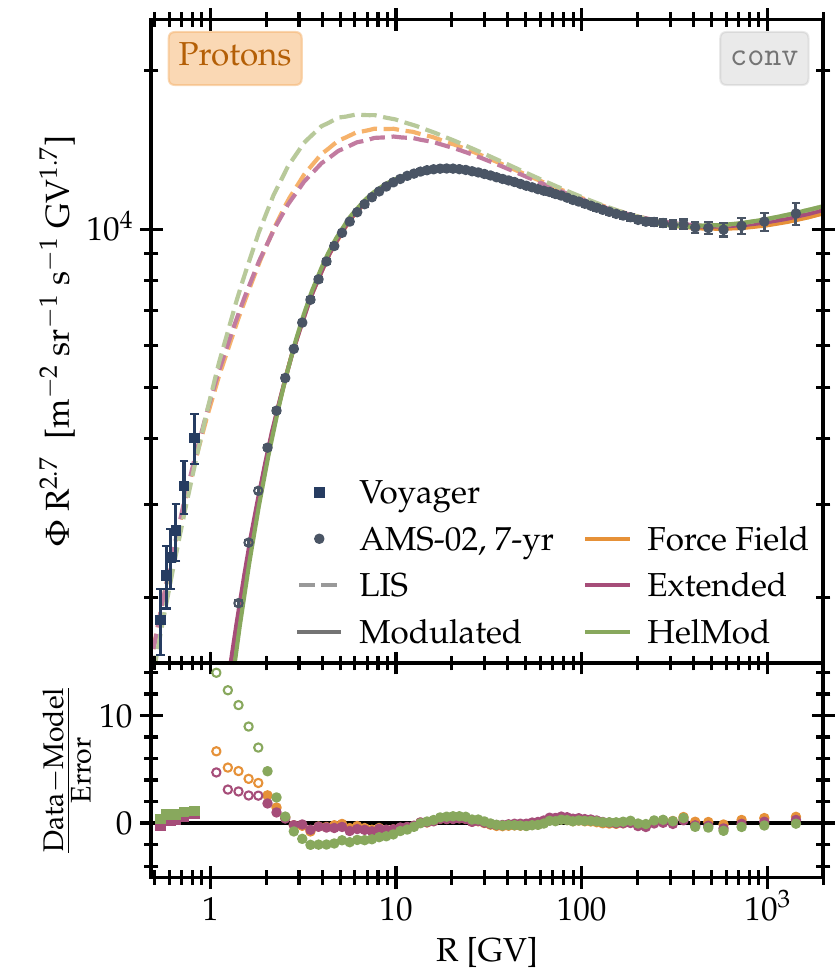}
    \end{minipage}
    \hfill
    \begin{minipage}[t]{0.48\linewidth}
        \centering
        \includegraphics[width=1\linewidth]{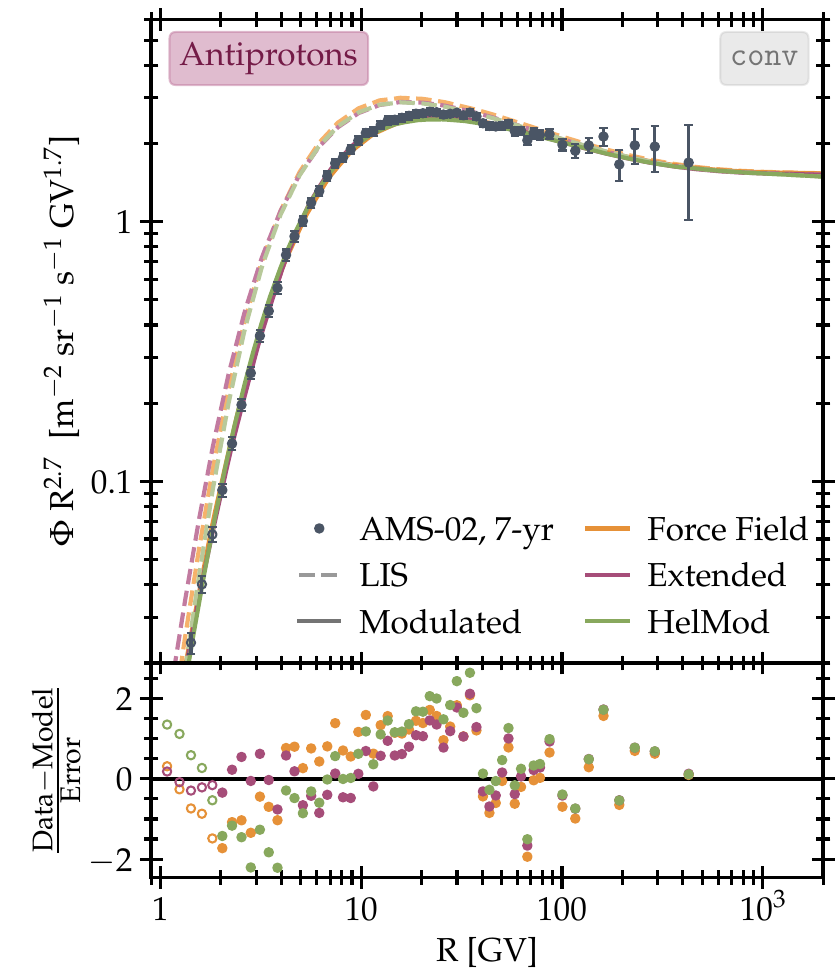}
    \end{minipage}
    \caption{The spectra for protons (left panel) and antiprotons (right panel) for the \texttt{conv} propagation model. In each plot, the LIS (dashed line) and modulated flux (solid) line are given for a different solar-modulation model: the force-field potential (orange), the extended force-field potential (burgundy) and the \texttt{HelMod} model (green). The residuals in each plot are compared to the 7-yr AMS-02 data (circle markers), as well as the Voyager data for protons (square markers)}.
    \label{fig: LIS conv time-independent}
\end{figure*}

In the following, we mainly focus on the consequences of our results for solar modulation. The full results for each fit are reported in Appendix~\ref{app: propagation models}, both in tabular form and in graphical form. We choose to focus on the results for protons and antiprotons, but provide further plots for the other cosmic-ray nuclei in Appendix~\ref{app: fluxes} and~\ref{app: SM potential time-independent fit}.

\begin{table*}[tbp]
\centering
\begin{tabular}{rr||c|c|c|c||c|c|c|c}
   & & \multicolumn{4}{c||}{Force-Field Potential} & \multicolumn{4}{c}{Extended Model} \\
   &  & \verb|preMAX| & \verb|MAX| & \verb|postMAX| & 7-yr t.-i. & \verb|preMAX| & \verb|MAX| & \verb|postMAX| & 7-yr t.-i. \\
    \cmidrule{2-10}
  \multirow{7}{*}{\texttt{conv} $\left.\begin{array}{l}
                \\
                \\
                \\
                \\
                \\
                \\
                \\
                \end{array}\right\lbrace $} 
    &  $\phi_p$ [GV]           & 0.52$^{+0.01}_{-0.01}$  & 0.68$^{+0.01}_{-0.02}$  & 0.31$^{+0.01}_{-0.01}$  & 0.47$^{+0.01}_{-0.01}$  & --    & --    & --    & -- \\
    & $\phi_{\bar{p}}$ [GV]    & 0.59$^{+0.03}_{-0.02}$  & 0.75$^{+0.03}_{-0.04}$  & 0.54$^{+0.03}_{-0.02}$  & 0.77$^{+0.03}_{-0.02}$  & --    & --    & --    & -- \\
    \cmidrule{2-10}
    & $\phi_{l, p}$ [GV]       & -- & --    & --    & -- & 0.52$^{+0.01}_{-0.01}$  & 0.67$^{+0.01}_{-0.01}$  & 0.31$^{+0.01}_{-0.01}$  & 0.47$^{+0.01}_{-0.01}$ \\
    & $\phi_{h, p}$ [GV]       & -- & --    & --    & -- & 0.50$^{+0.03}_{-0.03}$  & 0.60$^{+0.03}_{-0.03}$  & 0.28$^{+0.03}_{-0.03}$  & 0.41$^{+0.03}_{-0.03}$ \\
    & $\phi_{l, \bar{p}}$ [GV] & -- & --    & --    & -- & 0.82$^{+0.07}_{-0.07}$  & 1.02$^{+0.05}_{-0.09}$  & 0.81$^{+0.06}_{-0.07}$  & 0.95$^{+0.05}_{-0.01}$ \\
    & $\phi_{h, \bar{p}}$ [GV] & -- & --    & --    & -- & 0.34$^{+0.04}_{-0.04}$  & 0.47$^{+0.03}_{-0.04}$  & 0.27$^{+0.04}_{-0.04}$  & 0.54$^{+0.03}_{-0.02}$ \\
    & $R_{br}$ [GV]            & -- & --    & --    & -- & 4.86$^{+0.54}_{-0.51}$  & 5.03$^{+0.62}_{-0.43}$  & 4.64$^{+0.41}_{-0.38}$  & 3.95$^{+0.31}_{-0.32}$ \\
  \cmidrule{2-10}
  
  \cmidrule{2-10}
  \multirow{7}{*}{\texttt{reacc} $\left.\begin{array}{l}
                \\
                \\
                \\
                \\
                \\
                \\
                \\
                \end{array}\right\lbrace $} 
   & $\phi_p$ [GV]            & 0.65$^{+0.01}_{-0.01}$  & 0.81$^{+0.01}_{-0.01}$  & 0.40$^{+0.01}_{-0.01}$  & 0.65$^{+0.01}_{-0.01}$  & --    & --    & --    & -- \\
   & $\phi_{\bar{p}}$ [GV]    & 0.36$^{+0.03}_{-0.03}$  & 0.49$^{+0.05}_{-0.04}$  & 0.38$^{+0.03}_{-0.03}$  & 0.55$^{+0.06}_{-0.04}$  & --    & --    & --    & -- \\
   \cmidrule{2-10}
   & $\phi_{l, p}$ [GV]       & -- & --    & --    & -- & 0.55$^{+0.01}_{-0.02}$  & 0.70$^{+0.02}_{-0.02}$  & 0.34$^{+0.01}_{-0.01}$  & 0.54$^{+0.02}_{-0.02}$ \\
   & $\phi_{h, p}$ [GV]       & -- & --    & --    & -- & 0.28$^{+0.03}_{-0.04}$  & 0.37$^{+0.04}_{-0.03}$  & 0.12$^{+0.01}_{-0.02}$  & 0.29$^{+0.03}_{-0.04}$ \\
   & $\phi_{l, \bar{p}}$ [GV] & -- & --    & --    & -- & 1.11$^{+0.07}_{-0.06}$  & 1.45$^{+0.05}_{-0.02}$  & 1.12$^{+0.07}_{-0.09}$  & 1.00$^{+0.00}_{-0.00}$ \\
   & $\phi_{h, \bar{p}}$ [GV] & -- & --    & --    & -- & 0.42$^{+0.04}_{-0.05}$  & 0.58$^{+0.04}_{-0.05}$  & 0.40$^{+0.05}_{-0.04}$  & 0.64$^{+0.03}_{-0.04}$ \\
   & $R_{br}$ [GV]            & -- & --    & --    & -- & 3.27$^{+0.13}_{-0.13}$  & 2.96$^{+0.13}_{-0.13}$  & 3.13$^{+0.15}_{-0.18}$  & 3.37$^{+0.12}_{-0.10}$ \\
\end{tabular}
\caption{Best-fit solar-modulation parameters for the force-field potential ($\phi_p$, $\phi_{\bar{p}}$) and the extended model ($\phi_{l, p}$, $\phi_{h, p}$, $\phi_{l, \bar{p}}$, $\phi_{h, \bar{p}}$, $R_{br}$) for the 3 time periods, as well as the time-independent fits of the 7-yr time-integrated (7-yr t.-i.) data. Upper rows represent the values for the \conv, while lower rows for the \reacc model. Note that there are no free solar-modulation parameters in the \texttt{HelMod} fits.}
\label{tab: solar modulation parameters}
\end{table*}

Figure~\ref{fig: LIS conv time-independent} shows results for the three solar modulation models in the 7-yr time-integrated fits for the \conv propagation model. The left panel shows the proton fluxes and the right panel the antiproton fluxes. Dashed lines represent LIS fluxes, while solid lines are fluxes that are modulated using the force-field potential (orange), the extended model (burgundy) and \texttt{HelMod} (green). The corresponding  AMS-02 data is given in dark grey, where empty markers correspond to data points that are not included in the fit. The bottom of each plot shows the residuals between the data and the modulated fluxes. For protons, both the force-field and extended model fit the data very well to within $\sim 1\,\sigma$, only showing some worsening of the fit below $\sim 3$~GV, and provide very similar LISes. The \texttt{HelMod} case, instead, provides a significantly worse fit and a sizeable different LIS. For the antiproton fluxes, the results are more consistent between the three modulation models, both in the LIS fluxes, which are very similar, and the residuals. This is somewhat expected as cosmic-ray antiprotons are produced as secondaries in interactions of cosmic-ray protons with energies of $\sim 8$~GeV or higher, where the proton LIS fluxes are only weakly affected by solar modulation. We also recover an antiproton excess at $\sim 20 - 30$~GV, as in previous studies~\cite{Cuoco:2016eej, Cuoco:2019kuu, Cholis:2019ejx, Calore:2022stf, Cui:2016ppb, Duan:2025ead, Cirelli:2014lwa}. We do not attempt here to quantify the significance of this excess, a task that requires a specific analysis and is postponed to a dedicated study.

It follows from Fig.~\ref{fig: LIS conv time-independent}, that in principle, \texttt{HelMod} provides a detailed modelling of cosmic-ray propagation in the heliosphere and is able to capture many processes that are overlooked in the effective solar modulation models. However, in our scenario the \texttt{HelMod} fit shows the strongest discrepancy below about 10~GV in the proton fluxes, significantly worse than the force-field and extended model. We believe this is due to the fact that the \texttt{HelMod} modulation computation provided by the \texttt{HelMod} team is based on an underlying LIS that was optimised together with \texttt{Galprop}~\cite{Boschini:2022jwz}. In particular, in~\cite{Boschini:2022jwz} the injection spectra of the primary nuclei are parametrized differently than in our study and allow for a larger freedom. This highlights again that the proper implementation the \texttt{HelMod} model into our framework would require to re-derive simultaneously solar modulation and cosmic-ray propagation in a self-consistent way. This is, however, a non-trivial task which is left for future studies.

In Figure~\ref{fig: LIS reacc time-independent} in Appendix~\ref{app: fluxes}, we present the 7-yr fits also for the \reacc propagation model, where we reach similar conclusions.

Table~\ref{tab: solar modulation parameters} summarises the best-fit values of the solar modulation parameters from our analysis. As expected, across all time-dependent scenarios, solar modulation is strongest during the \texttt{MAX} period, and weakest during the \texttt{postMAX} phase, which includes the minimum of solar cycle 24 in 2020. The \texttt{preMAX} phase, only covering a couple of years before the maximum phase, shows intermediate modulation, accordingly. 

Looking at the values of the potential for the \texttt{conv} case and extended solar modulation in Tab.~\ref{tab: solar modulation parameters}, it can be seen that during the \texttt{preMAX} and \texttt{postMAX} phases, the lower and higher potential for protons ($\phi_{l, p}$ and $\phi_{h, p}$) are very similar, while they are incompatible during the \texttt{MAX} period. In practice, this implies that the simple force-field is a good approximation during phases of lower solar activity, while the more complex extended modulation is required during the solar maximum phase. This picture, however, is not confirmed for antiprotons, where the lower potential, $\phi_{l, \bar{p}}$, is significantly stronger than the higher potential, $\phi_{h, \bar{p}}$, across all fits, with a rigidity break $R_{br} \sim 5$~GV, that remains relatively consistent throughout the three time periods.

\begin{figure}[tbp]
    \centering
    \includegraphics[width=1\linewidth]{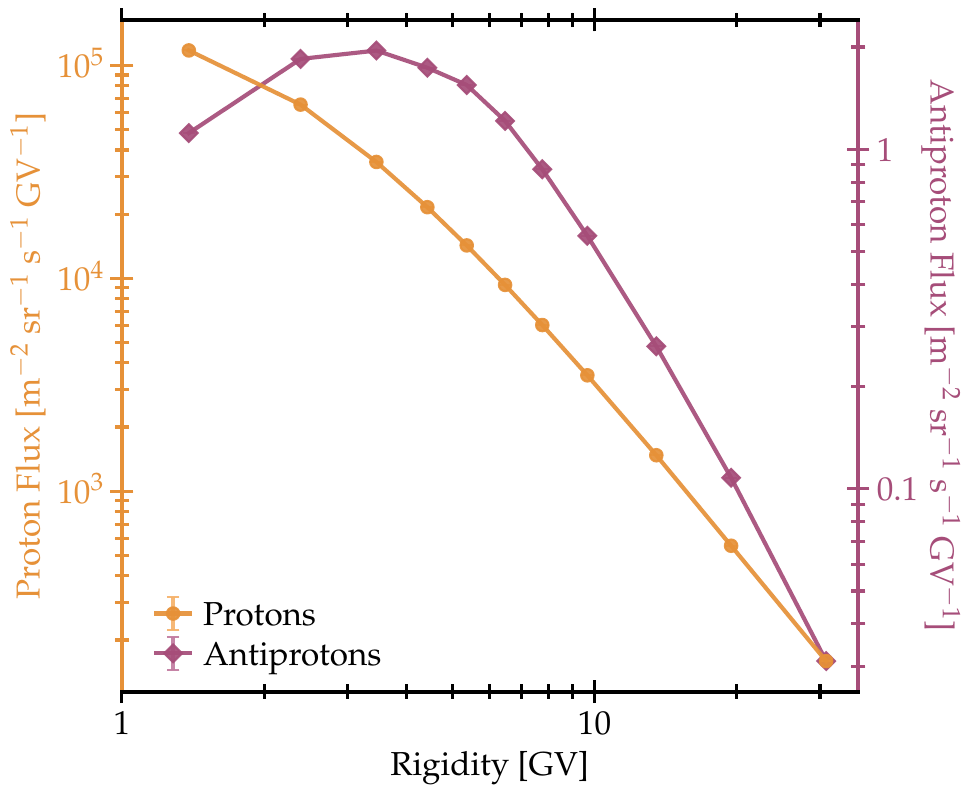}
    \caption{The cosmic-ray proton (orange circle markers) and antiproton (burgundy diamond markers) fluxes from \mbox{AMS-02}~\cite{AMS:2025npj, AMS:2025pgu}. The combined flux from Bartels rotation 2426 to 2560 is shown.}
    \label{fig: flux vs rigidity}
\end{figure}

We consider two possible explanations for this behaviour: (1) A first possibility is that the antiproton LIS is incorrect. This, indeed, can be due to an incorrect propagation model, or more likely to an incorrect modelling of the antiproton production cross-section. Although we include the overall normalization of the cross-section as a nuisance parameter in the fit, other sizeable uncertainties related to the shape in energy remain~\cite{diMauro:2014zea, Donato:2017ywo, Korsmeier:2018gcy, Maurin:2025gsz, Genolini:2018ekk, Genolini:2023kcj,Cuoco:2019kuu}. (2) A second possibility, which we believe is more likely, is a failure of the force-field approximation. This could be related to the peculiar LIS of the antiprotons, which has a peaked shape at few GVs, unlike other species which instead have a power-law-like spectrum. This is illustrated in Figure~\ref{fig: flux vs rigidity}, which shows the proton and antiproton fluxes as measured by AMS-02~\cite{AMS:2025npj, AMS:2025pgu}. This peculiar antiproton spectrum is due to the secondary nature of the species which is produced mainly via the reaction ${p \, + \, H \rightarrow p \, + \, p \, + \,  p \, + \, \bar{p} }$, which, with a high threshold of about 8~GeV, can only produce small amounts of low-energy antiprotons. A possibility is thus that the force-field approximation does not work well when applied to this peaked spectrum. Indeed, this seems to be confirmed by various considerations: (1) the very close coincidence between the position of the modulation break $R_{br}$ and the peak of the spectrum at about 5~GV, (2) the fact that the force field gives poor antiproton residuals below $\sim 10$~GV of $\sim 1.5\sigma$, while the extended modulation model improves the quality of the fit significantly with $\sim 0.7\sigma$ (see Fig.~\ref{fig: LIS conv time-independent}). On the other hand, the \texttt{Helmod} model, which, in principle, does include a correct treatment of the antiproton modulation, gives the worst residuals, at the level of $\sim 2.0\sigma$, which somewhat weakens this argument. Finally (3)  
a model-independent analysis by the AMS-02 collaboration~\cite{AMS:2025npj} has found that below $\sim 5$~GV the modulation of antiprotons has different characteristics than protons, electrons and positrons. In particular, the amplitude of the modulation of the latter three grows at lower rigidities, while it becomes constant for antiprotons below $\sim 5$~GV. These hints, thus, seem to point toward a failure of the force-field potential to describe antiproton modulation well, although a definitive confirmation will require further studies. 

The above picture is vastly different in the \reacc propagation case. In the extended modulation model, as can be seen in Tab.~\ref{tab: solar modulation parameters}, the potentials above and below the break are very different both for protons and antiprotons during all three time periods. In principle, this implies that the force-field approximation is always inaccurate, even during phases of low solar activity, at least 
when re-acceleration is included in propagation models.
We believe, however, that this is rather due to an unphysical result returned by the fit, as we describe in more detail further below.

\begin{figure*}[tbp]
    \begin{minipage}[t]{0.49\linewidth}
        \centering
        \includegraphics[width=1\linewidth]{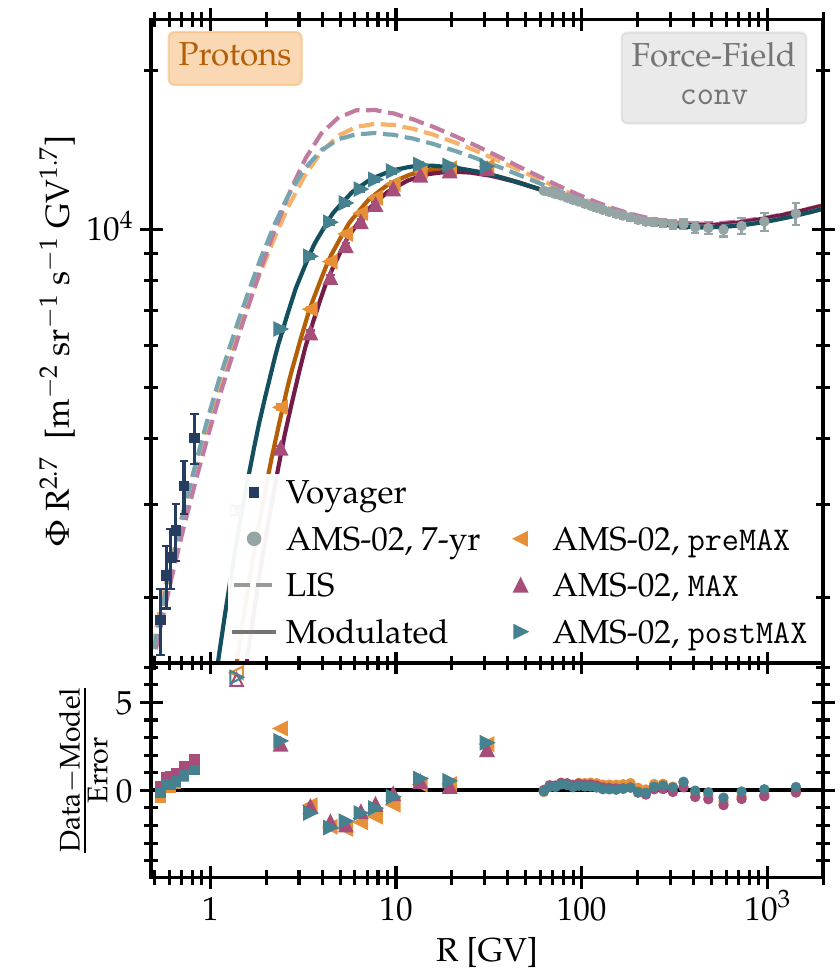}
    \end{minipage}
    \hfill
    \begin{minipage}[t]{0.49\linewidth}
        \centering
        \includegraphics[width=1\linewidth]{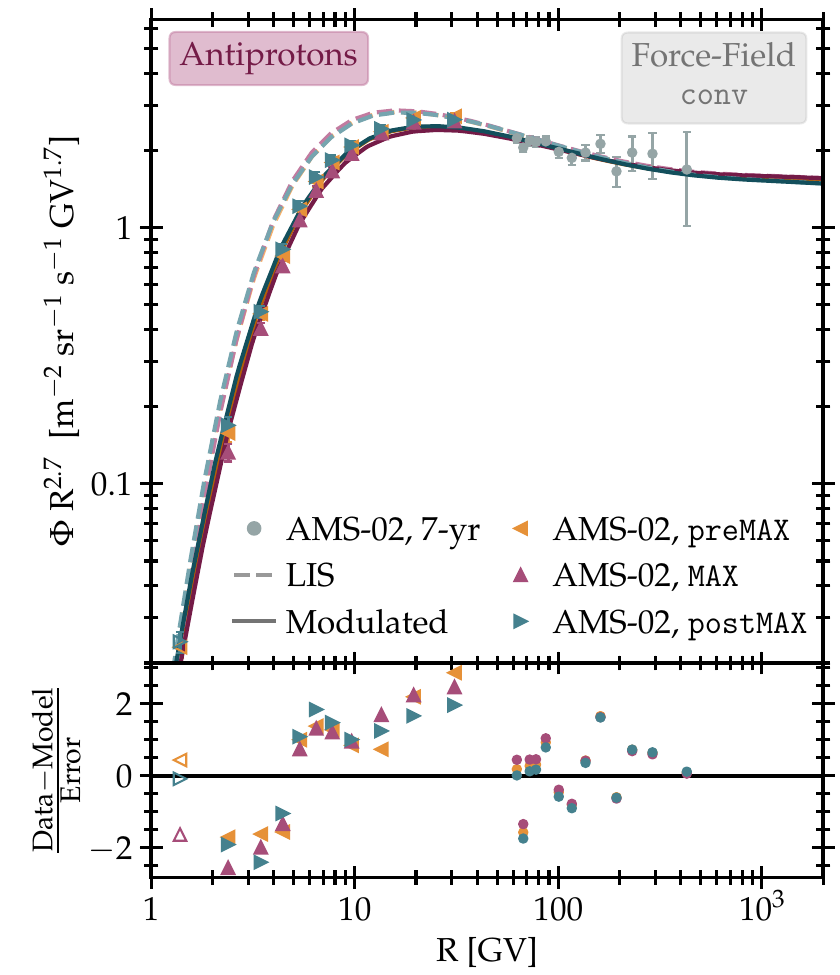}
        
    \end{minipage}
    \vfill\vspace{0.5cm}
    \begin{minipage}[t]{0.49\linewidth}
        \centering
        \includegraphics[width=1\linewidth]{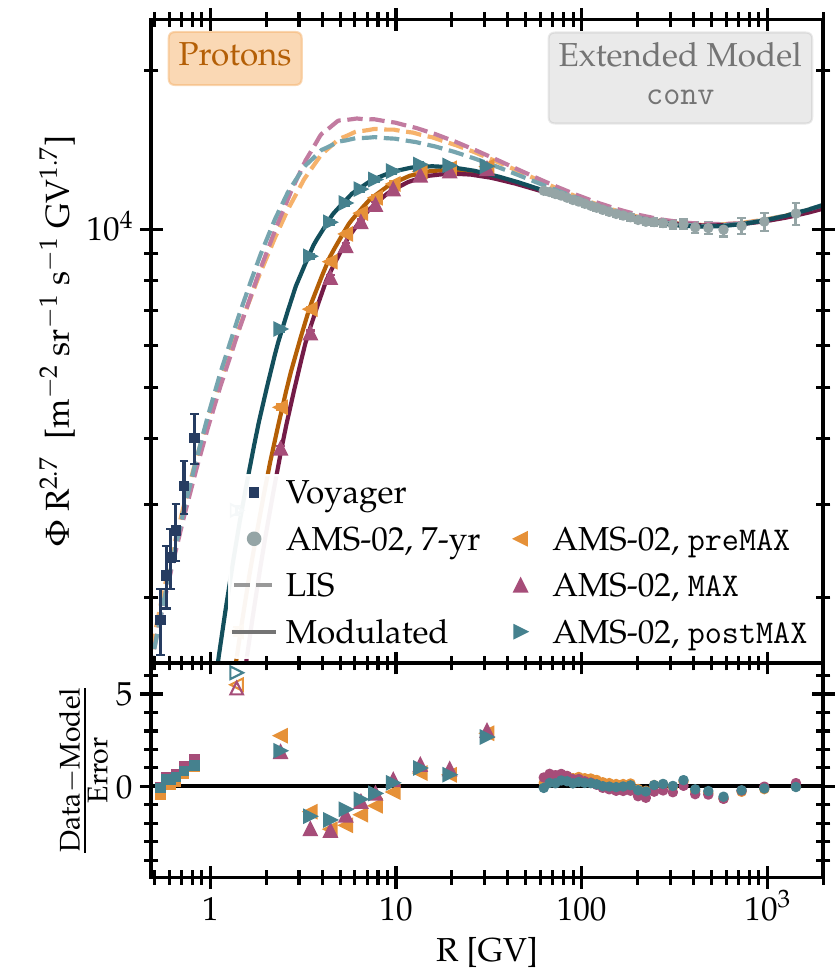}
    \end{minipage}
    \hfill
    \begin{minipage}[t]{0.49\linewidth}
        \centering
        \includegraphics[width=1\linewidth]{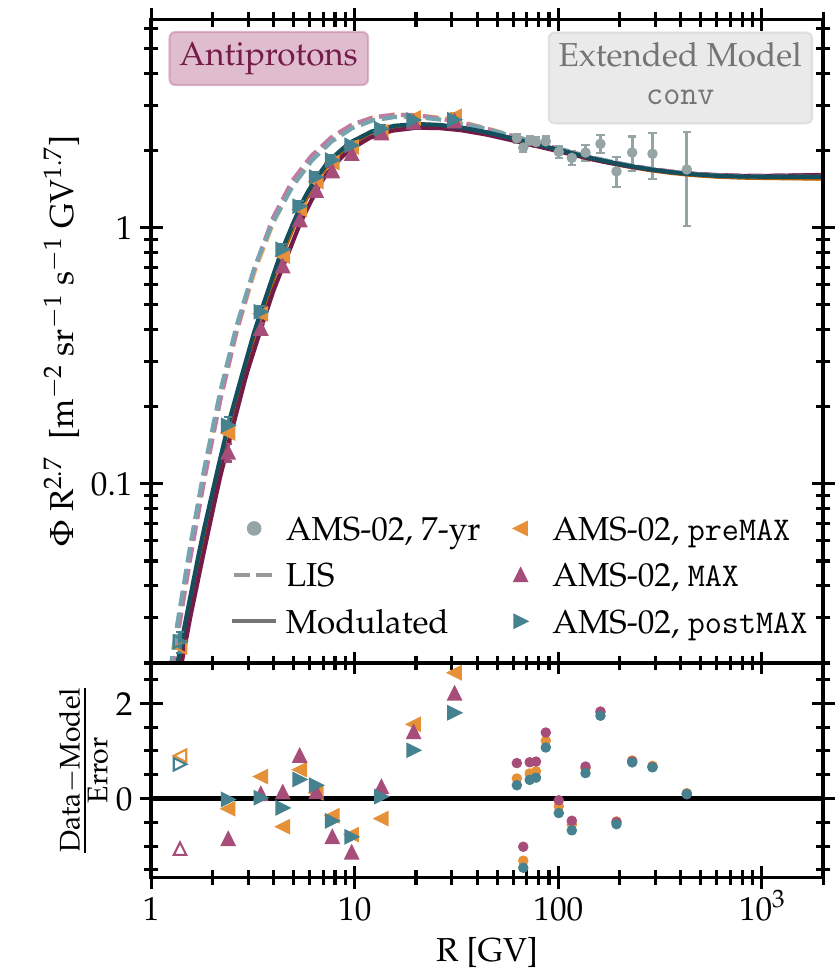}
    \end{minipage}
    \caption{The LIS spectra for protons (left panels) and antiprotons (right panels) using the force-field model (top panels) and extended force-field model (bottom panels) for solar modulation. The LISes are based on the \texttt{conv} propagation model.}
    \label{fig: LIS conv}
\end{figure*}

Figure~\ref{fig: LIS conv} shows the cosmic-ray proton and antiproton fluxes from our fits to the three time periods in the \texttt{conv} propagation model for the force-field and extended solar modulation. It can be noticed that the proton fit below 40~GV is significantly worse than what is achieved in the 7-yr dataset fit shown in Figure~\ref{fig: LIS conv time-independent}. This is mainly due to the fact that the time-dependent dataset for protons has only a few data points in rigidity, \textit{i.e.}, only 11 between 1 and 41~GV, which is insufficient given the large statistics and percent rigidity resolution of AMS-02. This is opposed to the 7-yr dataset, which has 40 data points in the same rigidity range. This issue fortunately is not present for the other nuclei, which instead have an adequate number of data points. For the future, we advocate the release of homogenous time-dependent datasets for all the species from the AMS-02 collaboration.

\begin{figure*}[tbp]
    \begin{minipage}[t]{0.49\linewidth}
        \centering
        \includegraphics[width=1\linewidth]{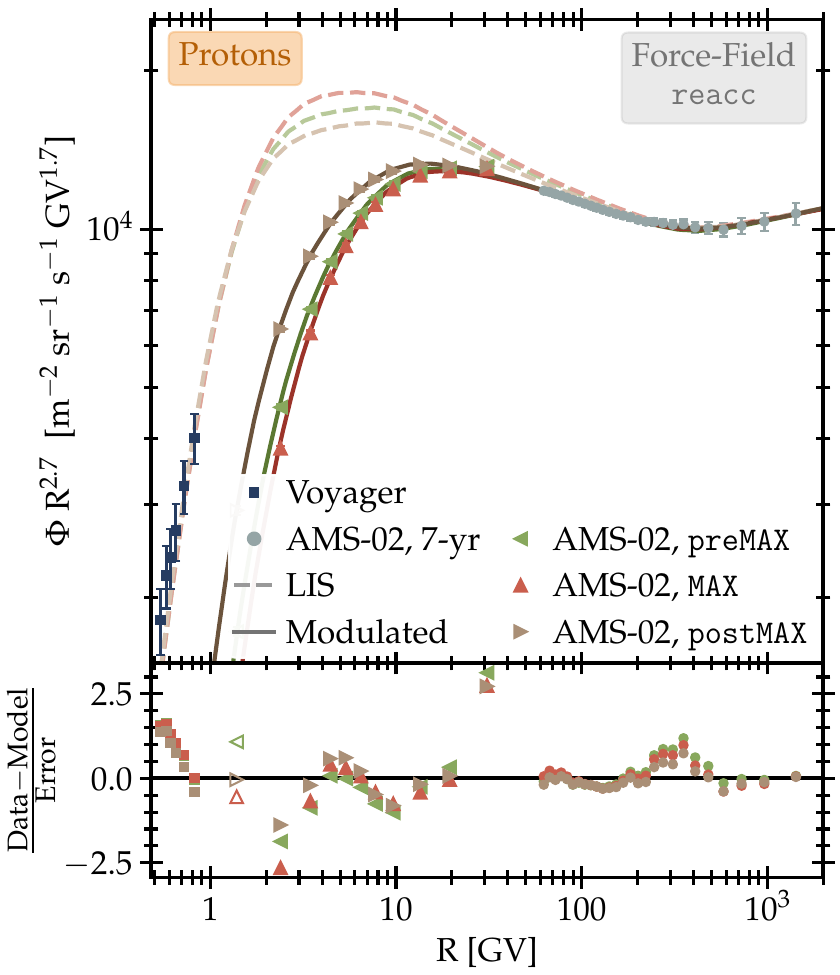}
    \end{minipage}
    \hfill
    \begin{minipage}[t]{0.49\linewidth}
        \centering
        \includegraphics[width=1\linewidth]{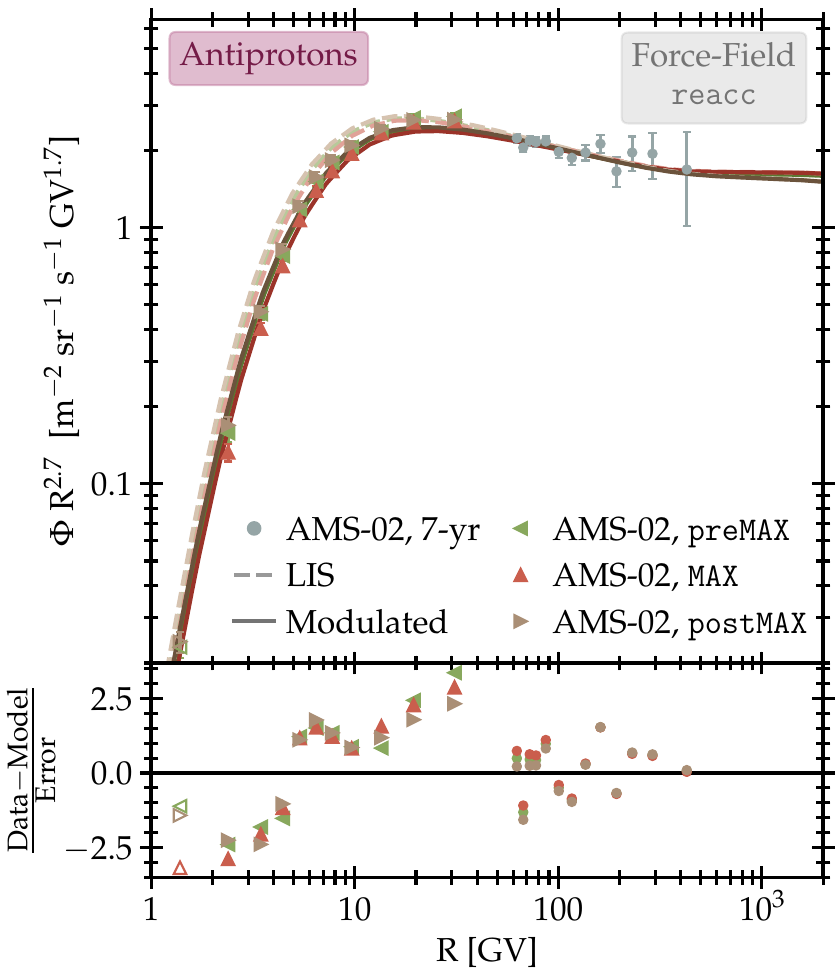}
        
    \end{minipage}
    \vfill\vspace{0.5cm}
    \begin{minipage}[t]{0.49\linewidth}
        \centering
        \includegraphics[width=1\linewidth]{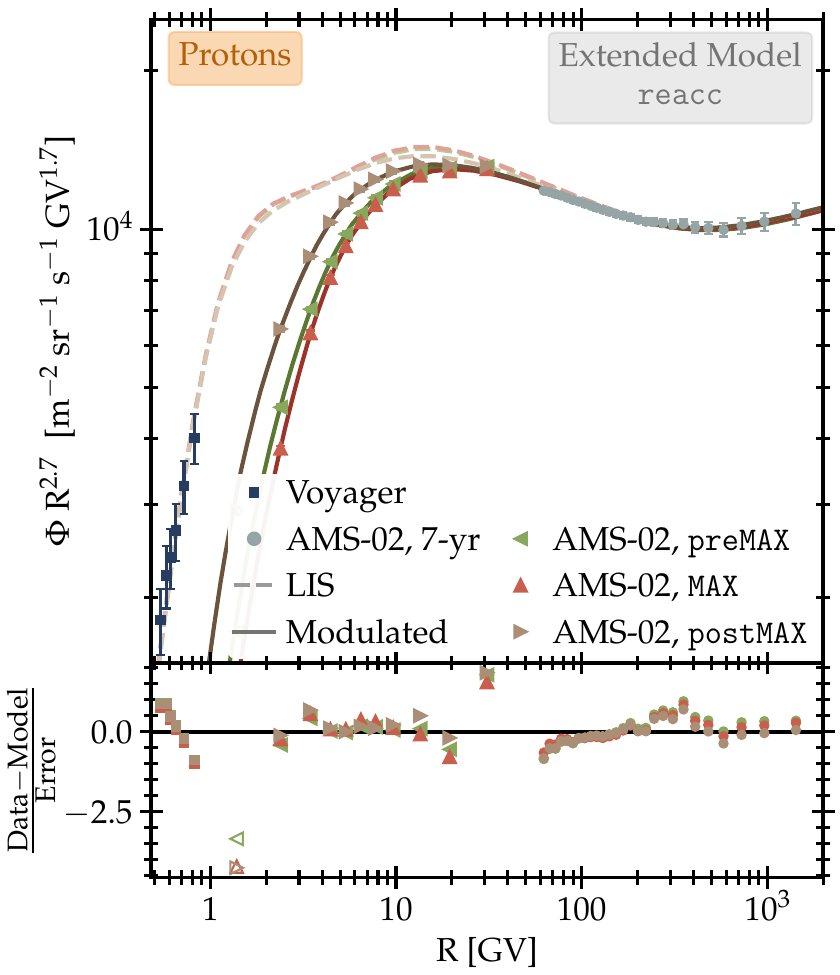}
    \end{minipage}
    \hfill
    \begin{minipage}[t]{0.49\linewidth}
        \centering
        \includegraphics[width=1\linewidth]{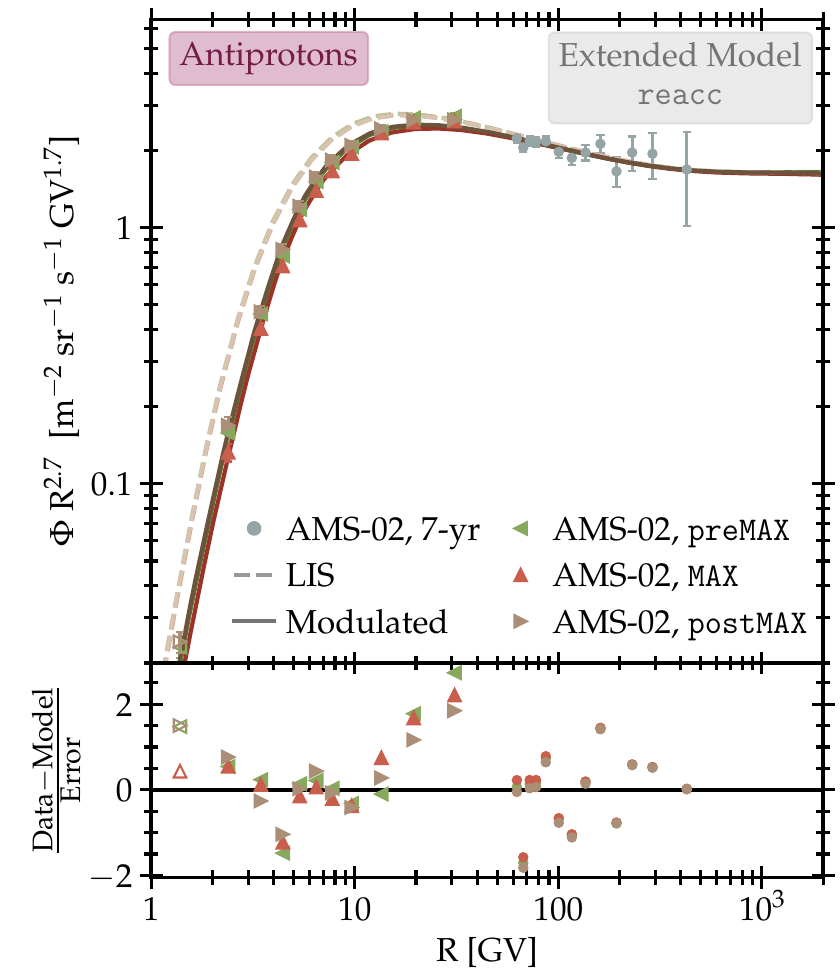}
    \end{minipage}
    \caption{The LIS spectra for protons (left panels) and antiprotons (right panels) using the force-field model (top panels) and extended force-field model (bottom panels) for solar modulation. The LISes are based on the \texttt{reacc} propagation model.}
    \label{fig: LIS reacc}
\end{figure*}

Besides the above issue, the most important observation from Fig.~\ref{fig: LIS conv} is that the proton LIS fluxes inferred in the three time periods show some discrepancy at a level of about 10\%. This is contrary to expectations, where the LIS fluxes should be exactly the same, as we do not expect any temporal changes of the LIS fluxes on comparable time scales. This discrepancy is more severe for the pure force-field modulation, while the agreement improves for the extended modulation. Although 10\% is not large in absolute terms, it is significantly larger than the percent-level statistical errors achieved by AMS-02. This discrepancy can be interpreted in two different ways: On one hand, this indicates that the current description of solar modulation is insufficient to self-consistently describe solar modulation effects over different phases of solar activity. In this respect, however, the extended models clearly provide a better modelling and should be preferred. On the other hand, this level of discrepancy can be understood as the current level of true, systematic uncertainty in the reconstructed LIS. Thus, the systematic uncertainties in propagation models are dominated by our limitation of solar modulation modelling. Finally, from Fig.~\ref{fig: LIS conv} it can also be seen that the antiproton LIS fluxes are, instead, more stable and less affected by solar modulation systematic uncertainties, in agreement with the discussion for Fig.~\ref{fig: LIS conv time-independent}.

Figure~\ref{fig: LIS reacc} is similar to Fig.~\ref{fig: LIS conv}, but instead shows the LIS fluxes obtained from the time-dependent fits to the \reacc propagation model. For protons in the force-field modulation case, the Fig.~\ref{fig: LIS reacc} confirms the same trend observed for the \conv case, \textit{i.e.}, a disagreement of the LIS fluxes reconstructed in the three time periods, at the level of 10-15\%. On the other hand, the LIS fluxes from the extended modulation model show a remarkable agreement. This, in principle, would indicate the ability of the extended modulation model to correctly describe solar modulation both in periods of high and low solar activity. However, the reconstructed LIS has a peculiar shape with a kink at around $\sim 4$~GV. Furthermore, this LIS also strongly differs from the LIS inferred through the force-field modulation. Additionally, the overall level of modulation is relatively low, even during the solar maximum, as can be seen from Tab.~\ref{tab: solar modulation parameters}. Overall, we believe this is an unphysical result likely due to the extra freedom present in the \reacc model for the injection spectra, pointing towards difficulties in separating solar modulation effects from low-rigidity cosmic-ray transport effects, such as re-acceleration. This is illustrated in more detail at the end of this section.

We provide the LIS and modulated fluxes for the time-integrated fits over the 7-year dataset, as well as for the other cosmic-ray nuclei and ratios included in our fit in Appendix~\ref{app: fluxes}.

\begin{figure*}[tbp]
\begin{minipage}[t]{0.48\linewidth}
    \centering
    \begin{minipage}[t]{0.96\linewidth}
        \centering
        \includegraphics[width=\linewidth]{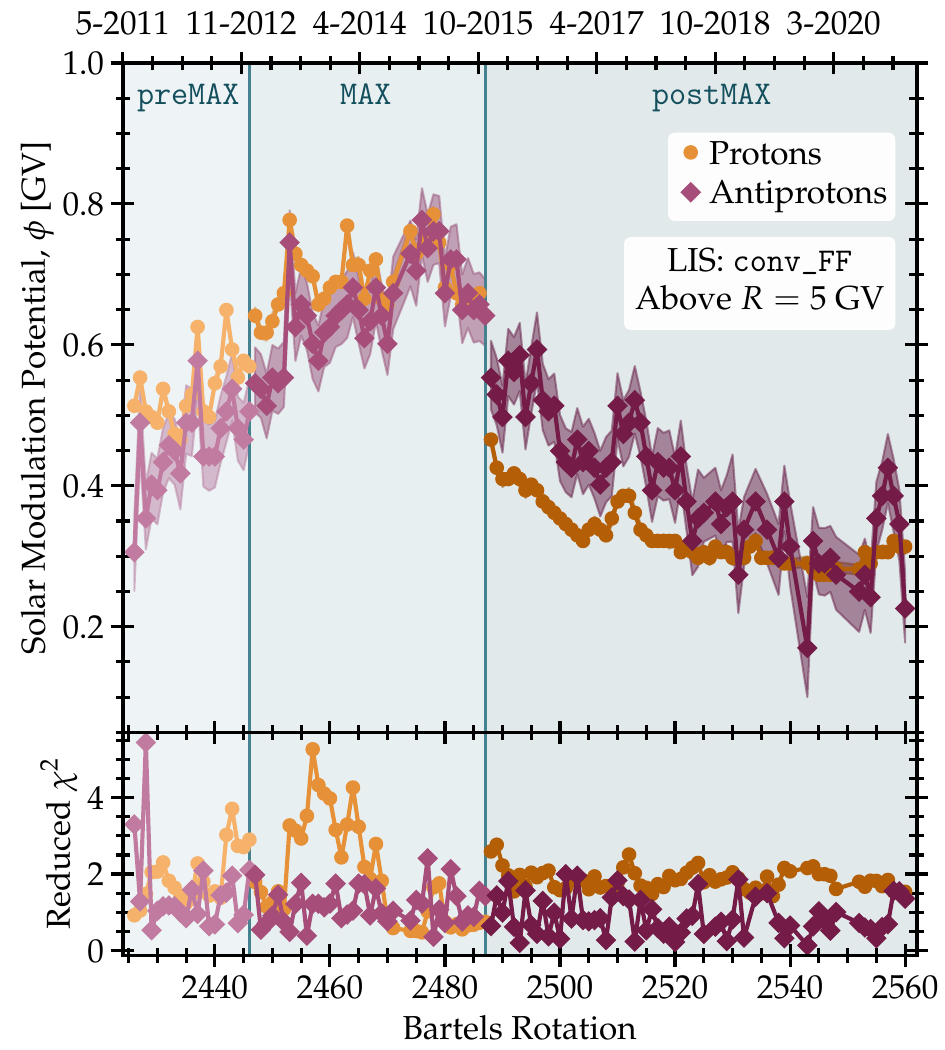}
    \end{minipage}
    \vfill\vspace{0.5cm}
    \begin{minipage}[t]{0.96\linewidth}
        \centering
        \includegraphics[width=\linewidth]{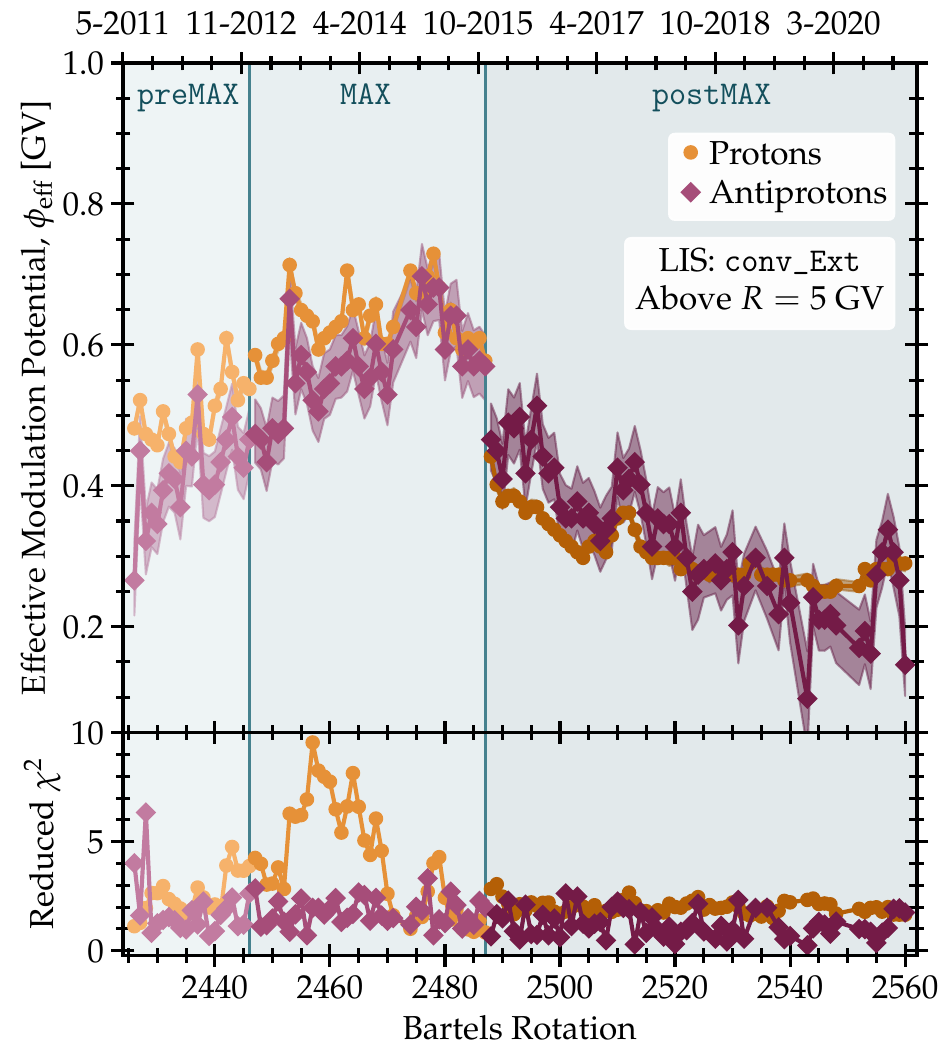}
    \end{minipage}
    \caption{The effective force-field solar modulation potential for each Bartels rotation for protons (shades of orange, circle markers) and antiprotons (shades of burgundy, diamond markers) for the three periods in the \texttt{conv} propagation model. Lighter, medium and darker shades indicate the \preMAX, \MAX and \postMAX phase, respectively; note that the LIS differs in each period. Shaded bands around the data correspond to the $1\,\sigma$ uncertainty bands. The top panel employs the force-field (\texttt{FF}) fit LISes and the bottom panel the extended model (\texttt{Ext}) LISes. The effective potential is determined using only data above 5~GV in order to reduces biases for the \texttt{Ext} case (see text for more details). The bottom panel of each plot shows the reduced $\chi^2$ for protons and antiprotons.}
    \label{fig: SM potential conv time-dependent}
\end{minipage}
\hfill
\begin{minipage}[t]{0.48\linewidth}
    \centering
    \begin{minipage}[t]{0.96\linewidth}
        \centering
        \includegraphics[width=\linewidth]{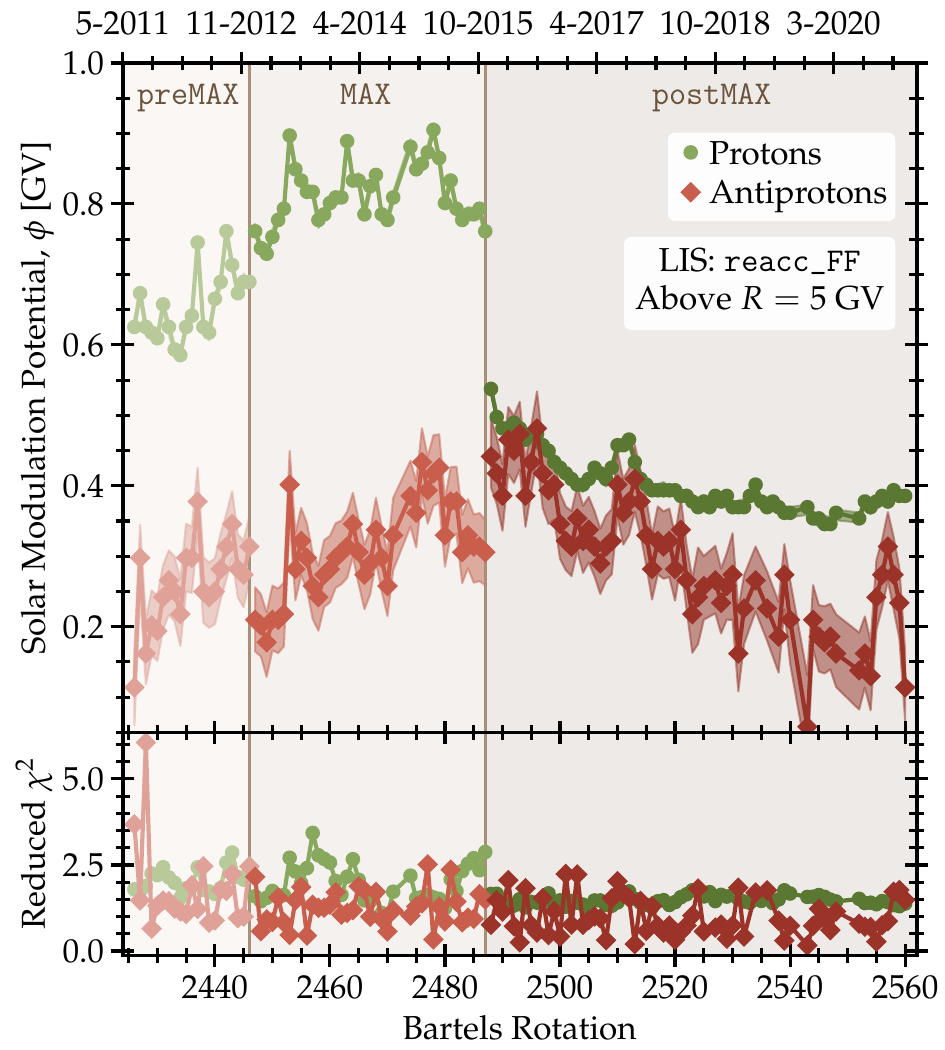}
    \end{minipage}
    \vfill\vspace{0.5cm}
    \begin{minipage}[t]{0.96\linewidth}
        \centering
        \includegraphics[width=\linewidth]{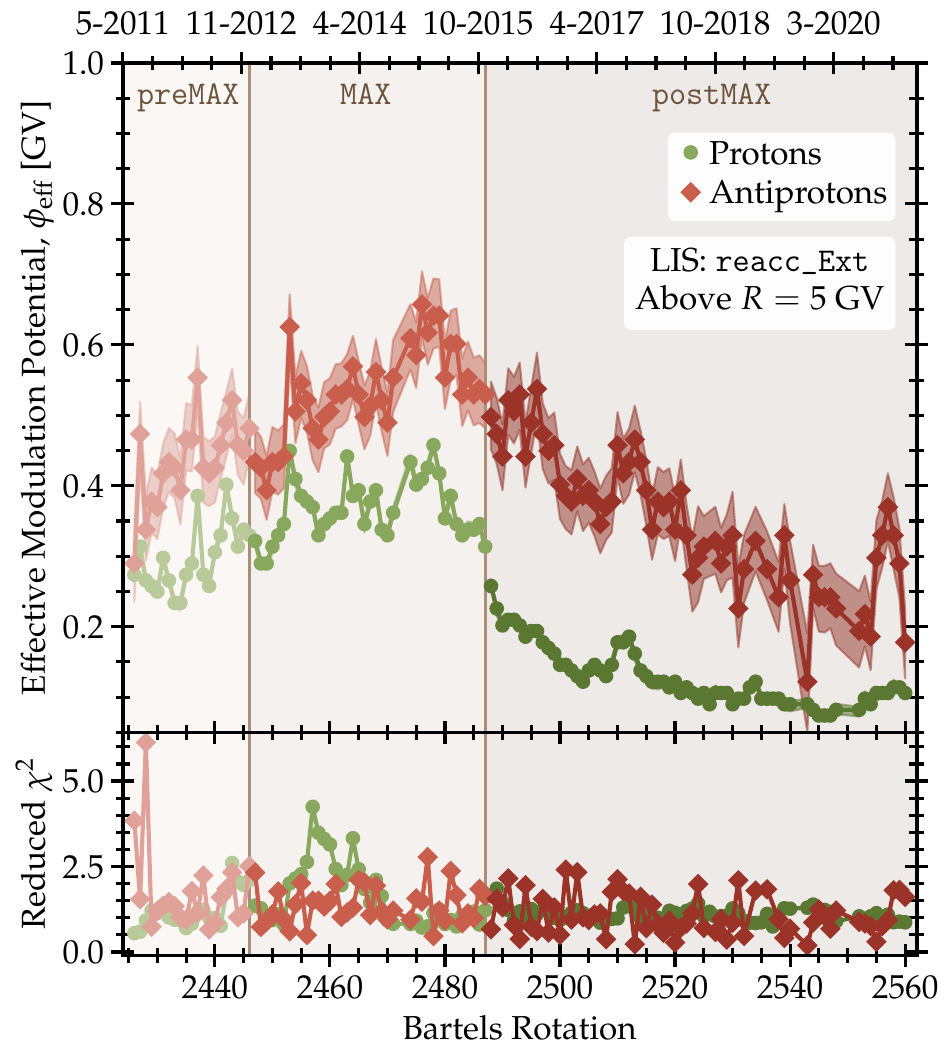}
    \end{minipage}
    \caption{Same as Figure \ref{fig: SM potential conv time-dependent} but for the \texttt{reacc} propagation model. Protons are given in shades of green, circle markers and antiprotons in shades of red, diamond markers. }
    \label{fig: SM potential reacc time-dependent}
\end{minipage}
\end{figure*}

Figures~\ref{fig: SM potential conv time-dependent} and~\ref{fig: SM potential reacc time-dependent} show the temporal changes of the solar modulation potential throughout our fit period in the \conv and \reacc propagation model, respectively. For each of the three time periods, we use the best-fit LIS obtained from the fit for that time period, and compare to the proton and antiproton data to obtain the effective force-field solar modulation potential at each Bartels rotation. We do this both for the LISes obtained with the force-field (\texttt{FF}) and extended modulation (\texttt{Ext}) models. Furthermore, to determine the effective potential, we only include fluxes above rigidities of 5~GV, which lies above the break in the extended model throughout all propagation models (see Tab.~\ref{tab: solar modulation parameters}). This choice minimises the bias in the \Ext model, as it now only includes a single potential (\textit{i.e.}, above the break). Thus, this method offers a good illustration of the temporal changes in the intensity of solar modulation for both our models.

Generally, as discussed in Sec.~\ref{sec: solar modulation models}, we expect that the modulation potential should be similar for protons and antiprotons during the period of the solar maximum, when solar activity is highest and the heliosphere most turbulent. Before the maximum of solar cycle 24, the heliospheric magnetic field polarity is such that protons experience stronger solar modulation effects, and after the maximum, when the magnetic field polarity has flipped, the antiproton flux should be stronger suppressed. This can be observed in the bottom part of Fig.~\ref{fig: ssn}. As the flux intensity is anti-correlated with the solar modulation potential, we expect the potential for protons to be slightly stronger than the potential for anti-protons in the \preMAX phase, similar in the \MAX phase, and weaker in the \postMAX phase.

In Fig.~\ref{fig: SM potential conv time-dependent}, which shows the \conv propagation model, we can see that the above expectations are roughly met, in particular in the force-field fit. In the extended modulation, the situation is somewhat borderline, as the proton and antiproton potentials are similar in the \postMAX phase, although we expect the antiproton potential to be larger. Nonetheless, both cases seem to overall follow the expected behaviour reasonably well. The bottom panels of each plot show the reduced $\chi^2$ for protons and antiprotons, obtained for the modulation potential at each Bartels rotation. In both the force-field and extended modulation case, the reduced $\chi^2$ strongly worsens for protons during the solar maximum, further emphasising the complexity of the heliospheric processes during this solar configuration, as well as the observations that the force-field approximation is insufficient during the solar maximum, even above 5~GV.

Fig.~\ref{fig: SM potential reacc time-dependent} is similar to Fig.~\ref{fig: SM potential conv time-dependent} but shows the effective solar modulation potential for the \reacc propagation model. It can be seen that in this case the results are completely at odds with expectations. In the force-field case the modulation potential for protons is always larger than that of antiprotons, in any time period, by large factors ranging from 2 to 3, even in the \MAX phase. For the extended modulation scenario the situation is reversed with antiprotons being more strongly modulated than protons during the entire time period, in contradiction with expectations. Overall, both the force-field and extended modulation results seem problematic within the \reacc propagation model. The proton modulation potential for the \conv and \reacc case are actually similar in the \postMAX phase, which indicates that, at least in this period, the force-field modulation might still be a reasonable approximation, at least for protons. Overall, however, the force-field model seems inadequate when used in conjunction with the \reacc propagation model.
Furthermore, these unexpected results for the extended modulation model also confirm that the achieved fit is probably unphysical, as discussed above. We note nonetheless that in both cases the quality of the fit is good, as can be seen from the bottom panels showing the reduced $\chi^2$. The unphysical result thus does not seem to come from a poor quality of the fit but from a complex interplay and degeneracy between several processes dominant at GeV energies, in particular solar modulation and re-acceleration, which the fit is not able to separate well.

For comparison, we also show the time-evolution of the solar modulation potentials for the 7-yr time-integrated models in Appendix~\ref{app: SM potential time-independent fit}.

\begin{figure}[tbp]
    \centering
    \includegraphics[width=\linewidth]{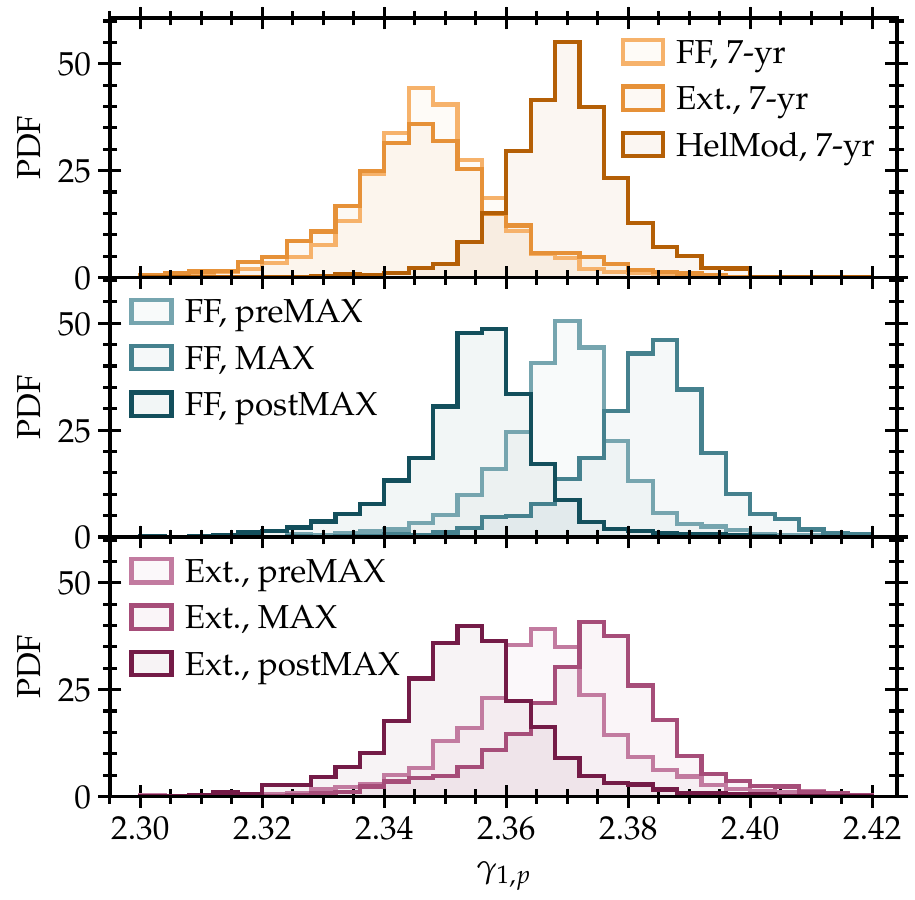}
    \caption{Probability distribution from the Bayesian fit for the proton injection index $\gamma_{1, p}$, in the \texttt{conv} propagation model. Top panel shows the 7-yr time-integrated fits for the force-field (FF), extended (Ext.) and \texttt{HelMod} solar modulation models. Middle and bottom panels show the fits to the force-field and extended model for the three time periods, respectively.}
    \label{fig: pdf conv nuc_g_1_01_001}
\end{figure}

\begin{figure}[tbp]
    \centering
    \includegraphics[width=\linewidth]{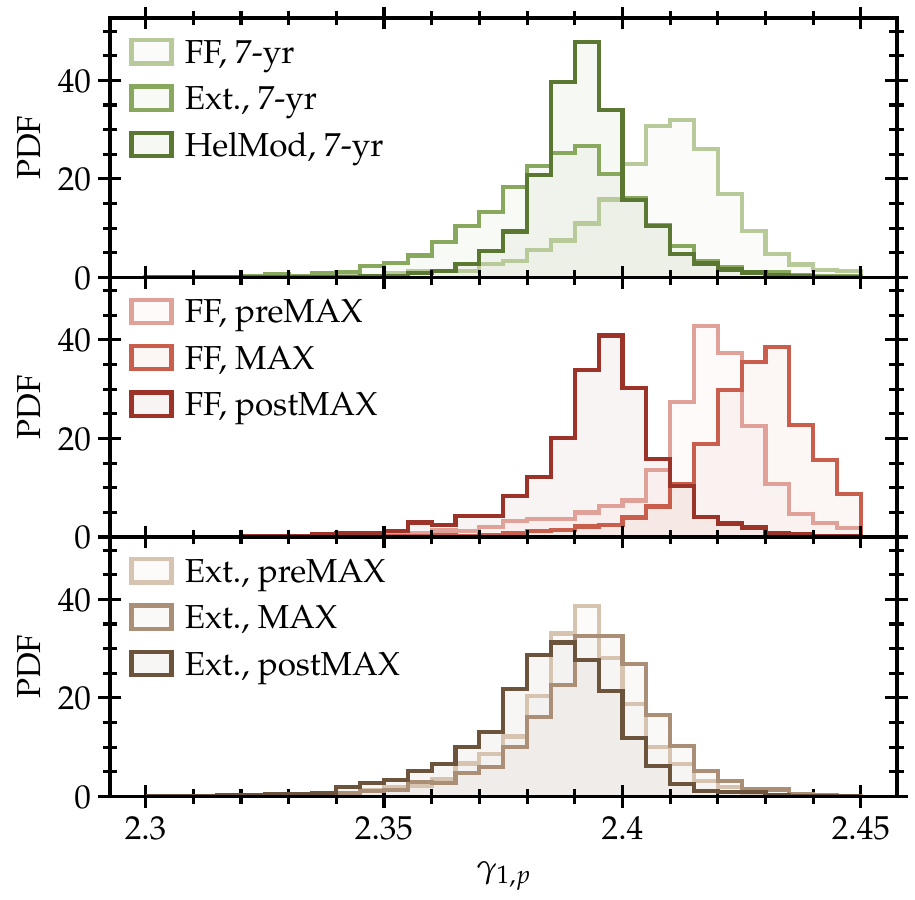}
    \caption{Same as Fig.~\ref{fig: pdf conv nuc_g_1_01_001}, but for the \texttt{reacc} propagation model.}
    \label{fig: pdf reacc nuc_g_1_01_001}
\end{figure}

\begin{figure}[tbp]
    \centering
    \includegraphics[width=\linewidth]{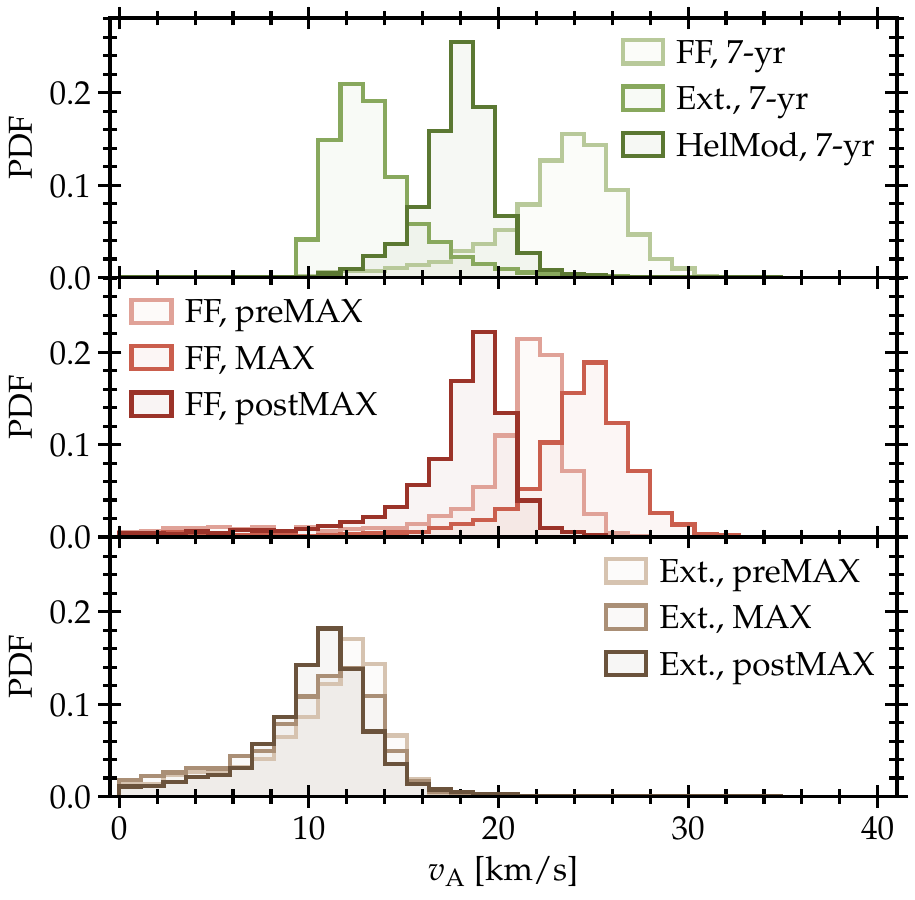}
    \caption{Probability distribution from the Bayesian fit for the Alfv\'en velocity, $v_A$, in the \texttt{reacc} propagation model. Top panel shows the 7-yr time-integrated fits for the force-field (FF), extended (Ext.) and \texttt{HelMod} solar modulation models. Middle and bottom panels show the fits to the force-field and extended model for the 3 time periods, respectively.}
    \label{fig: pdf reacc v_Alfven}
\end{figure}

To further illustrate the above issues, Figures~\ref{fig: pdf conv nuc_g_1_01_001} to~\ref{fig: pdf reacc v_Alfven} show the probability distribution functions (PDFs) of a few select parameters that are especially inconsistent between the propagation models and time periods. Each figure shows the PDF for the three 7-year time-integrated solar modulation models (top panels), the three time-dependent fits with the force-field (\FF) model (middle panels), and the three time-dependent fits with the extended (\Ext) model (bottom panels). We note that in the 7-year time-integrated models, it is not necessarily expected that the parameters are consistent as the underlying solar modulation models are different. However, for the time-dependent fits, which use the same propagation and modulation models, parameters should be consistent, and a failure of this points towards systematic inconsistencies with the underlying modulation model and/or propagation model. In Table~\ref{tab: sigmas}, we list the difference between the PDFs from Figs.~\ref{fig: pdf conv nuc_g_1_01_001} to~\ref{fig: pdf reacc v_Alfven} in terms of $\sigma$. We calculate the significances by approximating each distribution as a Gaussian and checking the compatibility between the differences of a set of two distributions with zero, using error propagation.

Fig.~\ref{fig: pdf conv nuc_g_1_01_001} presents the proton injection spectral index, $\gamma_{1,p}$ in the \conv model. The force-field fit shows a significant inconsistency, at a level of $4.22\,\sigma$ (see Tab.~\ref{tab: sigmas}), between the \MAX and \postMAX values, confirming that the force-field approximation breaks down during the \MAX phase. The values of the \preMAX and \postMAX phases, when the force-field approximation works better, are in a more reasonable agreement, although still in tension at a $2\,\sigma$ level. For the extended modulation case, the agreement is generally better, although the \MAX and \postMAX values are still in tension at a $2.8\,\sigma$ level.

\begin{table}[tbp]
    \centering
    \aboverulesep=0ex
\belowrulesep=0ex

\begin{tabular}{|c|c|c|c c|}
\hline
\multirow{18}{*}{$\gamma_{1,p}$} & \multirow{9}{*}{\conv} & \multirow{3}{*}{7-yr} & FF $\leftrightarrow$ Ext:     & $0.11\,\sigma$ \\
                                & & & FF $\leftrightarrow$ HelMod:         & $2.96\,\sigma$ \\
                                & & & Ext $\leftrightarrow$ HelMod:        & $3.03\,\sigma$ \\
                                \cmidrule{3-5}
                                & & \multirow{3}{*}{FF} &  \preMAX $\leftrightarrow$ \MAX:     & $2.27\,\sigma$ \\
                                & & &  \MAX $\leftrightarrow$ \postMAX:    & $4.22\,\sigma$ \\
                                & & &  \preMAX $\leftrightarrow$ \postMAX: & $1.98\,\sigma$ \\
                                \cmidrule{3-5}
                                & & \multirow{3}{*}{Ext} &  \preMAX $\leftrightarrow$ \MAX:     & $1.18\,\sigma$ \\
                                & & &  \MAX $\leftrightarrow$ \postMAX:    & $2.82\,\sigma$ \\
                                & & &  \preMAX $\leftrightarrow$ \postMAX: & $1.63\,\sigma$ \\

                                \cmidrule{2-5}
                                & \multirow{9}{*}{\reacc} & \multirow{3}{*}{7-yr} & FF $\leftrightarrow$ Ext:     & $1.53\,\sigma$ \\
                                & & & FF $\leftrightarrow$ HelMod:         & $0.03\,\sigma$ \\
                                & & & Ext $\leftrightarrow$ HelMod:        & $2.25\,\sigma$ \\
                                \cmidrule{3-5}
                                & & \multirow{3}{*}{FF} &  \preMAX $\leftrightarrow$ \MAX:     & $1.17\,\sigma$ \\
                                & & &  \MAX $\leftrightarrow$ \postMAX:    & $3.83\,\sigma$ \\
                                & & &  \preMAX $\leftrightarrow$ \postMAX: & $2.81\,\sigma$ \\
                                \cmidrule{3-5}
                                & & \multirow{3}{*}{Ext} &  \preMAX $\leftrightarrow$ \MAX:     & $0.37\,\sigma$ \\
                                & & &  \MAX $\leftrightarrow$ \postMAX:    & $0.67\,\sigma$ \\
                                & & &  \preMAX $\leftrightarrow$ \postMAX: & $0.33\,\sigma$ \\
\hline
\hline
\multirow{9}{*}{$v_A$} & \multirow{9}{*}{\reacc} & \multirow{3}{*}{7-yr} & FF $\leftrightarrow$ Ext:     & $6.15\,\sigma$ \\
                                & & & FF $\leftrightarrow$ HelMod:         & $3.59\,\sigma$ \\
                                & & & Ext $\leftrightarrow$ HelMod:        & $3.69\,\sigma$ \\
                                \cmidrule{3-5}
                                & & \multirow{3}{*}{FF} &  \preMAX $\leftrightarrow$ \MAX:     & $1.76\,\sigma$ \\
                                & & &  \MAX $\leftrightarrow$ \postMAX:    & $3.48\,\sigma$ \\
                                & & &  \preMAX $\leftrightarrow$ \postMAX: & $2.20\,\sigma$ \\
                                \cmidrule{3-5}
                                & & \multirow{3}{*}{Ext} &  \preMAX $\leftrightarrow$ \MAX:     & $0.19\,\sigma$ \\
                                & & &  \MAX $\leftrightarrow$ \postMAX:    & $0.56\,\sigma$ \\
                                & & &  \preMAX $\leftrightarrow$ \postMAX: & $0.82\,\sigma$ \\
\hline
\end{tabular}
    \caption{Summary of the differences in PDFs of Figs.~\ref{fig: pdf conv nuc_g_1_01_001},~\ref{fig: pdf reacc nuc_g_1_01_001} and~\ref{fig: pdf reacc v_Alfven}. The first column indicates the parameter, second column the propagation model, third column the solar modulation model, \textit{i.e.} either for time-integrated 7-yr fits or force-field (FF) and extended model (Ext), and last column lists the significance $\sigma$ of the tension between two distributions as indicated.}
    \label{tab: sigmas}
\end{table}

Figs.~\ref{fig: pdf reacc nuc_g_1_01_001} and \ref{fig: pdf reacc v_Alfven} show the PDFs of the proton injection spectral index, $\gamma_{1,p}$, and the Alfv\'en velocity, $v_A$, for the \reacc model, respectively. A similar trend as above can be observed. In the extended modulation scenario, the PDFs show a very good consistency among the different time periods. However, this is at the price of unphysical values for the modulation potential and the LIS, as discussed previously. Conversely, for the force-field case, we again see a divergence among the values in the different time periods at a level up to $3.5\,\sigma$ for the \postMAX and \MAX cases. This once more confirms that in the \reacc propagation model, the force-field solar modulation model fails to provide a consistent picture.

\section{Summary and Conclusions}
\label{sec: summary and conclusions}

We conduct detailed fits of cosmic-ray propagation to the latest time-dependent AMS-02 data and study the impact of the chosen solar modulation model on the parameters of the propagation model. The AMS-02 data provides cosmic-ray fluxes over a full solar cycle at each Bartels rotation. Crucially, this includes cosmic-ray antiprotons, which differ from other cosmic-ray nuclei in two ways: They are the only negatively charged cosmic-ray nuclei, which makes them an important and complementary probe of the charge sign-dependence of solar modulation. Furthermore, the spectral shape of their fluxes is unique, as they are only produced as secondaries in inelastic scattering processes. 

We consider two propagation scenarios, one focused on convection, dubbed \conv, and one on re-acceleration, dubbed \reacc, and three solar modulation models: the standard force-field, an extended force-field, where the modulation potential differs above and below a rigidity break at a few GV, and \texttt{HelMod}. Furthermore, we divide the solar cycle into three time periods: (1) before the solar maximum of cycle 24 in 2014 -- 2015 (\preMAX), (2) during the maximum (\MAX), and (3) after the solar maximum (\postMAX). We study these periods separately, allowing us to examine the consistency between the individual results. Our findings can be summarized as follows:

\begin{itemize}
    \item In the \conv propagation model for positively charged particles, the extended force-field modulation provides consistent modulation potentials below and above the break for phases of moderate to low solar activity, resembling a force field. This indicates that the force-field approximation can describe the modulation of positively charged nuclei relatively well during off-maximum phases of solar activity. On the other hand, during solar maximum, we detect a significant difference in the low and high potential in the extended model (see Tab.~\ref{tab: solar modulation parameters}), indicating that the force-field approximation is not appropriate during this period. This is also confirmed by a strong worsening in the fit quality during the solar maximum when the spectra are compared to the data for each Bartels rotation (see more below).
    
    \item For antiprotons, we detect strong differences in the low and high potential in the extended model during all solar phases, \textit{i.e.}, during both the maximum and minimum solar activity phases. We ascribe this result to two possible causes: (1) An incorrect antiproton LIS due to a combination of incorrect propagation models and/or uncertainties in the antiproton production cross section, with solar modulation trying to compensate for this. (2) A true failure of the force-field approximation, perhaps related to the peculiar, peak-shaped spectrum of antiprotons, as opposed to the power-law-like shape for the other nuclei (see Fig.~\ref{fig: flux vs rigidity}).
    
    \item For the \reacc propagation scenario, the conclusions are very different. The low and high potential of the extended modulation model strongly differ throughout each solar period, both for protons and antiprotons. This, in principle, would indicate that the force-field approximation is always inadequate, even during low solar activity periods. However, we believe that this is rather due to an unphysical result achieved by the best-fit. This seems to be confirmed by the shape of the LIS, which is very different with respect to the case of the \conv model (see Figs.~\ref{fig: LIS conv} and~\ref{fig: LIS reacc}), as well as by the properties of the time-dependent modulation potentials (see Figs.~\ref{fig: SM potential conv time-dependent} and~\ref{fig: SM potential reacc time-dependent}), which is at odds with expectations, as further discussed below. This seems to point towards difficulties of the fitting procedure to disentangle solar modulation and re-acceleration, which both act in a similar way at similar rigidities.

    \item For a given propagation and solar modulation model, the LIS fluxes must always be the same, independently of solar activity, as we do not expect time-dependent changes in the (nearby) Galactic propagation on short time scales. In the force-field approximation, for both propagation scenarios, we find that the LIS fluxes obtained in each of the three time periods show discrepancies by about $10 - 20\%$. When excluding the solar maximum and comparing only periods of low or moderate solar activity this discrepancy is still at the level of $10 - 15\%$. This persists when using the extended solar modulation model, indicating that, at the moment, the solar modulation-related systematic uncertainty in the reconstructed LIS and propagation parameters is at least at the level of $10 - 15\%$.

    Looking at specific parameters confirms the above conclusions. For example, the proton injection spectral index, $\gamma_{1, p}$, and the Alfv\'en velocity, $v_A$, show differences at the $\sim 4\,\sigma$ level between the values determined in the three time periods within the force-field approximation. Conversely, the discrepancy goes down to a more reasonable $\sim 1-2\,\sigma$ when the solar maximum period is excluded, or when the extended solar modulation is employed (see Figs.~\ref{fig: pdf conv nuc_g_1_01_001} to~\ref{fig: pdf reacc v_Alfven} and Tab.~\ref{tab: sigmas}).

    \item A further tool to evaluate the systematic uncertainties related to solar modulation is the temporal behaviour of the modulation itself. Given the LIS derived for each fit that we performed, we have calculated the effective force-field modulation potential as function of time for each Bartels rotation (see Figs.~\ref{fig: SM potential conv time-dependent} and~\ref{fig: SM potential reacc time-dependent}). Based on our current understanding of cosmic-ray propagation in the solar system, and from the behaviour of the time-dependent AMS-02 data, we expect that solar modulation for protons and antiprotons should be comparable during times of maximum solar activity, larger for protons before the maximum of solar cycle 24 and larger for antiprotons after the maximum. 

    While the temporal behaviour of the solar modulation potential in the \conv propagation model is somewhat consistent with these expectations, the results in the \reacc model are incompatible with our understanding of heliospheric propagation, for both the force-field and the extended modulation case. Thus, these effective solar modulation models are generally unable to disentangle the degeneracies between solar modulation itself and other low-energy processes, such as re-acceleration, which might only be achievable with more sophisticated heliospheric propagation models such as \texttt{HelMod}.

   \item Using the \texttt{HelMod} modulation, we obtain similar LIS fluxes (see Figs.~\ref{fig: LIS conv time-independent} and~\ref{fig: LIS reacc time-independent}) and time-dependent potentials (see Figs.~\ref{fig: SM potential conv time-independent} and~\ref{fig: SM potential reacc time-independent}) for both the \conv and \reacc scenarios. This is probably not surprising since we use a fixed modulation without free parameters as tuned in \cite{Boschini:2017fxq,Boschini:2018baj,Boschini:2020jty}, which means that the LIS is basically fixed. We stress however, that a self-consistent analysis would require to re-derive and re-tune the \texttt{HelMod} modulation together with Galactic propagation. This is, however, a task that requires a dedicated study which we leave for future works. 

\end{itemize}

Our work shows that our current modelling of solar modulation introduces significant systematic uncertainties of at least at a level of 10 -- 15\%, in the reconstruction of cosmic-ray LIS fluxes and Galactic propagation models. These uncertainties originate both from an interplay with processes at low energies, such as re-acceleration, that are degenerate with solar modulation effects, and from difficulties in correctly modelling solar modulation in the different phases of the solar activity cycle, in particular during the solar maximum. 

Together with the uncertainties related to nuclear cross sections, this is significantly limiting our ability to perform precise cosmic-ray physics at the percent-level precision which is otherwise achieved by the AMS-02 measurements. Future data, in particular time-dependent data spanning a full 22-year solar cycle covering both polarity configurations of the Sun's magnetic field will be crucial to improve our knowledge of cosmic-ray propagation in the solar system, and reduce these uncertainties.

\section*{Acknowledgements}
Many thanks to Stefano Della Torre for helpful discussions on \texttt{HelMod}. 
I.J. and A.C. acknowledge support from the Research grant TAsP (Theoretical Astroparticle Physics) funded by INFN, and Research grant ``Addressing systematic uncertainties in searches for dark matter'', Grant No.\ 2022F2843L, CUP D53D23002580006 funded by the Italian Ministry of University and Research (\textsc{mur}). 
M.D.M. acknowledges support from the Italian Ministry of University and Research (MUR), PRIN 2022 ``EXSKALIBUR – Euclid-Cross-SKA: Likelihood Inference Building for Universe’s Research'', Grant No. 20222BBYB9, CUP I53D23000610 0006, and from the European Union -- Next Generation EU.

\bibliography{main}

@article{diMauro:2014zea,
    author = "di Mauro, Mattia and Donato, Fiorenza and Goudelis, Andreas and Serpico, Pasquale Dario",
    title = "{New evaluation of the antiproton production cross section for cosmic ray studies}",
    eprint = "1408.0288",
    archivePrefix = "arXiv",
    primaryClass = "hep-ph",
    reportNumber = "LAPTH-051-14",
    doi = "10.1103/PhysRevD.90.085017",
    journal = "Phys. Rev. D",
    volume = "90",
    number = "8",
    pages = "085017",
    year = "2014",
    note = "[Erratum: Phys.Rev.D 98, 049901 (2018)]"
}

@article{Donato:2017ywo,
    author = "Donato, Fiorenza and Korsmeier, Michael and Di Mauro, Mattia",
    title = "{Prescriptions on antiproton cross section data for precise theoretical antiproton flux predictions}",
    eprint = "1704.03663",
    archivePrefix = "arXiv",
    primaryClass = "astro-ph.HE",
    reportNumber = "TTK-17-08",
    doi = "10.1103/PhysRevD.96.043007",
    journal = "Phys. Rev. D",
    volume = "96",
    number = "4",
    pages = "043007",
    year = "2017"
}

@article{Korsmeier:2018gcy,
    author = "Korsmeier, Michael and Donato, Fiorenza and Di Mauro, Mattia",
    title = "{Production cross sections of cosmic antiprotons in the light of new data from the NA61 and LHCb experiments}",
    eprint = "1802.03030",
    archivePrefix = "arXiv",
    primaryClass = "astro-ph.HE",
    reportNumber = "TTK-18-06",
    doi = "10.1103/PhysRevD.97.103019",
    journal = "Phys. Rev. D",
    volume = "97",
    number = "10",
    pages = "103019",
    year = "2018"
}

@article{Kappl:2015hxv,
    author = "Kappl, Rolf",
    title = "{SOLARPROP: Charge-sign Dependent Solar Modulation for Everyone}",
    eprint = "1511.07875",
    archivePrefix = "arXiv",
    primaryClass = "astro-ph.SR",
    doi = "10.1016/j.cpc.2016.05.025",
    journal = "Comput. Phys. Commun.",
    volume = "207",
    pages = "386--399",
    year = "2016"
}

@article{Maccione:2012cu,
    author = "Maccione, Luca",
    title = "{Low energy cosmic ray positron fraction explained by charge-sign dependent solar modulation}",
    eprint = "1211.6905",
    archivePrefix = "arXiv",
    primaryClass = "astro-ph.HE",
    reportNumber = "LMU-ASC-83-12, MPP-2012-155",
    doi = "10.1103/PhysRevLett.110.081101",
    journal = "Phys. Rev. Lett.",
    volume = "110",
    number = "8",
    pages = "081101",
    year = "2013"
}

@article{Tomassetti:2017hbe,
    author = "Tomassetti, Nicola",
    title = "{Solar and nuclear physics uncertainties in cosmic-ray propagation}",
    eprint = "1707.06917",
    archivePrefix = "arXiv",
    primaryClass = "astro-ph.HE",
    doi = "10.1103/PhysRevD.96.103005",
    journal = "Phys. Rev. D",
    volume = "96",
    number = "10",
    pages = "103005",
    year = "2017"
}

@article{Tomassetti:2018tnl,
    author = "Tomassetti, Nicola and Bar{\~a}o, Fernando and Bertucci, Bruna and Fiandrini, Emanuele and Figueiredo, Jos{\'e} and Barreira, Jo{\~a}o and Orcinha, Miguel",
    title = "{Testing diffusion of cosmic rays in the heliosphere with proton and helium data from AMS}",
    eprint = "1811.08909",
    archivePrefix = "arXiv",
    primaryClass = "astro-ph.HE",
    doi = "10.1103/PhysRevLett.121.251104",
    journal = "Phys. Rev. Lett.",
    volume = "121",
    number = "25",
    pages = "251104",
    year = "2018"
}

@article{Tomassetti:2025nna,
    author = "Tomassetti, Nicola and Bertucci, Bruna and Fiandrini, Emanuele and Khiali, Behrouz",
    title = "{Propagation Times and Energy Losses of Cosmic Protons and Antiprotons in Interplanetary Space}",
    eprint = "2503.14025",
    archivePrefix = "arXiv",
    primaryClass = "astro-ph.HE",
    doi = "10.3390/galaxies13020023",
    journal = "Galaxies",
    volume = "13",
    number = "2",
    pages = "23",
    year = "2025"
}

@article{Maurin:2025gsz,
    author = "Maurin, D. and others",
    title = "{Precision cross-sections for advancing cosmic-ray physics and other applications: A comprehensive programme for the next decade}",
    eprint = "2503.16173",
    archivePrefix = "arXiv",
    primaryClass = "astro-ph.HE",
    doi = "10.1016/j.physrep.2025.11.002",
    journal = "Phys. Rept.",
    volume = "1161",
    pages = "1--81",
    year = "2026"
}

@article{Genolini:2023kcj,
    author = "G{\'e}nolini, Yoann and Maurin, David and Moskalenko, Igor V. and Unger, Michael",
    title = "{Current status and desired precision of the isotopic production cross sections~relevant to astrophysics of cosmic rays. II. Fluorine to silicon and updated results for Li, Be, and B}",
    eprint = "2307.06798",
    archivePrefix = "arXiv",
    primaryClass = "astro-ph.HE",
    reportNumber = "LAPTH-043/23",
    doi = "10.1103/PhysRevC.109.064914",
    journal = "Phys. Rev. C",
    volume = "109",
    number = "6",
    pages = "064914",
    year = "2024"
}

@article{Genolini:2018ekk,
    author = "Genolini, Yoann and Maurin, David and Moskalenko, Igor V. and Unger, Michael",
    title = "{Current status and desired precision of the isotopic production cross sections relevant to astrophysics of cosmic rays: Li, Be, B, C, and N}",
    eprint = "1803.04686",
    archivePrefix = "arXiv",
    primaryClass = "astro-ph.HE",
    doi = "10.1103/PhysRevC.98.034611",
    journal = "Phys. Rev. C",
    volume = "98",
    number = "3",
    pages = "034611",
    year = "2018"
}

@article{Zhang:2026ohu,
    author = "Zhang, Hui-Ming and Lin, Su-Jie and Feng, Jie and Jiang, Jie-Teng and Yang, Li-Li",
    title = "{A Unified Charge-Dependent Modulation Model for AMS-02 Proton and Antiproton Fluxes during Solar Minimum}",
    eprint = "2601.07649",
    archivePrefix = "arXiv",
    primaryClass = "astro-ph.HE",
    month = "1",
    year = "2026"
}

@article{Boschini:2017fxq,
    author = "Boschini, M. J. and others",
    title = "{Solution of heliospheric propagation: unveiling the local interstellar spectra of cosmic ray species}",
    eprint = "1704.06337",
    archivePrefix = "arXiv",
    primaryClass = "astro-ph.HE",
    doi = "10.3847/1538-4357/aa6e4f",
    journal = "Astrophys. J.",
    volume = "840",
    number = "2",
    pages = "115",
    year = "2017"
}

@article{Boschini:2018baj,
    author = "Boschini, M. J. and others",
    title = "{Deciphering the local Interstellar spectra of primary cosmic ray species with HelMod}",
    eprint = "1804.06956",
    archivePrefix = "arXiv",
    primaryClass = "astro-ph.HE",
    doi = "10.3847/1538-4357/aabc54",
    journal = "Astrophys. J.",
    volume = "858",
    number = "1",
    pages = "61",
    year = "2018"
}

@article{Boschini:2020jty,
    author = "Boschini, M. J. and others",
    title = "{Inference of the Local Interstellar Spectra of Cosmic-Ray Nuclei Z {\ensuremath{\leq}} 28 with the GalProp{\textendash}HelMod  Framework}",
    eprint = "2006.01337",
    archivePrefix = "arXiv",
    primaryClass = "astro-ph.HE",
    doi = "10.3847/1538-4365/aba901",
    journal = "Astrophys. J. Suppl.",
    volume = "250",
    number = "2",
    pages = "27",
    year = "2020"
}

@article{Korsmeier:2021bkw,
    author = "Korsmeier, Michael and Cuoco, Alessandro",
    title = "{Testing the universality of cosmic-ray nuclei from protons to oxygen with AMS-02}",
    eprint = "2112.08381",
    archivePrefix = "arXiv",
    primaryClass = "astro-ph.HE",
    doi = "10.1103/PhysRevD.105.103033",
    journal = "Phys. Rev. D",
    volume = "105",
    number = "10",
    pages = "103033",
    year = "2022"
}

@article{Parker1958,
  author  = {Parker, E. N.},
  title   = {Dynamics of the Interplanetary Gas and Magnetic Fields},
  journal = {The Astrophysical Journal},
  volume  = {128},
  pages   = {664--676},
  year    = {1958},
  doi     = {10.1086/146579}
}

@article{Parker1965,
  author  = {Parker, E. N.},
  title   = {The passage of energetic charged particles through interplanetary space},
  journal = {Planetary and Space Science},
  volume  = {13},
  number  = {1},
  pages   = {9--49},
  year    = {1965},
  doi     = {10.1016/0032-0633(65)90131-5}
}

@article{Jokipii1971,
  author  = {Jokipii, J. R.},
  title   = {Propagation of cosmic rays in the solar wind},
  journal = {Reviews of Geophysics and Space Physics},
  volume  = {9},
  number  = {1},
  pages   = {27--87},
  year    = {1971},
  doi     = {10.1029/RG009i001p00027}
}

@article{JokipiiLevyHubbard1977,
  author  = {Jokipii, J. R. and Levy, E. H. and Hubbard, W. B.},
  title   = {Effects of particle drift on cosmic-ray transport. I. General properties, application to solar modulation},
  journal = {The Astrophysical Journal},
  volume  = {213},
  pages   = {861--868},
  year    = {1977},
  doi     = {10.1086/155218}
}

@article{CaballeroLopezMoraal2004,
  author  = {Caballero-Lopez, R. A. and Moraal, H.},
  title   = {Limitations of the force field equation to describe cosmic ray modulation},
  journal = {Journal of Geophysical Research: Space Physics},
  volume  = {109},
  pages   = {A01101},
  year    = {2004},
  doi     = {10.1029/2003JA010098}
}

@article{VosPotgieter2015,
  author  = {Vos, E. E. and Potgieter, M. S.},
  title   = {New Modeling of Galactic Proton Modulation during the Minimum of Solar Cycle 23/24},
  journal = {The Astrophysical Journal},
  volume  = {815},
  number  = {2},
  pages   = {119},
  year    = {2015},
  doi     = {10.1088/0004-637X/815/2/119}
}

@article{GieselerHeberHerbst2017,
  author  = {Gieseler, J. and Heber, B. and Herbst, K.},
  title   = {An empirical modification of the force field approach to describe the modulation of galactic cosmic rays close to Earth in a broad range of rigidities},
  journal = {Journal of Geophysical Research: Space Physics},
  volume  = {122},
  number  = {11},
  pages   = {10964--10979},
  year    = {2017},
  doi     = {10.1002/2017JA024763}
}

@article{AdrianiEtAl2013PAMELAProtonTime,
  author  = {Adriani, O. and others},
  title   = {Time Dependence of the Proton Flux Measured by PAMELA during the 2006 July--2009 December Solar Minimum},
  journal = {The Astrophysical Journal},
  volume  = {765},
  number  = {2},
  pages   = {91},
  year    = {2013},
  doi     = {10.1088/0004-637X/765/2/91}
}

@article{AguilarEtAl2018AMSProtonHeliumTime,
  author  = {Aguilar, M. and others},
  title   = {Observation of Fine Time Structures in the Cosmic Proton and Helium Fluxes with the Alpha Magnetic Spectrometer on the International Space Station},
  journal = {Physical Review Letters},
  volume  = {121},
  pages   = {051101},
  year    = {2018},
  doi     = {10.1103/PhysRevLett.121.051101}
}

@article{DiMauro:2023oqx,
    author = "Di Mauro, Mattia and Donato, Fiorenza and Korsmeier, Michael and Manconi, Silvia and Orusa, Luca",
    title = "{Novel prediction for secondary positrons and electrons in the Galaxy}",
    eprint = "2304.01261",
    archivePrefix = "arXiv",
    primaryClass = "astro-ph.HE",
    reportNumber = "LAPTH-014/23, TTK-23-07",
    doi = "10.1103/PhysRevD.108.063024",
    journal = "Phys. Rev. D",
    volume = "108",
    number = "6",
    pages = "063024",
    year = "2023"
}

@article{DiMauro:2023jgg,
    author = "Di Mauro, Mattia and Korsmeier, Michael and Cuoco, Alessandro",
    title = "{Data-driven constraints on cosmic-ray diffusion: Probing self-generated turbulence in the Milky~Way}",
    eprint = "2311.17150",
    archivePrefix = "arXiv",
    primaryClass = "astro-ph.HE",
    doi = "10.1103/PhysRevD.109.123003",
    journal = "Phys. Rev. D",
    volume = "109",
    number = "12",
    pages = "123003",
    year = "2024"
}

@article{AMS:2025npj,
    author = "Aguilar, M. and others",
    collaboration = "AMS",
    title = "{Antiprotons and Elementary Particles over a Solar Cycle: Results from the Alpha Magnetic Spectrometer}",
    doi = "10.1103/PhysRevLett.134.051002",
    journal = "Phys. Rev. Lett.",
    volume = "134",
    number = "5",
    pages = "051002",
    year = "2025"
}

@article{AMS:2025pgu,
    author = "Aguilar, M. and others",
    collaboration = "AMS",
    title = "{Solar Modulation of Cosmic Nuclei over a Solar Cycle: Results from the Alpha Magnetic Spectrometer}",
    doi = "10.1103/PhysRevLett.134.051001",
    journal = "Phys. Rev. Lett.",
    volume = "134",
    number = "5",
    pages = "051001",
    year = "2025"
}

@ARTICLE{1968ApJ...154.1011G,
       author = {{Gleeson}, L.~J. and {Axford}, W.~I.},
        title = "{Solar Modulation of Galactic Cosmic Rays}",
      journal = {ApJ},
         year = 1968,
        month = dec,
       volume = {154},
        pages = {1011},
          doi = {10.1086/149822},
       adsurl = {https://ui.adsabs.harvard.edu/abs/1968ApJ...154.1011G}
}

@article{Cholis:2015gna,
    author = "Cholis, Ilias and Hooper, Dan and Linden, Tim",
    title = "{A Predictive Analytic Model for the Solar Modulation of Cosmic Rays}",
    eprint = "1511.01507",
    archivePrefix = "arXiv",
    primaryClass = "astro-ph.SR",
    reportNumber = "FERMILAB-PUB-15-477-A",
    doi = "10.1103/PhysRevD.93.043016",
    journal = "Phys. Rev. D",
    volume = "93",
    number = "4",
    pages = "043016",
    year = "2016"
}

@article{Cholis:2020tpi,
    author = "Cholis, Ilias and Hooper, Dan and Linden, Tim",
    title = "{Constraining the charge-sign and rigidity-dependence of solar modulation}",
    eprint = "2007.00669",
    archivePrefix = "arXiv",
    primaryClass = "astro-ph.HE",
    reportNumber = "FERMILAB-PUB-20-271-AE-T",
    doi = "10.1088/1475-7516/2022/10/051",
    journal = "JCAP",
    volume = "10",
    pages = "051",
    year = "2022"
}

@article{Kuhlen:2019hqb,
    author = "Kuhlen, Marco and Mertsch, Philipp",
    title = "{Time-dependent AMS-02 electron-positron fluxes in an extended force-field model}",
    eprint = "1909.01154",
    archivePrefix = "arXiv",
    primaryClass = "astro-ph.HE",
    reportNumber = "TTK-19-34",
    doi = "10.1103/PhysRevLett.123.251104",
    journal = "Phys. Rev. Lett.",
    volume = "123",
    number = "25",
    pages = "251104",
    year = "2019"
}

@article{Long:2024nty,
    author = "Long, Wei-Cheng and Wu, Juan",
    title = "{Probing solar modulation analytic models with cosmic ray periodic spectra}",
    eprint = "2403.20038",
    archivePrefix = "arXiv",
    primaryClass = "astro-ph.SR",
    doi = "10.1103/PhysRevD.109.083009",
    journal = "Phys. Rev. D",
    volume = "109",
    number = "8",
    pages = "083009",
    year = "2024"
}

@article{Zhu:2024ega,
    author = "Zhu, Cheng-Rui",
    title = "{Local Interstellar Spectra and Solar Modulation of Cosmic-Ray Proton and Helium}",
    eprint = "2409.18389",
    archivePrefix = "arXiv",
    primaryClass = "astro-ph.HE",
    doi = "10.3847/1538-4357/ad794b",
    journal = "Astrophys. J.",
    volume = "975",
    number = "2",
    pages = "270",
    year = "2024"
}

@article{Korsmeier:2021brc,
    author = "Korsmeier, Michael and Cuoco, Alessandro",
    title = "{Implications of Lithium to Oxygen AMS-02 spectra on our understanding of cosmic-ray diffusion}",
    eprint = "2103.09824",
    archivePrefix = "arXiv",
    primaryClass = "astro-ph.HE",
    reportNumber = "TTK-21-10",
    doi = "10.1103/PhysRevD.103.103016",
    journal = "Phys. Rev. D",
    volume = "103",
    number = "10",
    pages = "103016",
    year = "2021"
}

@article{Cuoco:2019kuu,
    author = {Cuoco, Alessandro and Heisig, Jan and Klamt, Lukas and Korsmeier, Michael and Kr{\"a}mer, Michael},
    title = "{Scrutinizing the evidence for dark matter in cosmic-ray antiprotons}",
    eprint = "1903.01472",
    archivePrefix = "arXiv",
    primaryClass = "astro-ph.HE",
    reportNumber = "LAPTH-052/18, TTK-19-09, CP3-19-08",
    doi = "10.1103/PhysRevD.99.103014",
    journal = "Phys. Rev. D",
    volume = "99",
    number = "10",
    pages = "103014",
    year = "2019"
}

@article{Duan:2025ead,
    author = "Duan, Kai-Kai and Wang, Xiao and Li, Wen-Hao and Xu, Zhi-Hui and Tsai, Yue-Lin Sming and Fan, Yi-Zhong",
    title = "{Scrutinizing the impact of the solar modulation on AMS-02 antiproton excess}",
    eprint = "2506.13352",
    archivePrefix = "arXiv",
    primaryClass = "astro-ph.CO",
    month = "6",
    year = "2025"
}

@article{AMS:2024idr,
    author = "Aguilar, M. and others",
    collaboration = "AMS",
    title = "{Properties of Cosmic Deuterons Measured by the Alpha Magnetic Spectrometer}",
    doi = "10.1103/PhysRevLett.132.261001",
    journal = "Phys. Rev. Lett.",
    volume = "132",
    number = "26",
    pages = "261001",
    year = "2024"
}

@article{Strong:2015zva,
    author = "Strong, A. W.",
    title = "{Recent extensions to GALPROP}",
    eprint = "1507.05020",
    archivePrefix = "arXiv",
    primaryClass = "astro-ph.HE",
    month = "7",
    year = "2015"
}

@article{Strong:1998pw,
    author = "Strong, A. W. and Moskalenko, I. V.",
    title = "{Propagation of cosmic-ray nucleons in the galaxy}",
    eprint = "astro-ph/9807150",
    archivePrefix = "arXiv",
    doi = "10.1086/306470",
    journal = "Astrophys. J.",
    volume = "509",
    pages = "212--228",
    year = "1998"
}

@article{Feroz:2008xx,
    author = "Feroz, F. and Hobson, M. P. and Bridges, M.",
    title = "{MultiNest: an efficient and robust Bayesian inference tool for cosmology and particle physics}",
    eprint = "0809.3437",
    archivePrefix = "arXiv",
    primaryClass = "astro-ph",
    doi = "10.1111/j.1365-2966.2009.14548.x",
    journal = "Mon. Not. Roy. Astron. Soc.",
    volume = "398",
    pages = "1601--1614",
    year = "2009"
}

@article{AMS:2021nhj,
    author = "Aguilar, M. and others",
    collaboration = "AMS",
    title = "{The Alpha Magnetic Spectrometer (AMS) on the international space station: Part II {\textemdash} Results from the first seven years}",
    doi = "10.1016/j.physrep.2020.09.003",
    journal = "Phys. Rept.",
    volume = "894",
    pages = "1--116",
    year = "2021"
}

@article{Alemanno:2021gpb,
    author = "Alemanno, F. and others",
    title = "{Measurement of the cosmic ray helium energy spectrum from 70 GeV to 80 TeV with the DAMPE space mission}",
    eprint = "2105.09073",
    archivePrefix = "arXiv",
    primaryClass = "astro-ph.HE",
    doi = "10.1103/PhysRevLett.126.201102",
    journal = "Phys. Rev. Lett.",
    volume = "126",
    pages = "201102",
    year = "2021"
}

@article{CALET:2022vro,
    author = "Adriani, O. and others",
    collaboration = "CALET",
    title = "{Observation of Spectral Structures in the Flux of Cosmic-Ray Protons from 50~GeV to 60~TeV with the Calorimetric Electron Telescope on the International Space Station}",
    eprint = "2209.01302",
    archivePrefix = "arXiv",
    primaryClass = "astro-ph.HE",
    doi = "10.1103/PhysRevLett.129.101102",
    journal = "Phys. Rev. Lett.",
    volume = "129",
    number = "10",
    pages = "101102",
    year = "2022"
}

@article{DAMPE:2022jgy,
    author = "Alemanno, Francesca and others",
    collaboration = "DAMPE",
    title = "{Detection of spectral hardenings in cosmic-ray boron-to-carbon and boron-to-oxygen flux ratios with DAMPE}",
    eprint = "2210.08833",
    archivePrefix = "arXiv",
    primaryClass = "astro-ph.HE",
    doi = "10.1016/j.scib.2022.10.002",
    journal = "Sci. Bull.",
    volume = "67",
    pages = "2162--2166",
    year = "2022"
}

@article{Stone:2013zlg,
    author = "Stone, E. C. and Cummings, A. C. and McDonald, F. B. and Heikkila, B. C. and Lal, N. and Webber, W. R.",
    title = "{Voyager 1 Observes Low-Energy Galactic Cosmic Rays in a Region Depleted of Heliospheric Ions}",
    doi = "10.1126/science.1236408",
    journal = "Science",
    volume = "341",
    number = "6142",
    pages = "1236408",
    year = "2013"
}

@article{Cirelli:2014lwa,
    author = {Cirelli, Marco and Gaggero, Daniele and Giesen, Ga{\"e}lle and Taoso, Marco and Urbano, Alfredo},
    title = "{Antiproton constraints on the GeV gamma-ray excess: a comprehensive analysis}",
    eprint = "1407.2173",
    archivePrefix = "arXiv",
    primaryClass = "hep-ph",
    doi = "10.1088/1475-7516/2014/12/045",
    journal = "JCAP",
    volume = "12",
    pages = "045",
    year = "2014"
}

@article{Cuoco:2016eej,
    author = {Cuoco, Alessandro and Kr{\"a}mer, Michael and Korsmeier, Michael},
    title = "{Novel Dark Matter Constraints from Antiprotons in Light of AMS-02}",
    eprint = "1610.03071",
    archivePrefix = "arXiv",
    primaryClass = "astro-ph.HE",
    doi = "10.1103/PhysRevLett.118.191102",
    journal = "Phys. Rev. Lett.",
    volume = "118",
    number = "19",
    pages = "191102",
    year = "2017"
}

@article{Boschini:2019ubh,
    author = "Boschini, Matteo Jerome and Della Torre, Stefano and Gervasi, Massimo and La Vacca, Giuseppe and Rancoita, Pier Giorgio",
    title = "{The HelMod Model in the Works for Inner and Outer Heliosphere: from AMS to Voyager Probes Observations}",
    eprint = "1903.07501",
    archivePrefix = "arXiv",
    primaryClass = "physics.space-ph",
    doi = "10.1016/j.asr.2019.04.007",
    journal = "Adv. SpaceRes.",
    volume = "64",
    number = "12",
    pages = "2459--2476",
    year = "2019"
}

@article{Boschini:2017gic,
    author = "Boschini, M. J. and Della Torre, S. and Gervasi, M. and La Vacca, G. and Rancoita, P. G.",
    title = "{Propagation of cosmic rays in heliosphere: The HELMOD model}",
    eprint = "1704.03733",
    archivePrefix = "arXiv",
    primaryClass = "astro-ph.SR",
    doi = "10.1016/j.asr.2017.04.017",
    journal = "Adv. Space Res.",
    volume = "62",
    pages = "2859--2879",
    year = "2018"
}

@article{Boschini:2022jwz,
    author = "Boschini, M. J. and Della Torre, S. and Gervasi, M. and La Vacca, G. and Rancoita, P. G.",
    title = "{Forecasting of cosmic rays intensities with HelMod Model}",
    doi = "10.1016/j.asr.2022.01.031",
    journal = "Adv. Space Res.",
    volume = "70",
    number = "9",
    pages = "2649--2657",
    year = "2022"
}

@article{Zhu:2025qqx,
    author = "Zhu, Cheng-Rui and Wang, Mei-Juan",
    title = "{Solar Modulation of AMS-02 Daily Proton and Helium Fluxes with Modified Force-field Approximation Models}",
    eprint = "2501.10922",
    archivePrefix = "arXiv",
    primaryClass = "astro-ph.HE",
    doi = "10.3847/1538-4357/adacd8",
    journal = "Astrophys. J.",
    volume = "980",
    number = "1",
    pages = "116",
    year = "2025"
}

@article{Zhu:2025lrc,
    author = "Zhu, Cheng-Rui and Wang, Mei-Juan",
    title = "{Probing Solar Modulation of AMS-02 Time-dependent D, $^{3}$He, and $^{4}$He Fluxes with Modified Force-field Approximation Models}",
    eprint = "2502.10016",
    archivePrefix = "arXiv",
    primaryClass = "astro-ph.HE",
    doi = "10.3847/1538-4357/adc12c",
    journal = "Astrophys. J.",
    volume = "983",
    number = "2",
    pages = "156",
    year = "2025"
}

@article{Tomassetti:2017gkx,
    author = "Tomassetti, Nicola and Orcinha, Miguel and Bar{\~a}o, Fernando and Bertucci, Bruna",
    title = "{Evidence for a Time Lag in Solar Modulation of Galactic Cosmic Rays}",
    eprint = "1707.06916",
    archivePrefix = "arXiv",
    primaryClass = "astro-ph.HE",
    doi = "10.3847/2041-8213/aa9373",
    journal = "Astrophys. J. Lett.",
    volume = "849",
    number = "2",
    pages = "L32",
    year = "2017"
}

@article{Tomassetti:2022fal,
    author = "Tomassetti, Nicola and Bertucci, Bruna and Fiandrini, Emanuele",
    title = "{Temporal evolution and rigidity dependence of the solar modulation lag of Galactic cosmic rays}",
    eprint = "2210.05693",
    archivePrefix = "arXiv",
    primaryClass = "astro-ph.SR",
    doi = "10.1103/PhysRevD.106.103022",
    journal = "Phys. Rev. D",
    volume = "106",
    number = "10",
    pages = "103022",
    year = "2022"
}

@misc{SILSO_Sunspot_Number,
    author = {{Clette}, F. and {Lefèvre}, L.},
    title = {SILSO Sunspot Number V2.0},
    doi = {10.24414/qnza-ac80},
    howpublished = {https://doi.org/10.24414/qnza-ac80},
    month = {07},
    year = {2015},
    note = {Published by WDC SILSO - Royal Observatory of Belgium (ROB)}
}

@article{Corti:2015bqi,
    author = "Corti, C. and Bindi, Veronica and Consolandi, Cristina and Whitman, Kathryn",
    title = "{Solar Modulation of the Local Interstellar Spectrum with Voyager 1, AMS-02, PAMELA, and BESS}",
    eprint = "1511.08790",
    archivePrefix = "arXiv",
    primaryClass = "astro-ph.HE",
    doi = "10.3847/0004-637X/829/1/8",
    journal = "Astrophys. J.",
    volume = "829",
    number = "1",
    pages = "8",
    year = "2016"
}

@article{Linden:2020lvz,
    author = "Linden, Tim and Beacom, John F. and Peter, Annika H. G. and Buckman, Benjamin J. and Zhou, Bei and Zhu, Guanying",
    title = "{First observations of solar disk gamma rays over a full solar cycle}",
    eprint = "2012.04654",
    archivePrefix = "arXiv",
    primaryClass = "astro-ph.HE",
    doi = "10.1103/PhysRevD.105.063013",
    journal = "Phys. Rev. D",
    volume = "105",
    number = "6",
    pages = "063013",
    year = "2022"
}

@article{Linden:2025xom,
    author = "Linden, Tim and Li, Jung-Tsung and Zhou, Bei and John, Isabelle and Crnogor{\v{c}}evi{\'c}, Milena and Peter, Annika H. G. and Beacom, John F.",
    title = "{First observations of solar halo gamma rays over a full solar cycle}",
    eprint = "2505.04625",
    archivePrefix = "arXiv",
    primaryClass = "astro-ph.HE",
    reportNumber = "FERMILAB-PUB-25-0301-T",
    doi = "10.1103/qm68-ng62",
    journal = "Phys. Rev. D",
    volume = "112",
    number = "10",
    pages = "103030",
    year = "2025"
}

@article{Acharyya:2025xya,
    author = "Acharyya, A. and others",
    title = "{Puzzling Variation of Gamma Rays from the Sun over the Solar Cycle Revealed with Fermi-LAT}",
    eprint = "2505.06348",
    archivePrefix = "arXiv",
    primaryClass = "astro-ph.HE",
    doi = "10.3847/2041-8213/adef4d",
    journal = "Astrophys. J. Lett.",
    volume = "989",
    number = "1",
    pages = "L16",
    year = "2025"
}

@article{Ng:2015gya,
    author = "Ng, Kenny C. Y. and Beacom, John F. and Peter, Annika H. G. and Rott, Carsten",
    title = "{First Observation of Time Variation in the Solar-Disk Gamma-Ray Flux with Fermi}",
    eprint = "1508.06276",
    archivePrefix = "arXiv",
    primaryClass = "astro-ph.HE",
    doi = "10.1103/PhysRevD.94.023004",
    journal = "Phys. Rev. D",
    volume = "94",
    number = "2",
    pages = "023004",
    year = "2016"
}

@article{Wang:2019xtu,
    author = "Wang, Bing-Bing and Bi, Xiao-Jun and Fang, Kung and Lin, Su-Jie and Yin, Peng-Fei",
    title = "{Time dependent solar modulation of cosmic rays from solar minimum to solar maximum}",
    eprint = "1904.03747",
    archivePrefix = "arXiv",
    primaryClass = "astro-ph.HE",
    doi = "10.1103/PhysRevD.100.063006",
    journal = "Phys. Rev. D",
    volume = "100",
    number = "6",
    pages = "063006",
    year = "2019"
}

@article{Usoskin:2017cli,
    author = "Usoskin, Ilya G. and Gil, Agnieszka and Kovaltsov, Gennady A. and Mishev, Alexander L. and Mikhailov, Vladimir V.",
    title = "{Heliospheric modulation of cosmic rays during the neutron monitor era: Calibration using PAMELA data for 2006{\textendash}2010}",
    eprint = "1705.07197",
    archivePrefix = "arXiv",
    primaryClass = "physics.space-ph",
    doi = "10.1002/2016JA023819",
    journal = "J. Geophys. Res. Space Phys.",
    volume = "122",
    number = "4",
    pages = "3875--3887",
    year = "2017"
}

@article{Li:2022zio,
    author = "Li, Jung-Tsung and Beacom, John F. and Peter, Annika H. G.",
    title = "{Galactic Cosmic-Ray Propagation in the Inner Heliosphere: Improved Force-field Model}",
    eprint = "2206.14815",
    archivePrefix = "arXiv",
    primaryClass = "astro-ph.SR",
    doi = "10.3847/1538-4357/ac8cf3",
    journal = "Astrophys. J.",
    volume = "937",
    pages = "27",
    year = "2022"
}

@article{Cui:2016ppb,
    author = "Cui, Ming-Yang and Yuan, Qiang and Tsai, Yue-Lin Sming and Fan, Yi-Zhong",
    title = "{Possible dark matter annihilation signal in the AMS-02 antiproton data}",
    eprint = "1610.03840",
    archivePrefix = "arXiv",
    primaryClass = "astro-ph.HE",
    doi = "10.1103/PhysRevLett.118.191101",
    journal = "Phys. Rev. Lett.",
    volume = "118",
    number = "19",
    pages = "191101",
    year = "2017"
}

@ARTICLE{2013LRSP...10....3P,
       author = {{Potgieter}, Marius S.},
        title = "{Solar Modulation of Cosmic Rays}",
      journal = {Living Reviews in Solar Physics},
         year = 2013,
        month = dec,
       volume = {10},
       number = {1},
          eid = {3},
        pages = {3},
          doi = {10.12942/lrsp-2013-3},
archivePrefix = {arXiv},
       eprint = {1306.4421},
 primaryClass = {physics.space-ph},
       adsurl = {https://ui.adsabs.harvard.edu/abs/2013LRSP...10....3P}
}

@article{Rankin:2022poh,
    author = "Rankin, Jamie S. and Bindi, Veronica and Bykov, Andrei M. and Cummings, Alan C. and Della Torre, Stefano and Florinski, Vladimir and Heber, Bernd and Potgieter, Marius S. and Stone, Edward C. and Zhang, Ming",
    title = "{Galactic Cosmic Rays Throughout the Heliosphere and in the Very Local Interstellar Medium}",
    doi = "10.1007/s11214-022-00912-4",
    journal = "Space Sci. Rev.",
    volume = "218",
    number = "5",
    pages = "42",
    year = "2022"
}

@article{Drury:2016ubm,
    author = "Drury, Luke O'C. and Strong, Andrew W.",
    title = "{Power requirements for cosmic ray propagation models involving diffusive reacceleration; estimates and implications for the damping of interstellar turbulence}",
    eprint = "1608.04227",
    archivePrefix = "arXiv",
    primaryClass = "astro-ph.HE",
    doi = "10.1051/0004-6361/201629526",
    journal = "Astron. Astrophys.",
    volume = "597",
    pages = "A117",
    year = "2017"
}

@article{Cholis:2019ejx,
    author = "Cholis, Ilias and Linden, Tim and Hooper, Dan",
    title = "{A Robust Excess in the Cosmic-Ray Antiproton Spectrum: Implications for Annihilating Dark Matter}",
    eprint = "1903.02549",
    archivePrefix = "arXiv",
    primaryClass = "astro-ph.HE",
    reportNumber = "FERMILAB-PUB-19-091-A",
    doi = "10.1103/PhysRevD.99.103026",
    journal = "Phys. Rev. D",
    volume = "99",
    number = "10",
    pages = "103026",
    year = "2019"
}

@article{Calore:2022stf,
    author = "Calore, Francesca and Cirelli, Marco and Derome, Laurent and Genolini, Yoann and Maurin, David and Salati, Pierre and Serpico, Pasquale Dario",
    title = "{AMS-02 antiprotons and dark matter: Trimmed hints and robust bounds}",
    eprint = "2202.03076",
    archivePrefix = "arXiv",
    primaryClass = "hep-ph",
    reportNumber = "LAPTH-003/22",
    doi = "10.21468/SciPostPhys.12.5.163",
    journal = "SciPost Phys.",
    volume = "12",
    number = "5",
    pages = "163",
    year = "2022"
}

@article{Bergstrom:2013jra,
    author = "Bergstrom, Lars and Bringmann, Torsten and Cholis, Ilias and Hooper, Dan and Weniger, Christoph",
    title = "{New Limits on Dark Matter Annihilation from AMS Cosmic Ray Positron Data}",
    eprint = "1306.3983",
    archivePrefix = "arXiv",
    primaryClass = "astro-ph.HE",
    reportNumber = "FERMILAB-PUB-13-202-A",
    doi = "10.1103/PhysRevLett.111.171101",
    journal = "Phys. Rev. Lett.",
    volume = "111",
    pages = "171101",
    year = "2013"
}

@article{John:2021ugy,
    author = "John, Isabelle and Linden, Tim",
    title = "{Cosmic-Ray Positrons Strongly Constrain Leptophilic Dark Matter}",
    eprint = "2107.10261",
    archivePrefix = "arXiv",
    primaryClass = "astro-ph.HE",
    doi = "10.1088/1475-7516/2021/12/007",
    journal = "JCAP",
    volume = "12",
    pages = "007",
    year = "2021"
}

@article{Adriani:2012paa,
    author = "Adriani, O. and others",
    title = "{Measurement of the flux of primary cosmic ray antiprotons with energies of 60-MeV to 350-GeV in the PAMELA experiment}",
    doi = "10.1134/S002136401222002X",
    journal = "Pisma Zh. Eksp. Teor. Fiz.",
    volume = "96",
    pages = "693--699",
    year = "2012"
}

@article{Adriani_2010,
   title={PAMELA Results on the Cosmic-Ray Antiproton Flux from 60 MeV to 180 GeV in Kinetic Energy},
   volume={105},
   ISSN={1079-7114},
   url={http://dx.doi.org/10.1103/PhysRevLett.105.121101},
   DOI={10.1103/physrevlett.105.121101},
   number={12},
   journal={Physical Review Letters},
   publisher={American Physical Society (APS)},
   author = "Adriani, O. and others",
   year={2010},
   month=sep }

@article{Adriani:2015kxa,
    author = "Adriani, O. and others",
    title = "{Time dependence of the e$^−$ flux measured by PAMELA during the 2006 july {\textendash}2009 december solar minimum}",
    eprint = "1512.01079",
    archivePrefix = "arXiv",
    primaryClass = "astro-ph.SR",
    doi = "10.1088/0004-637X/810/2/142",
    journal = "Astrophys. J.",
    volume = "810",
    number = "2",
    pages = "142",
    year = "2015"
}

@article{Adriani:2016uhu,
    author = "Adriani, O. and others",
    title = "{Time Dependence of the Electron and Positron Components of the Cosmic Radiation Measured by the PAMELA Experiment between July 2006 and December 2015}",
    eprint = "1606.08626",
    archivePrefix = "arXiv",
    primaryClass = "astro-ph.HE",
    doi = "10.1103/PhysRevLett.116.241105",
    journal = "Phys. Rev. Lett.",
    volume = "116",
    number = "24",
    pages = "241105",
    year = "2016"
}

@article{Nguyen:2024kwy,
    author = "Nguyen, Thong T. Q. and John, Isabelle and Linden, Tim and Tait, Tim M. P.",
    title = "{Strong Constraints on Dark Photon and Scalar Dark Matter Decay from INTEGRAL and AMS-02}",
    eprint = "2412.00180",
    archivePrefix = "arXiv",
    primaryClass = "hep-ph",
    reportNumber = "UCI-HEP-TR-2024-19",
    month = "11",
    year = "2024"
}

@article{Silver:2024ero,
    author = "Silver, Ethan and Orlando, Elena",
    title = "{Testing Cosmic-Ray Propagation Scenarios with AMS-02 and Voyager Data}",
    eprint = "2401.06242",
    archivePrefix = "arXiv",
    primaryClass = "astro-ph.HE",
    doi = "10.3847/1538-4357/ad1ce8",
    journal = "Astrophys. J.",
    volume = "963",
    number = "2",
    pages = "111",
    year = "2024"
}

@ARTICLE{2015ApJ...812..140S,
       author = {{Strub}, Peter and {Kr{\"u}ger}, Harald and {Sterken}, Veerle J.},
        title = "{Sixteen Years of Ulysses Interstellar Dust Measurements in the Solar System. II. Fluctuations in the Dust Flow from the Data}",
      journal = {\apj},
         year = 2015,
        month = oct,
       volume = {812},
       number = {2},
          eid = {140},
        pages = {140},
          doi = {10.1088/0004-637X/812/2/140},
archivePrefix = {arXiv},
       eprint = {1508.03242},
 primaryClass = {astro-ph.EP},
       adsurl = {https://ui.adsabs.harvard.edu/abs/2015ApJ...812..140S}
}

@ARTICLE{2010ApJ...722L..58S,
       author = {{Strong}, A.~W. and {Porter}, T.~A. and {Digel}, S.~W. and {J{\'o}hannesson}, G. and {Martin}, P. and {Moskalenko}, I.~V. and {Murphy}, E.~J. and {Orlando}, E.},
        title = "{Global Cosmic-ray-related Luminosity and Energy Budget of the Milky Way}",
      journal = {\apjl},
         year = 2010,
        month = oct,
       volume = {722},
       number = {1},
        pages = {L58-L63},
          doi = {10.1088/2041-8205/722/1/L58},
archivePrefix = {arXiv},
       eprint = {1008.4330},
 primaryClass = {astro-ph.HE},
       adsurl = {https://ui.adsabs.harvard.edu/abs/2010ApJ...722L..58S}
}

@article{DeLaTorreLuque:2024wtv,
    author = "De La Torre Luque, Pedro and Linden, Tim",
    title = "{Galactic gas models strongly affect the determination of the diffusive halo height}",
    eprint = "2408.05179",
    archivePrefix = "arXiv",
    primaryClass = "astro-ph.GA",
    doi = "10.1088/1475-7516/2025/02/062",
    journal = "JCAP",
    volume = "02",
    pages = "062",
    year = "2025"
}


\appendix

\setcounter{equation}{0}
\setcounter{figure}{0}
\setcounter{section}{0}
\setcounter{table}{0}
\makeatletter
\renewcommand{\theequation}{\thesection.\arabic{equation}}
\renewcommand{\thefigure}{\thesection.\arabic{figure}}
\renewcommand{\thetable}{\thesection.\arabic{table}}

\section{Further LIS and Modulated Fluxes of Cosmic-Ray Nuclei}
\label{app: fluxes}
\setcounter{figure}{0} 
\setcounter{table}{0} 
\setcounter{equation}{0}  

\begin{figure*}[tbp]
    \begin{minipage}[t]{0.48\linewidth}
        \centering
        \includegraphics[width=1\linewidth]{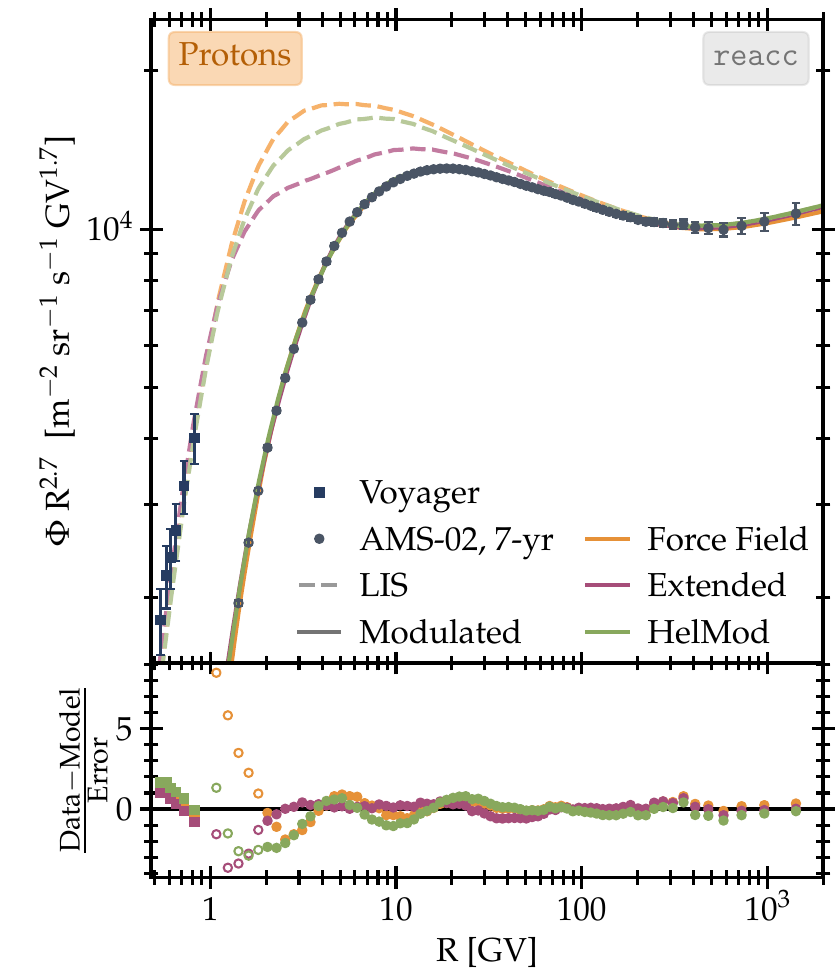}
    \end{minipage}
    \hfill
    \begin{minipage}[t]{0.48\linewidth}
        \centering
        \includegraphics[width=1\linewidth]{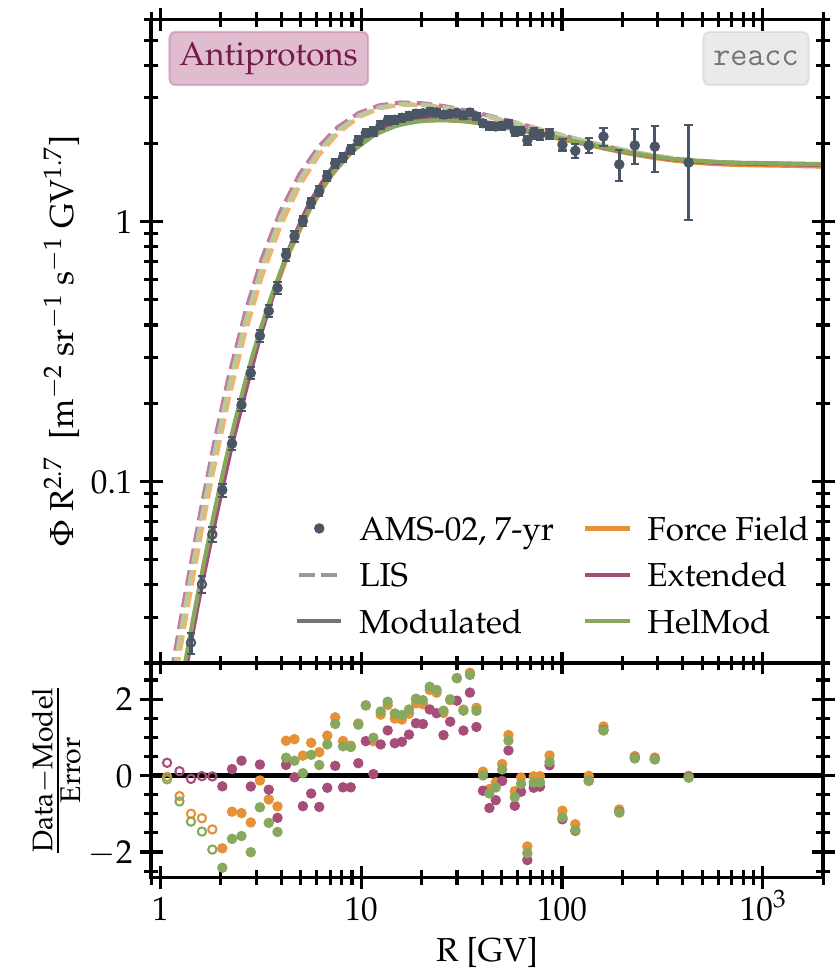}
    \end{minipage}
    \caption{The spectra for protons (left panel) and antiprotons (right panel) for the 7-year time-integrated fit in the \texttt{reacc} model. In each plot, the LIS (dashed line) and modulated flux (solid) line are given for a different solar-modulation model: the force-field potential (orange), the extended force-field potential (burgundy) and the \texttt{HelMod} model (green). The residuals in each plot are compared to the 7-yr time-integrated AMS-02 data (circle markers), as well as the Voyager data for protons (square markers).}
    \label{fig: LIS reacc time-independent}
\end{figure*}

Figure~\ref{fig: LIS reacc time-independent} is similar to Fig.~\ref{fig: LIS conv time-independent} but for the \reacc propagation model, showing the cosmic-ray fluxes for our 7-year time-integrated fits with the force-field (orange), extended (burgundy) and \texttt{HelMod} (green) solar modulations models. The agreement with the data is similar to the \conv case with the residuals at a level of $\sim 1-2\,\sigma$. The main difference is a larger spread in the proton LIS fluxes, mainly due to the LIS of the extended model differing the most from the other two. This LIS is similar to the one obtained in time-dependent fits and is further discussed in the main text.

\begin{figure*}[tbp]
    \begin{minipage}[t]{0.32\linewidth}
        \centering
        \includegraphics[width=1\linewidth]{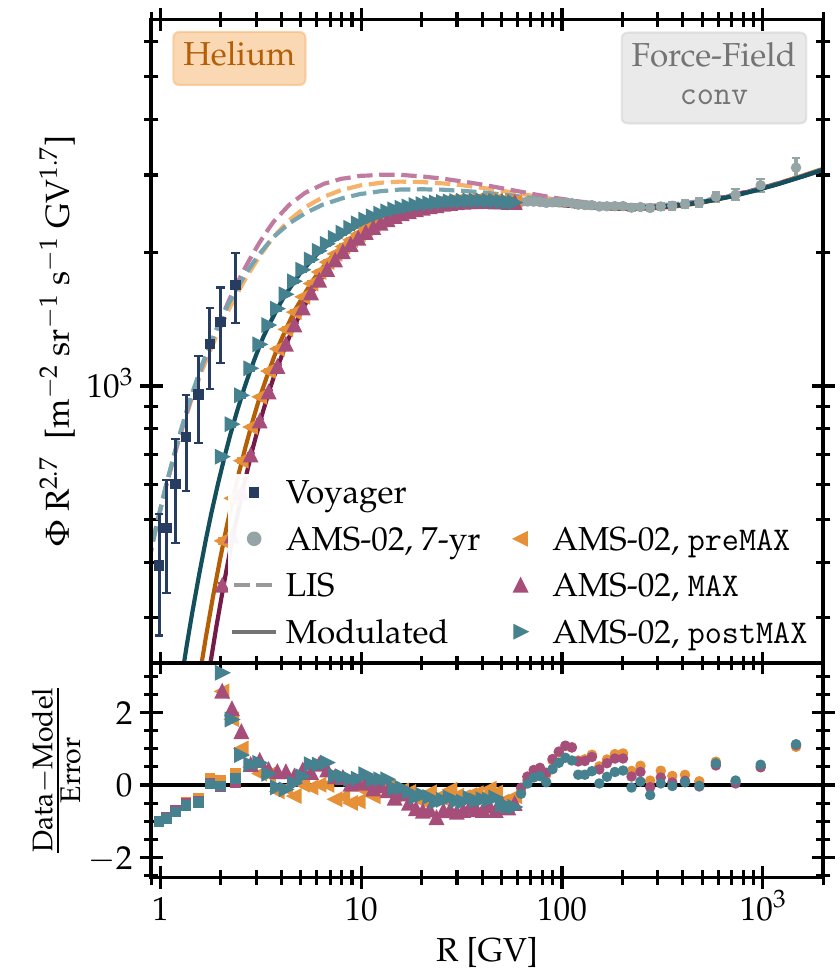}
    \end{minipage}
    \hfill
    \begin{minipage}[t]{0.32\linewidth}
        \centering
        \includegraphics[width=1\linewidth]{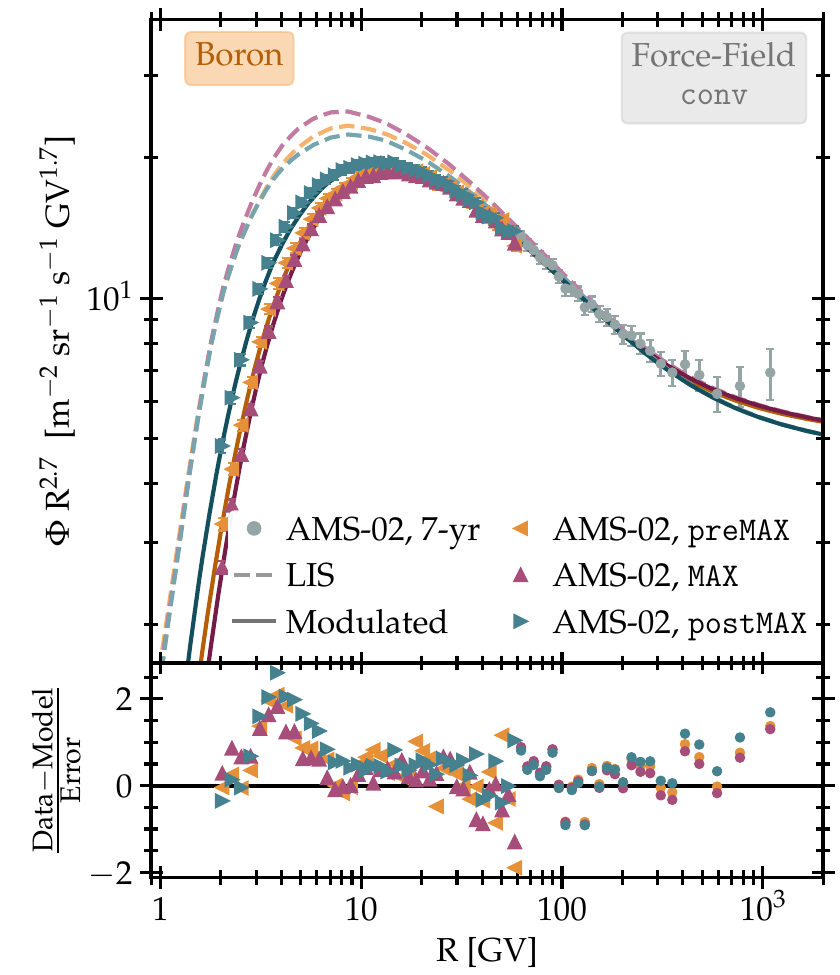}
    \end{minipage}
    \hfill
    \begin{minipage}[t]{0.32\linewidth}
        \centering
        \includegraphics[width=1\linewidth]{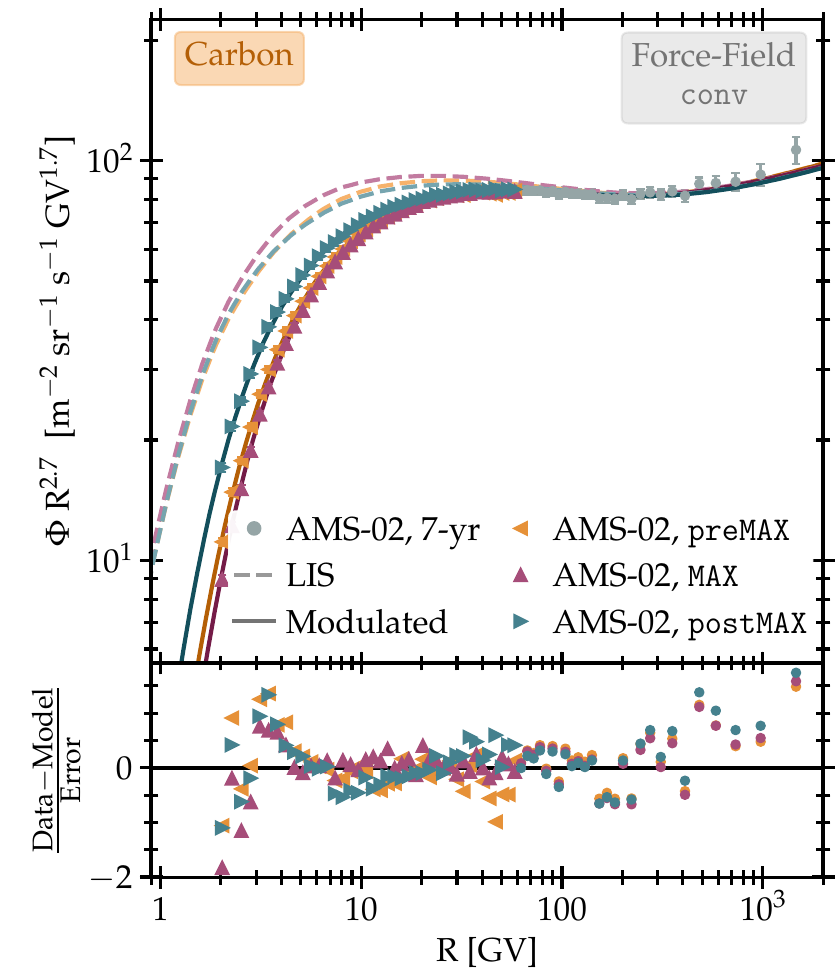}
    \end{minipage}
    \vfill\vspace{0.3cm}
    \begin{minipage}[t]{0.32\linewidth}
        \centering
        \includegraphics[width=1\linewidth]{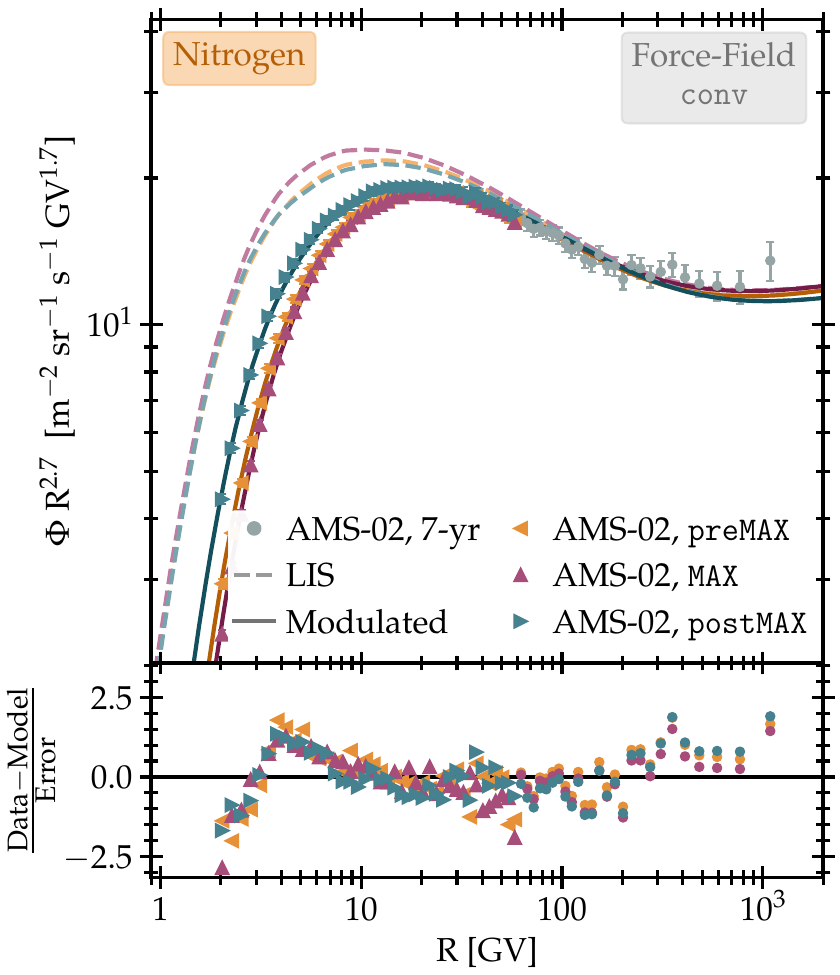}
    \end{minipage}
    \hfill
    \begin{minipage}[t]{0.32\linewidth}
        \centering
        \includegraphics[width=1\linewidth]{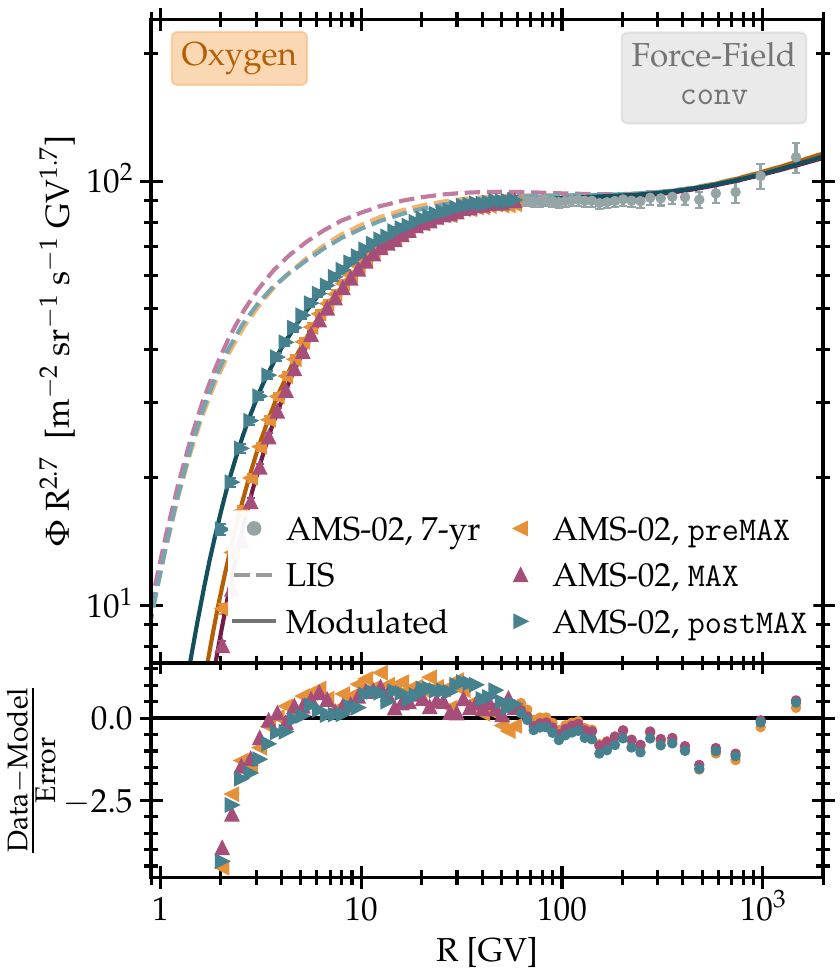}
    \end{minipage}
    \hfill
    \begin{minipage}[t]{0.32\linewidth}
        \centering
        \includegraphics[width=1\linewidth]{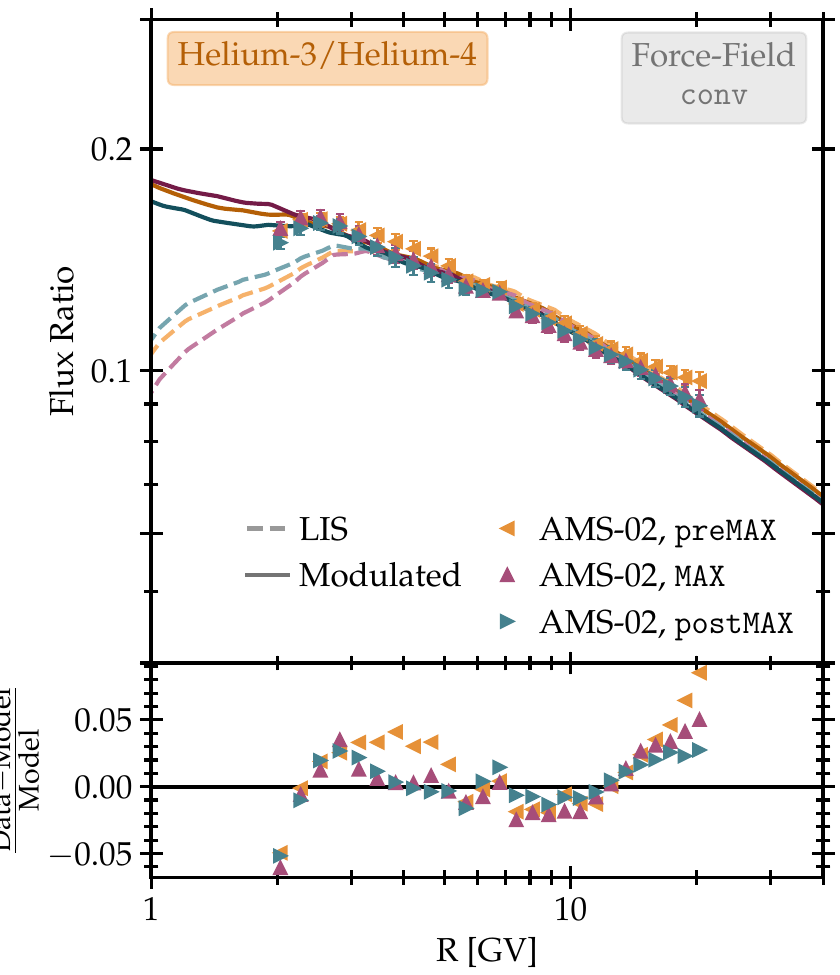}
    \end{minipage}
    \vfill\vspace{0.3cm}
    \begin{minipage}[t]{0.32\linewidth}
        \centering
        \includegraphics[width=1\linewidth]{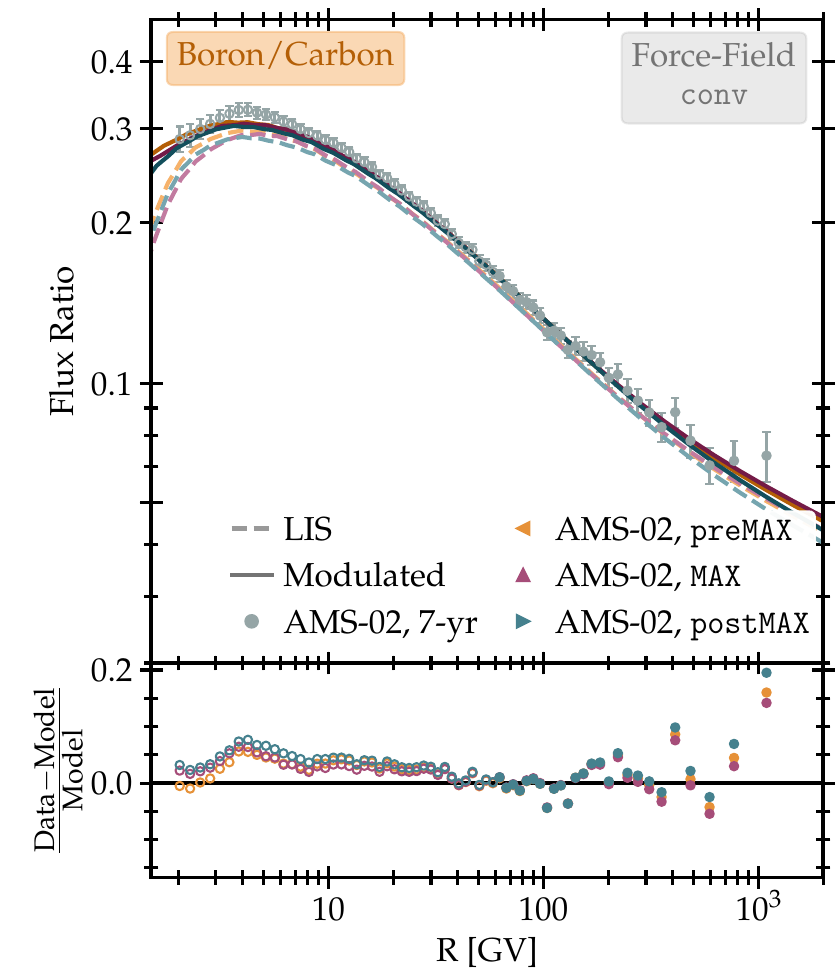}
    \end{minipage}
    \hfill
    \begin{minipage}[t]{0.32\linewidth}
        \centering
        \includegraphics[width=1\linewidth]{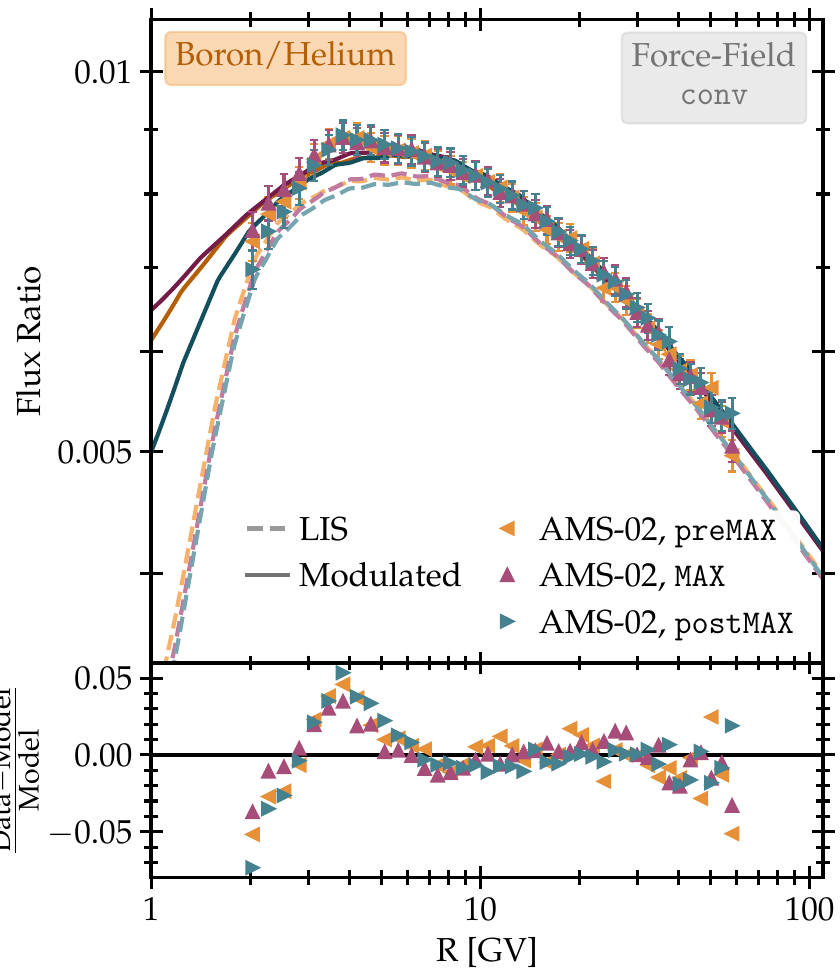}
    \end{minipage}
    \hfill
    \begin{minipage}[t]{0.32\linewidth}
        \centering
        \includegraphics[width=1\linewidth]{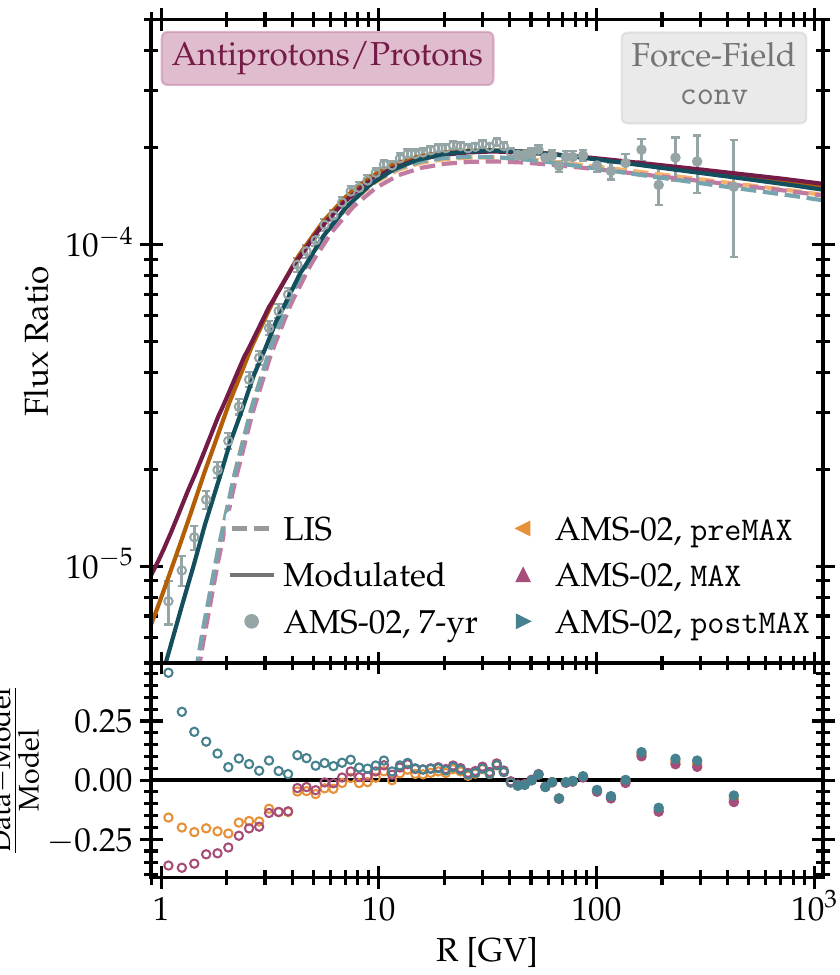}
    \end{minipage}
    \caption{Similar to Fig.~\ref{fig: LIS conv}, the LIS and modulated fluxes of the fits to the 3 time-periods for the \conv propagation and force-field modulation model. Note that for the Boron/Carbon and antiproton/proton no time-dependent data is available.}
    \label{fig: LIS nuclei conv FF}
\end{figure*}

\begin{figure*}[tbp]
    \begin{minipage}[t]{0.32\linewidth}
        \centering
        \includegraphics[width=1\linewidth]{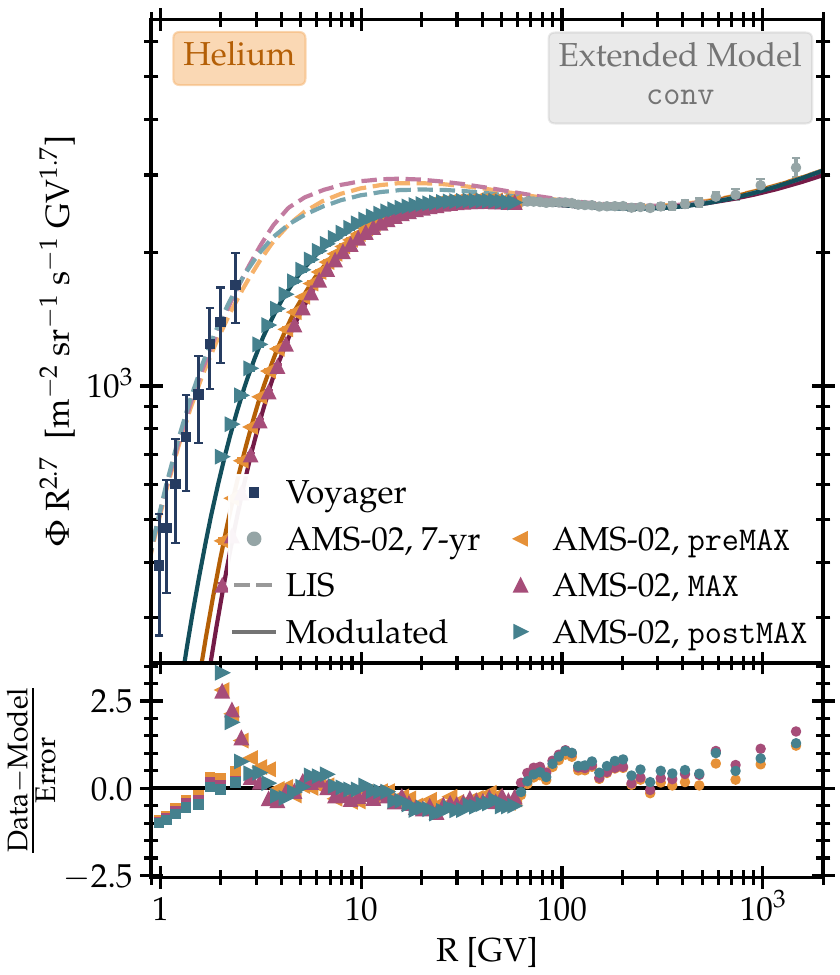}
    \end{minipage}
    \hfill
    \begin{minipage}[t]{0.32\linewidth}
        \centering
        \includegraphics[width=1\linewidth]{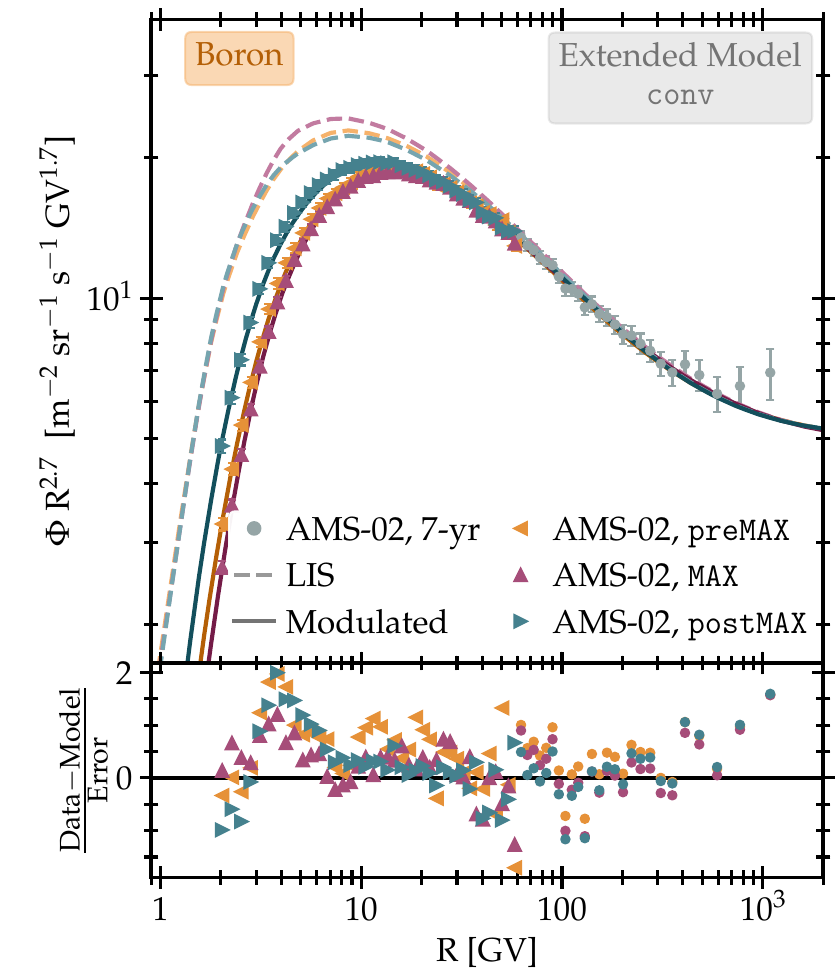}
    \end{minipage}
    \hfill
    \begin{minipage}[t]{0.32\linewidth}
        \centering
        \includegraphics[width=1\linewidth]{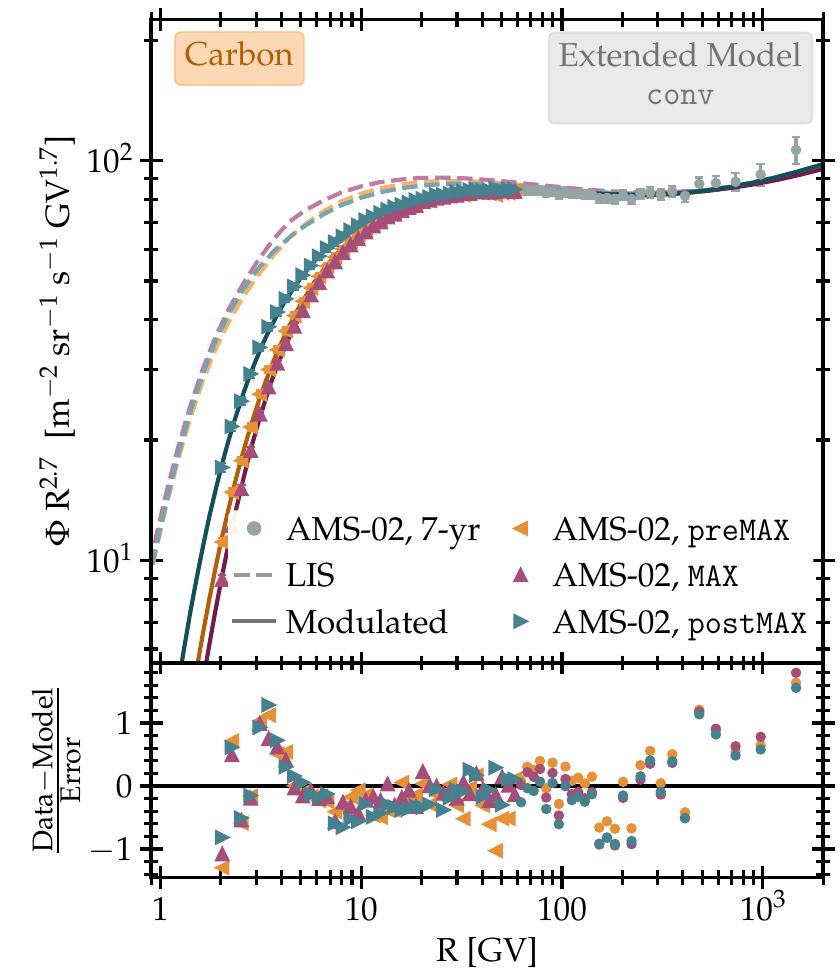}
    \end{minipage}
    \vfill\vspace{0.3cm}
    \begin{minipage}[t]{0.32\linewidth}
        \centering
        \includegraphics[width=1\linewidth]{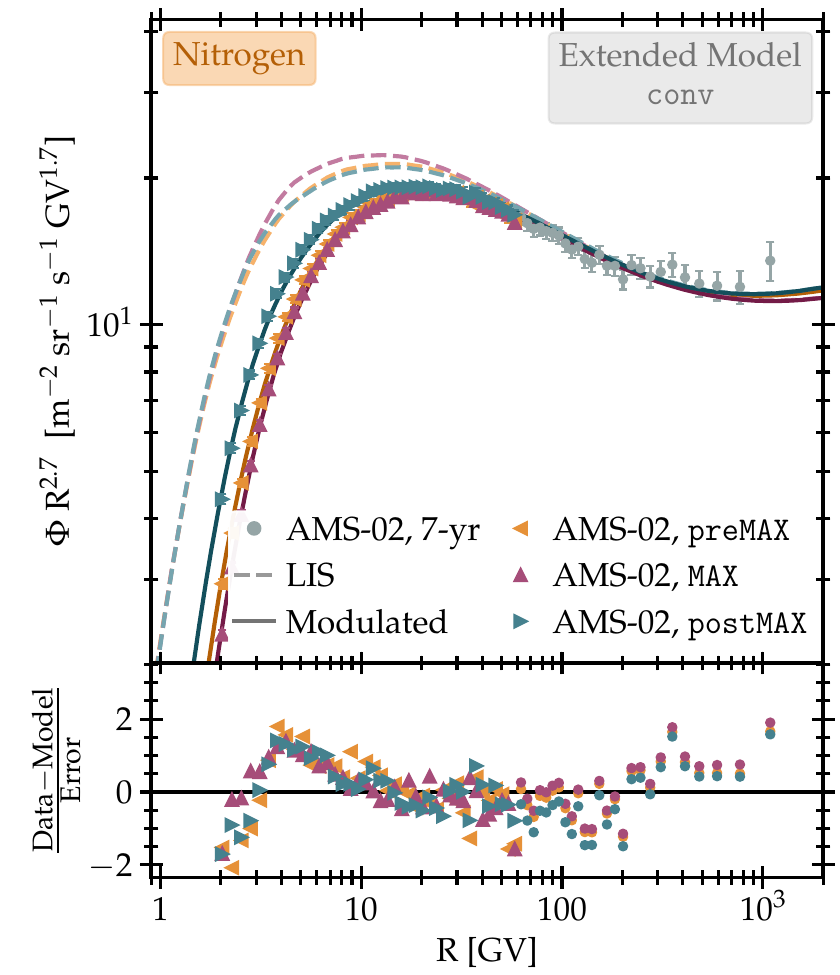}
    \end{minipage}
    \hfill
    \begin{minipage}[t]{0.32\linewidth}
        \centering
        \includegraphics[width=1\linewidth]{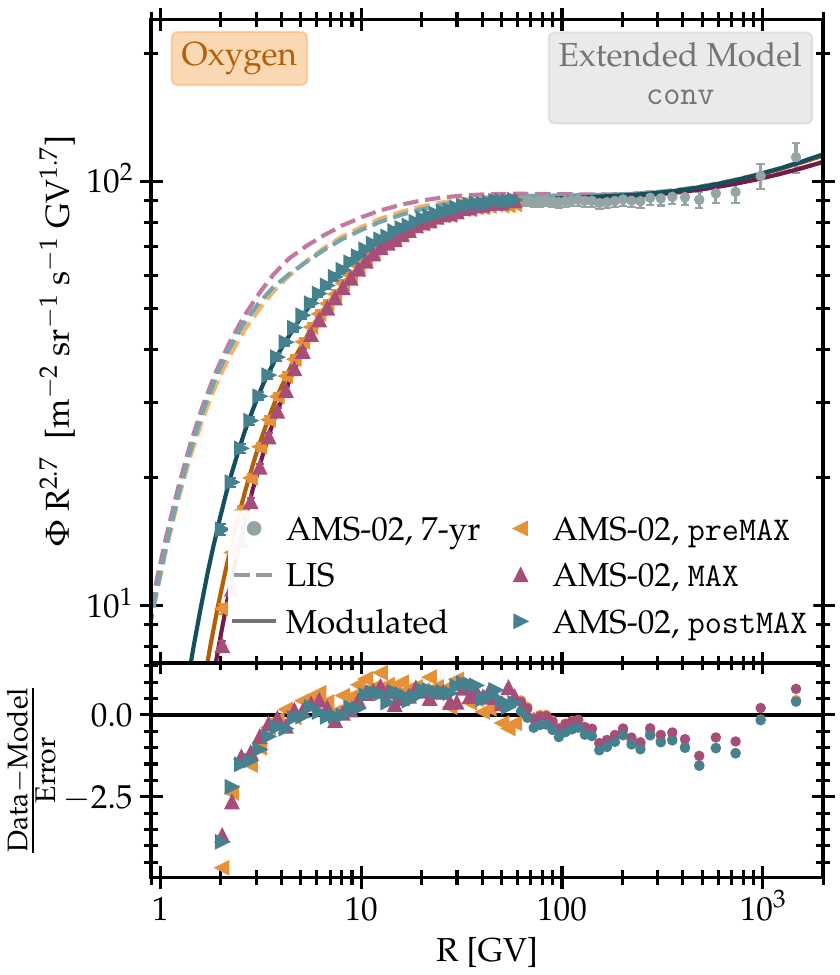}
    \end{minipage}
    \hfill
    \begin{minipage}[t]{0.32\linewidth}
        \centering
        \includegraphics[width=1\linewidth]{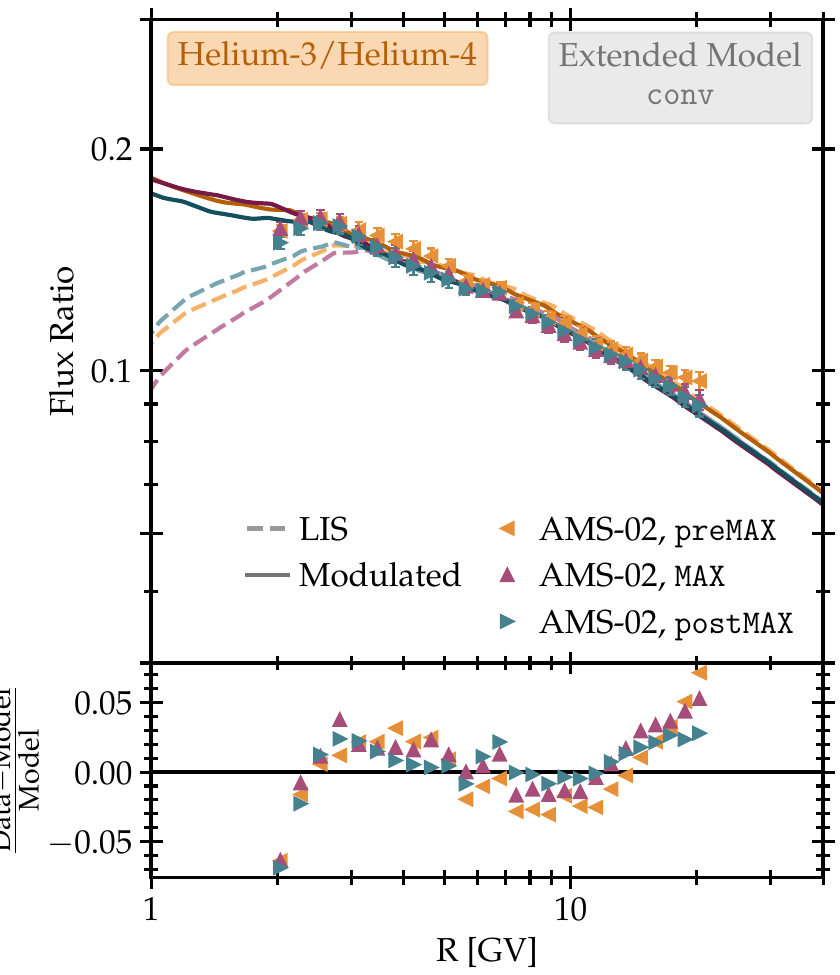}
    \end{minipage}
    \vfill\vspace{0.3cm}
    \begin{minipage}[t]{0.32\linewidth}
        \centering
        \includegraphics[width=1\linewidth]{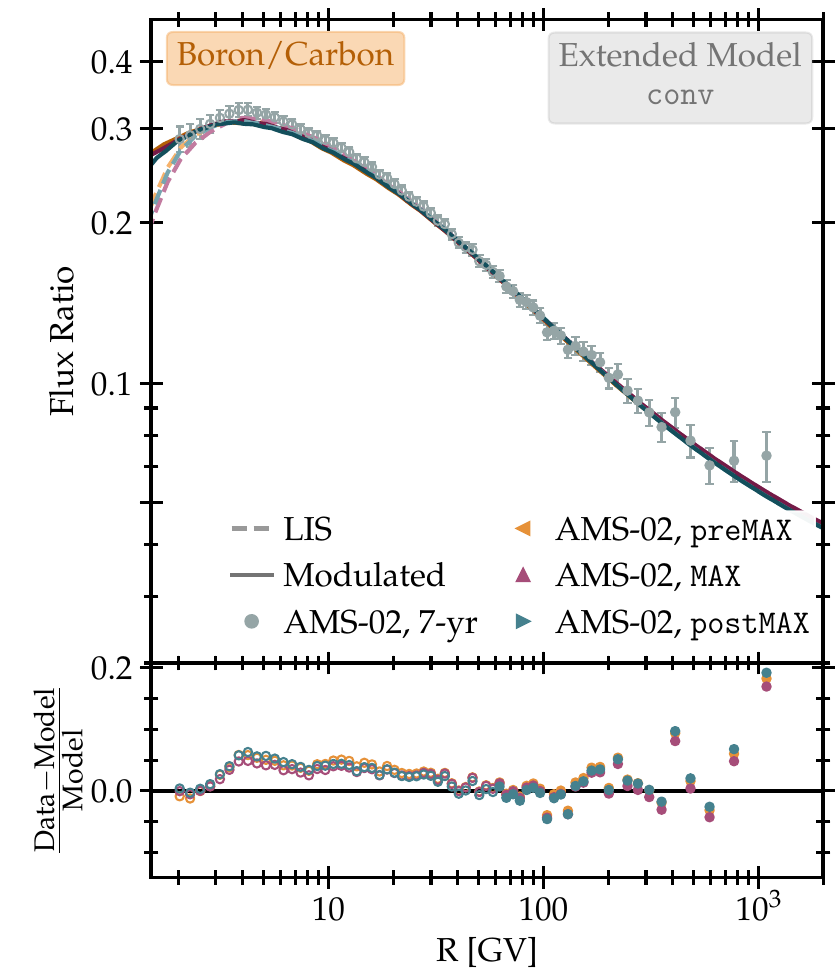}
    \end{minipage}
    \hfill
    \begin{minipage}[t]{0.32\linewidth}
        \centering
        \includegraphics[width=1\linewidth]{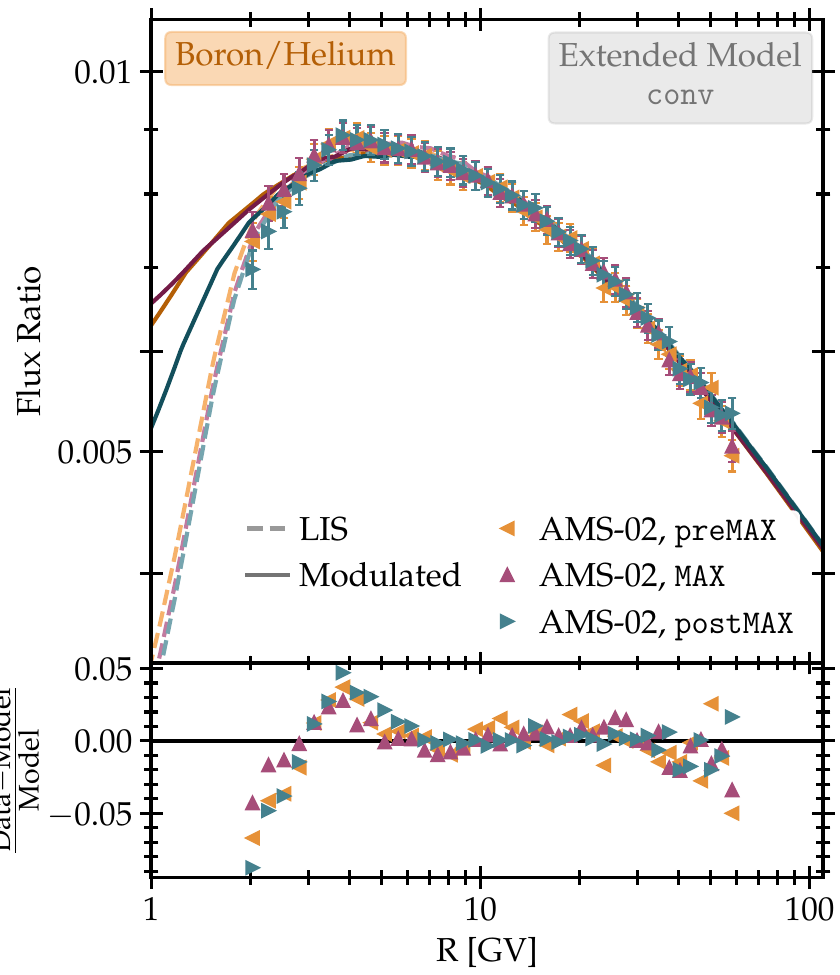}
    \end{minipage}
    \hfill
    \begin{minipage}[t]{0.32\linewidth}
        \centering
        \includegraphics[width=1\linewidth]{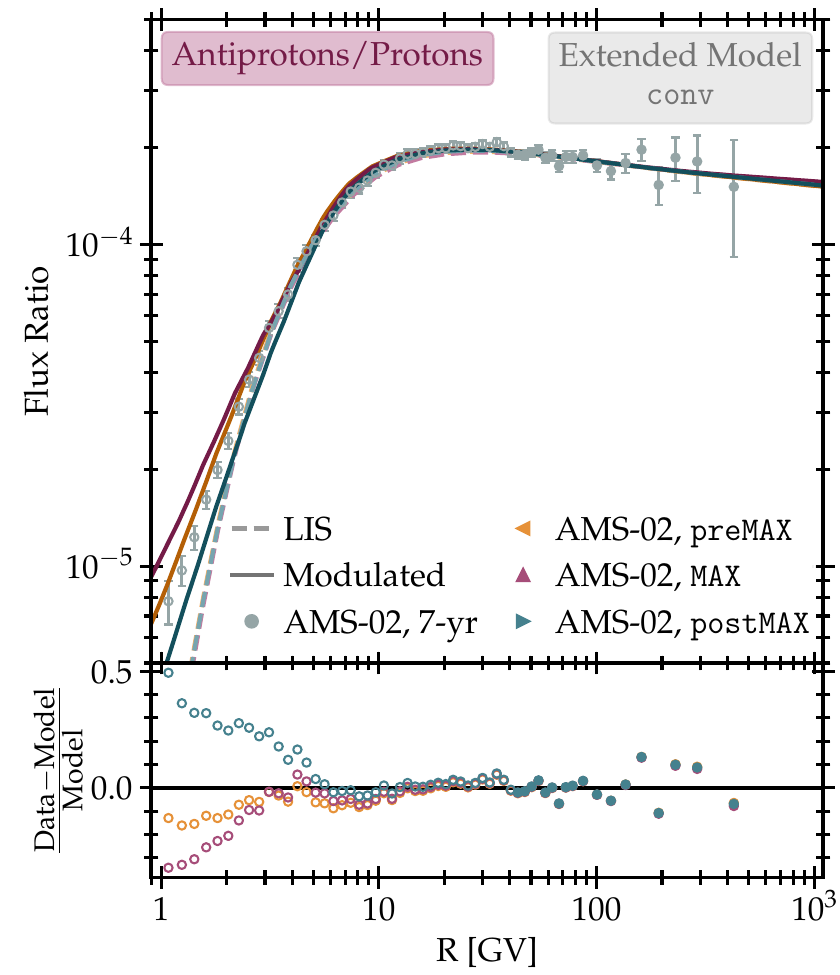}
    \end{minipage}
    \caption{Similar to Fig.~\ref{fig: LIS conv}, the LIS and modulated fluxes of the fits to the 3 time-periods for the \conv propagation and extended modulation model. Note that for the Boron/Carbon and antiproton/proton no time-dependent data is available.}
    \label{fig: LIS nuclei conv Zhu}
\end{figure*}

\begin{figure*}[tbp]
    \begin{minipage}[t]{0.32\linewidth}
        \centering
        \includegraphics[width=1\linewidth]{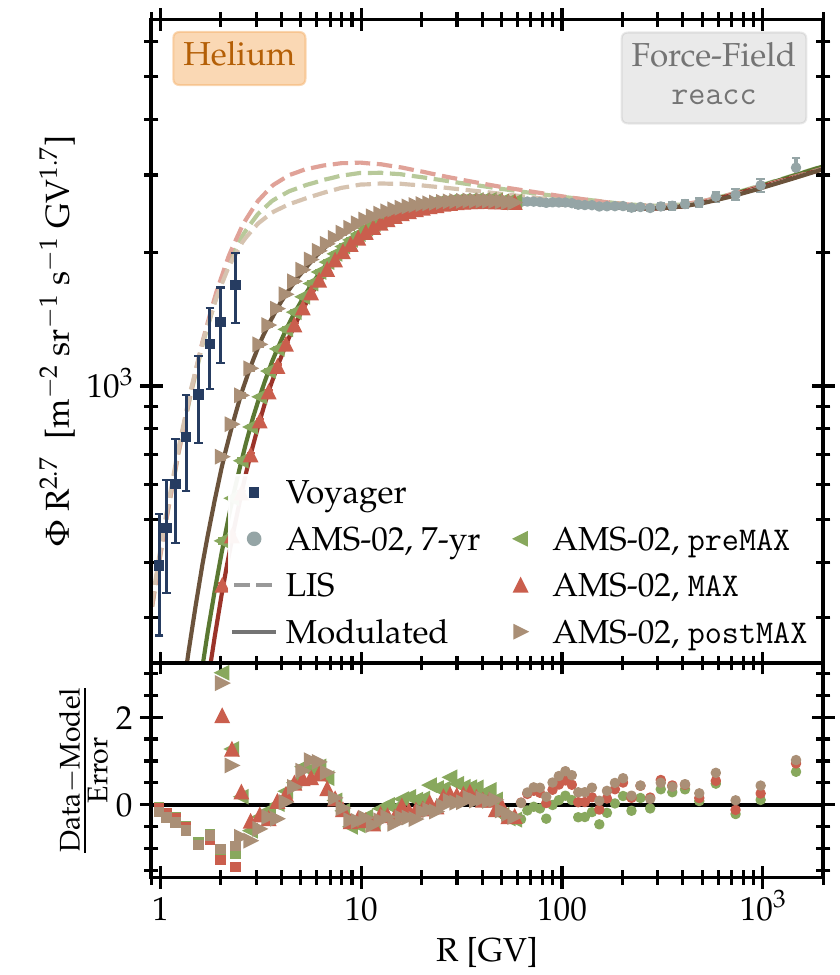}
    \end{minipage}
    \hfill
    \begin{minipage}[t]{0.32\linewidth}
        \centering
        \includegraphics[width=1\linewidth]{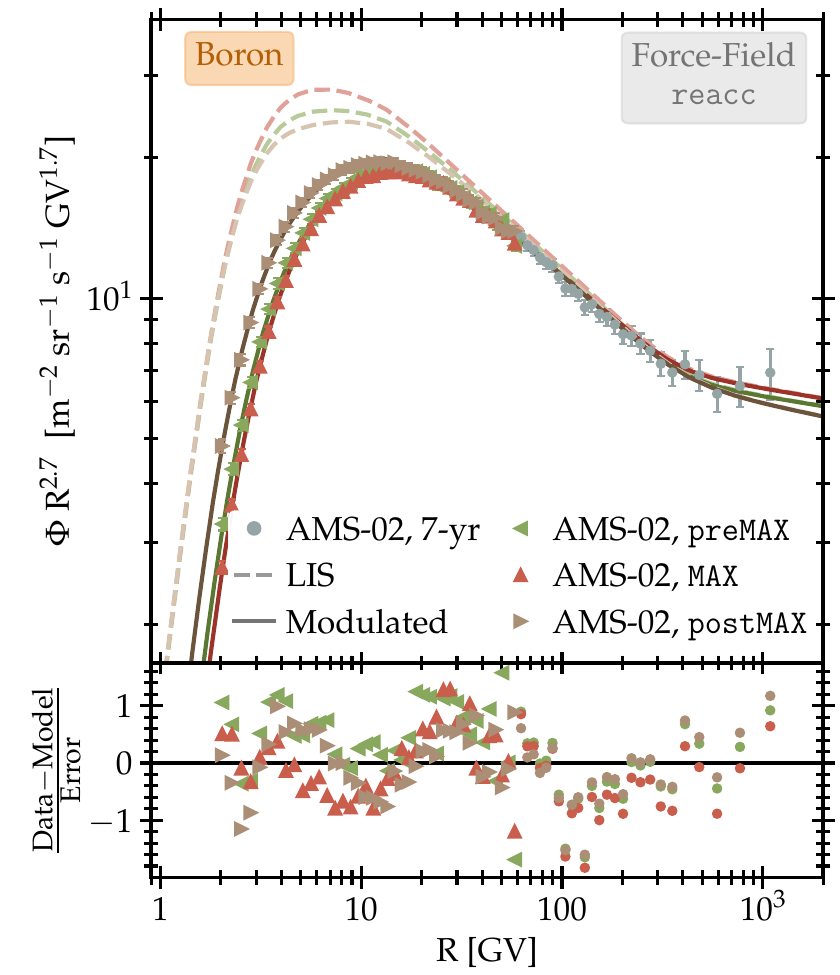}
    \end{minipage}
    \hfill
    \begin{minipage}[t]{0.32\linewidth}
        \centering
        \includegraphics[width=1\linewidth]{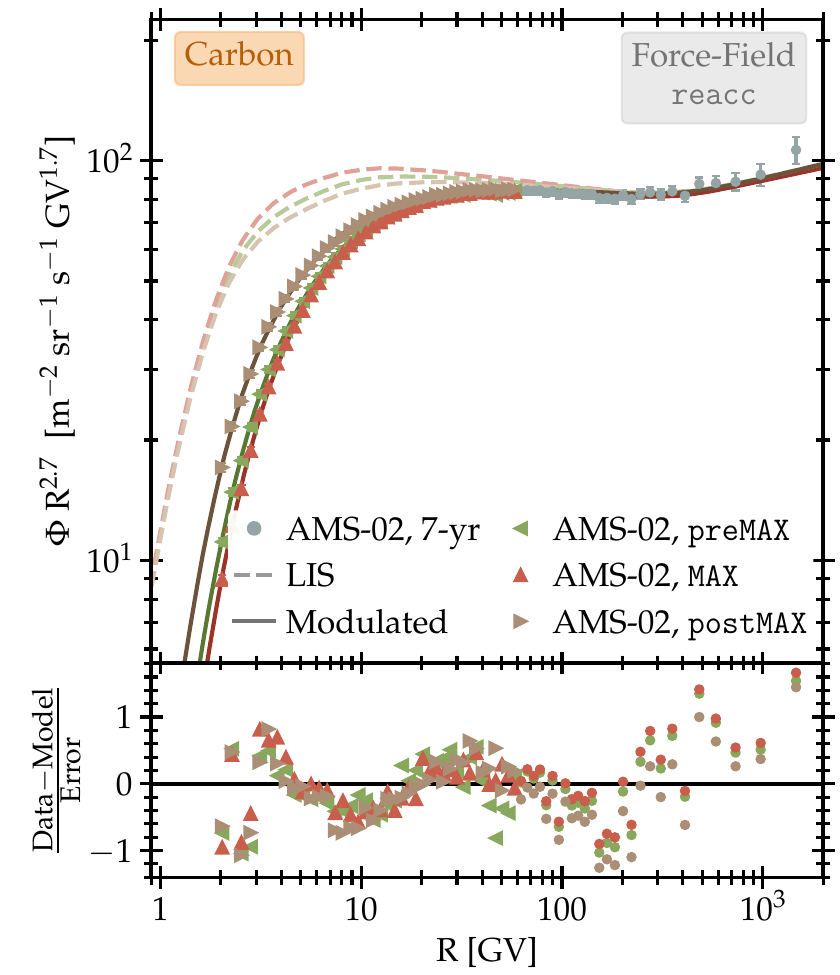}
    \end{minipage}
    \vfill\vspace{0.3cm}
    \begin{minipage}[t]{0.32\linewidth}
        \centering
        \includegraphics[width=1\linewidth]{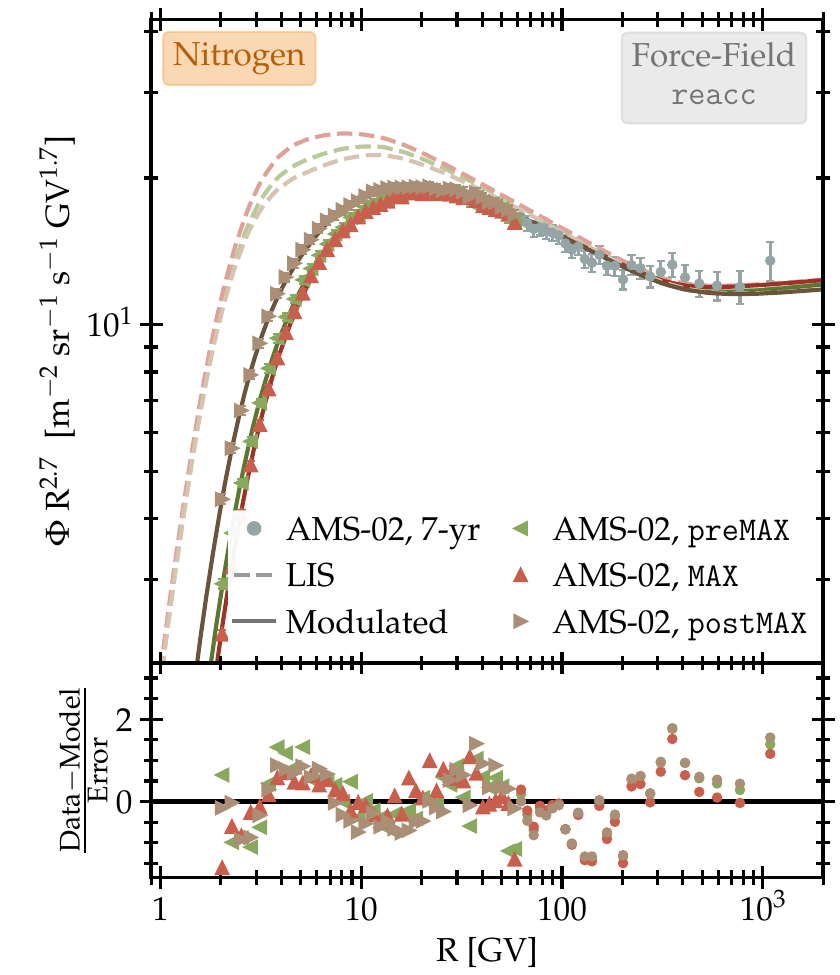}
    \end{minipage}
    \hfill
    \begin{minipage}[t]{0.32\linewidth}
        \centering
        \includegraphics[width=1\linewidth]{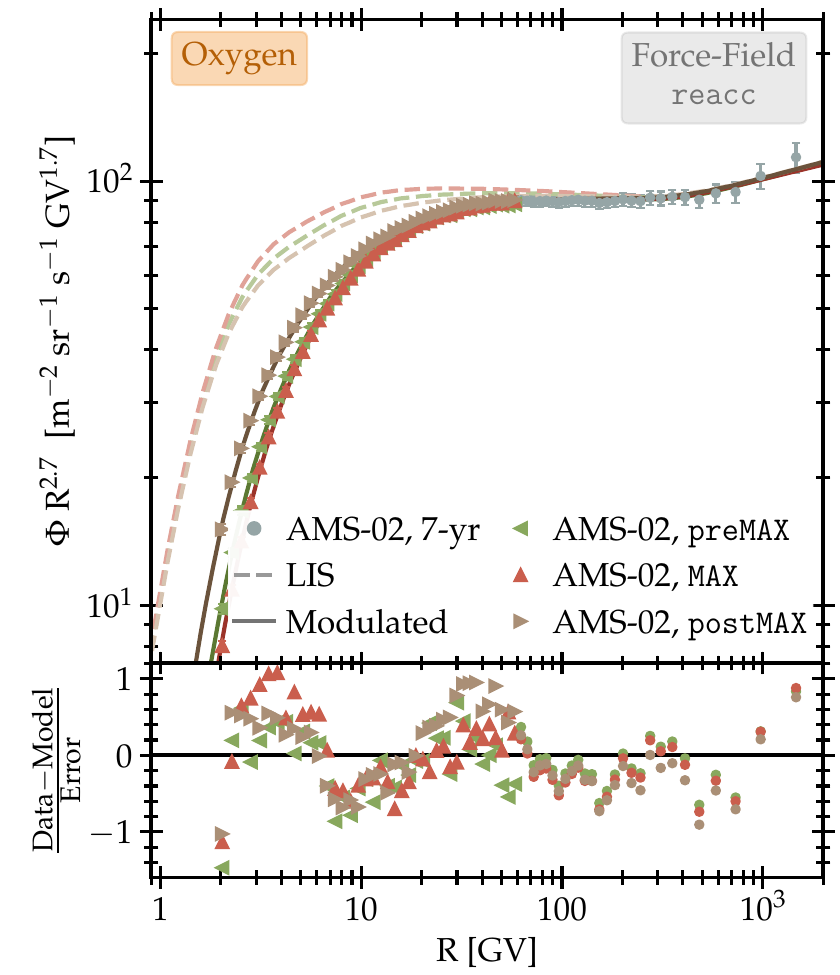}
    \end{minipage}
    \hfill
    \begin{minipage}[t]{0.32\linewidth}
        \centering
        \includegraphics[width=1\linewidth]{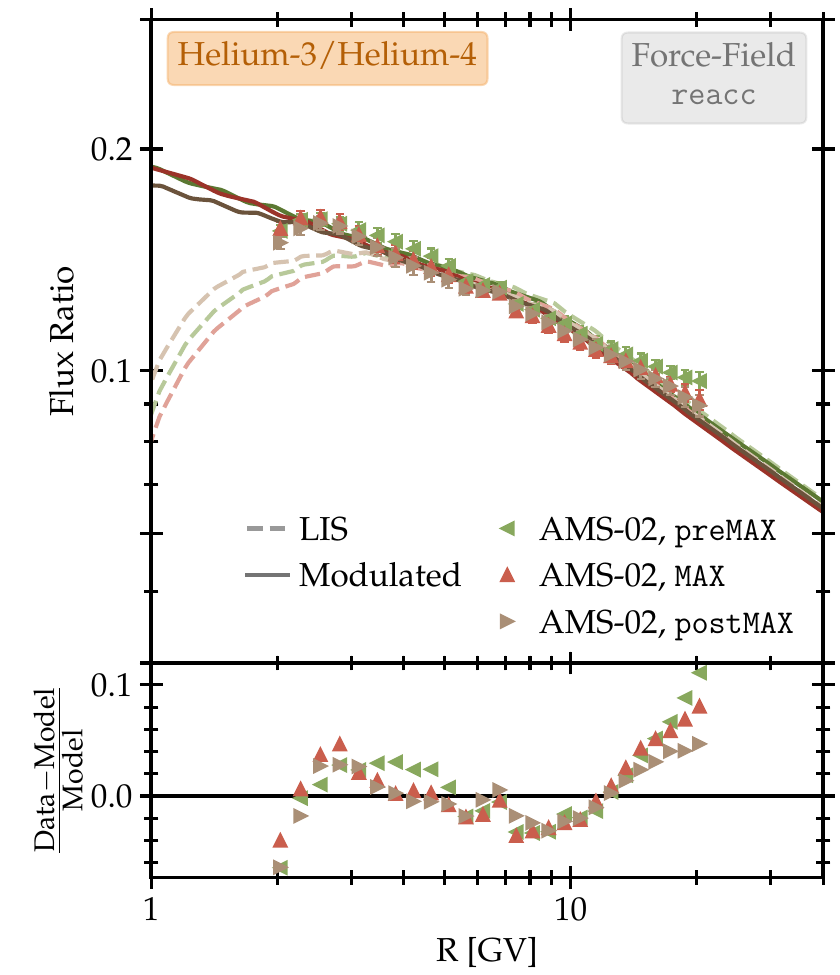}
    \end{minipage}
    \vfill\vspace{0.3cm}
    \begin{minipage}[t]{0.32\linewidth}
        \centering
        \includegraphics[width=1\linewidth]{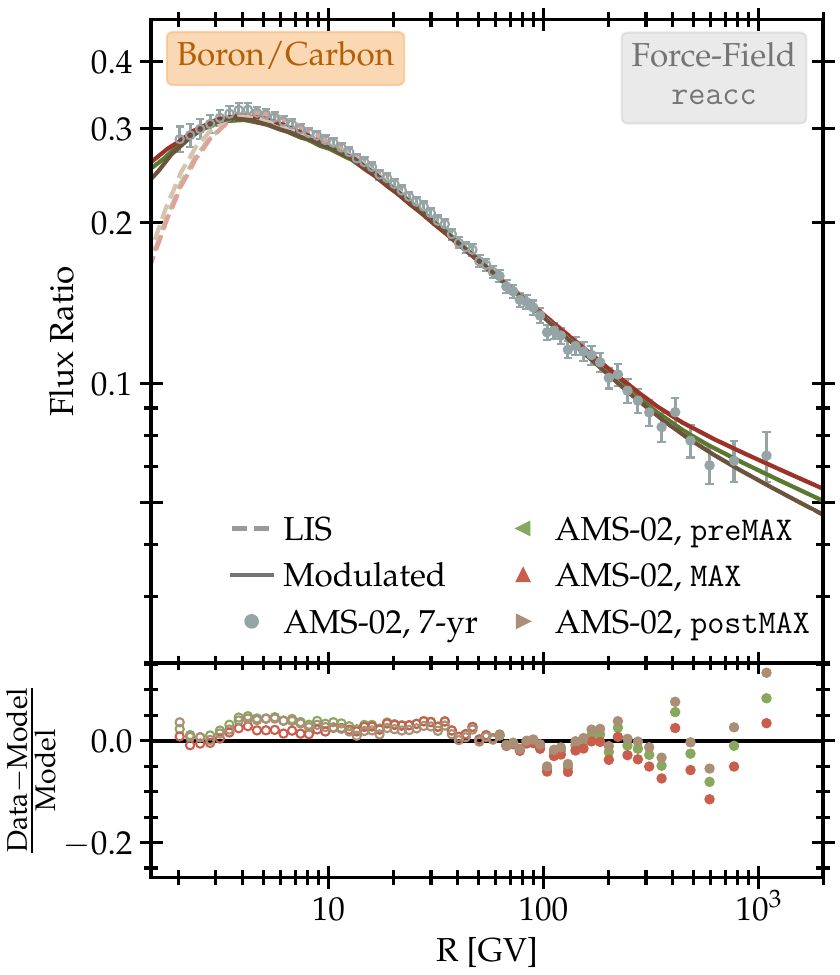}
    \end{minipage}
    \hfill
    \begin{minipage}[t]{0.32\linewidth}
        \centering
        \includegraphics[width=1\linewidth]{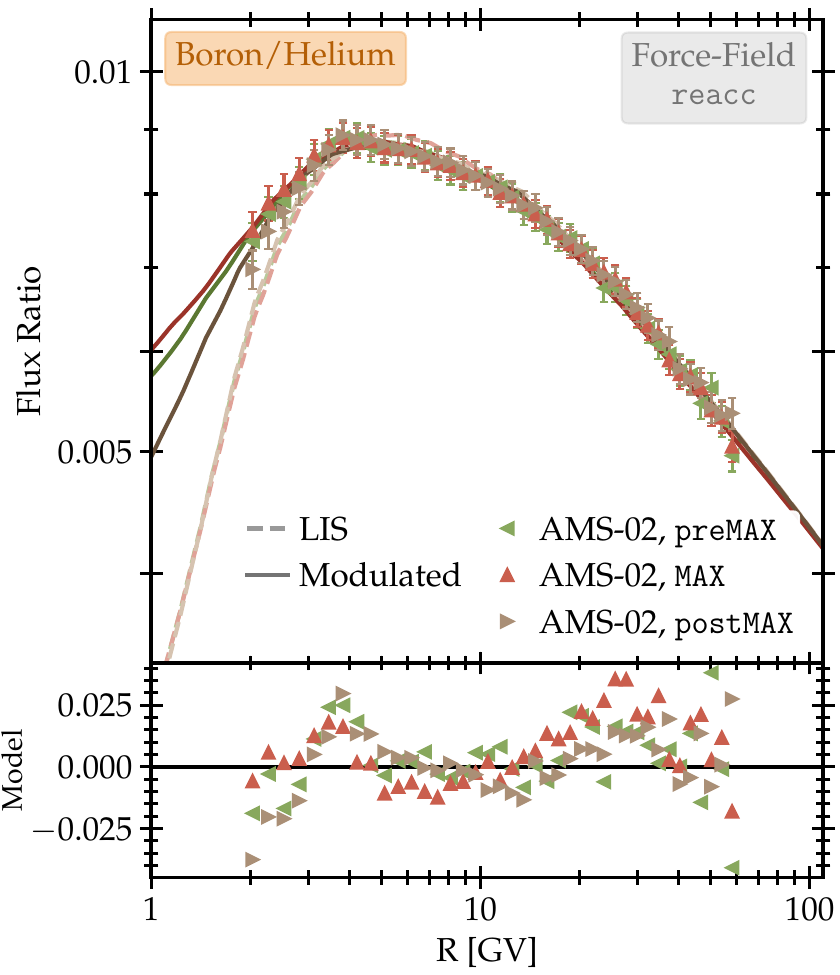}
    \end{minipage}
    \hfill
    \begin{minipage}[t]{0.32\linewidth}
        \centering
        \includegraphics[width=1\linewidth]{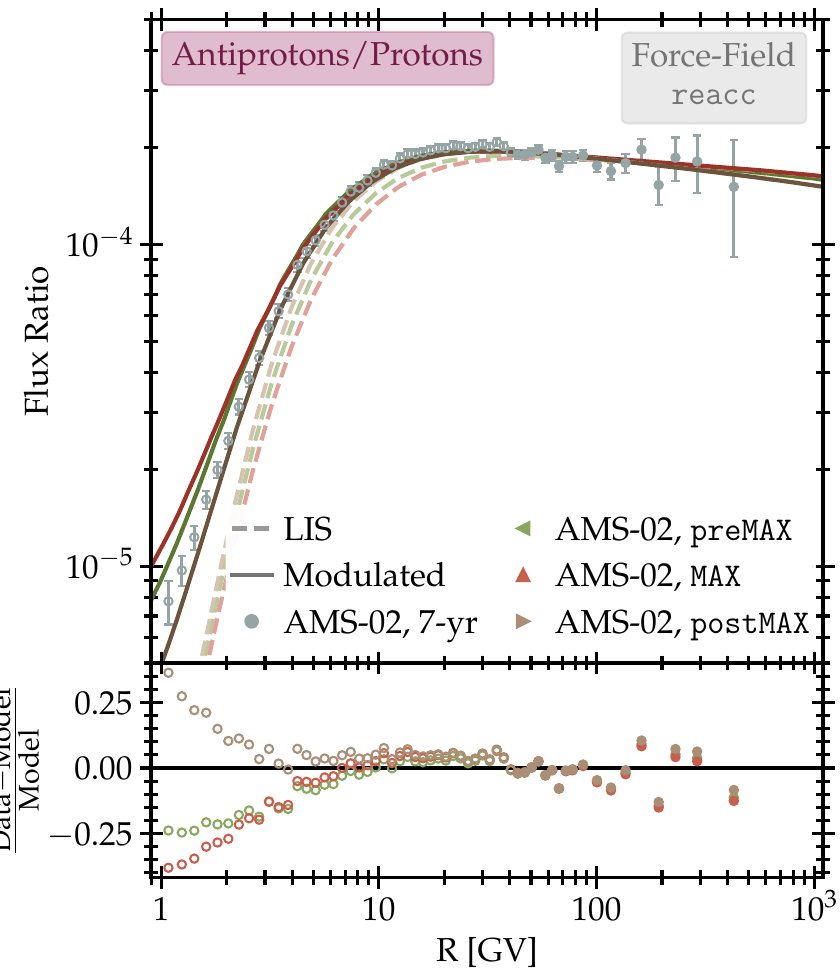}
    \end{minipage}
    \caption{Similar to Fig.~\ref{fig: LIS reacc}, the LIS and modulated fluxes of the fits to the 3 time-periods for the \reacc propagation and force-field modulation model. Note that for the Boron/Carbon and antiproton/proton no time-dependent data is available.}
    \label{fig: LIS nuclei reacc FF}
\end{figure*}

\begin{figure*}[tbp]
    \begin{minipage}[t]{0.32\linewidth}
        \centering
        \includegraphics[width=1\linewidth]{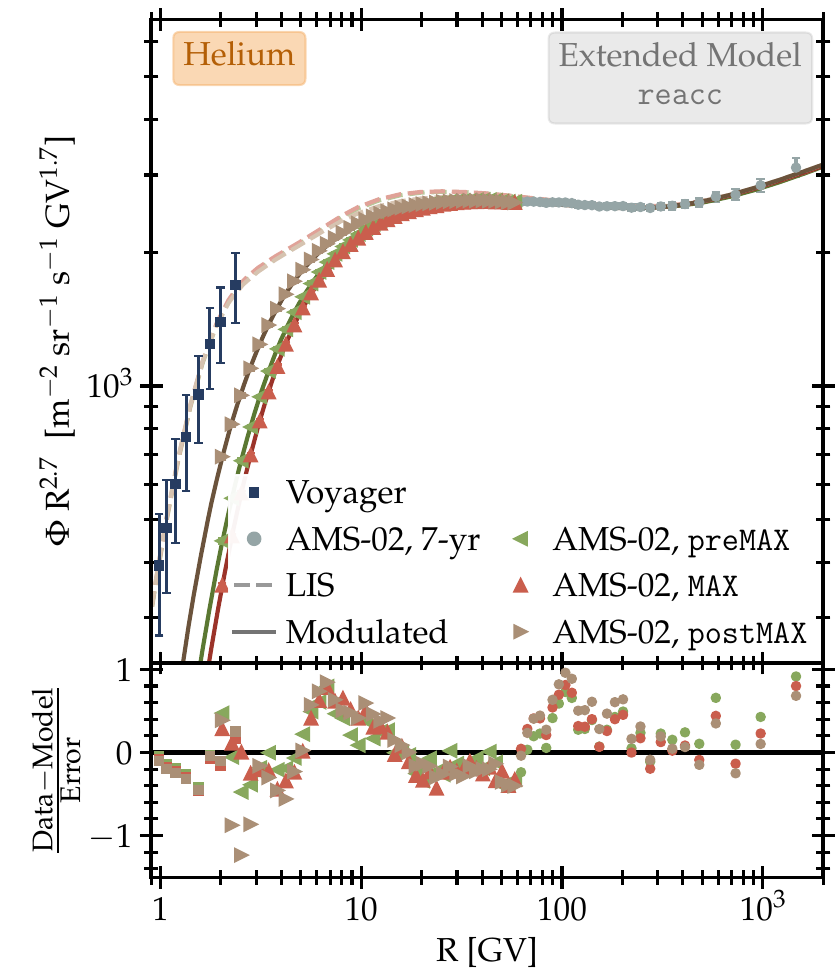}
    \end{minipage}
    \hfill
    \begin{minipage}[t]{0.32\linewidth}
        \centering
        \includegraphics[width=1\linewidth]{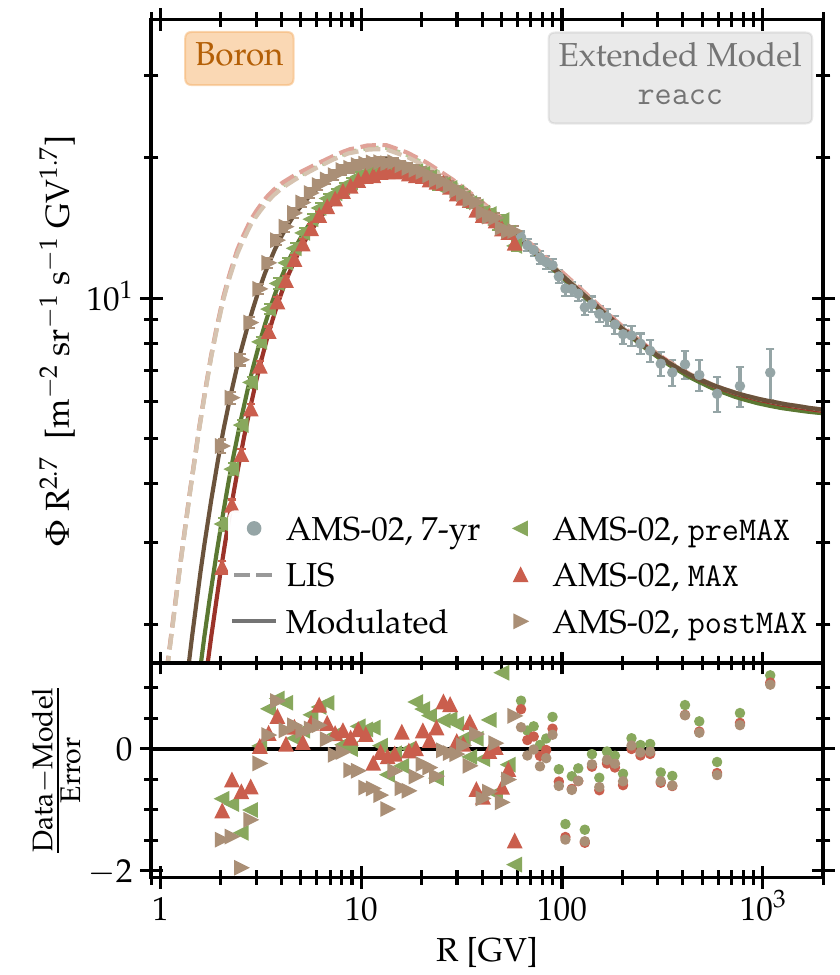}
    \end{minipage}
    \hfill
    \begin{minipage}[t]{0.32\linewidth}
        \centering
        \includegraphics[width=1\linewidth]{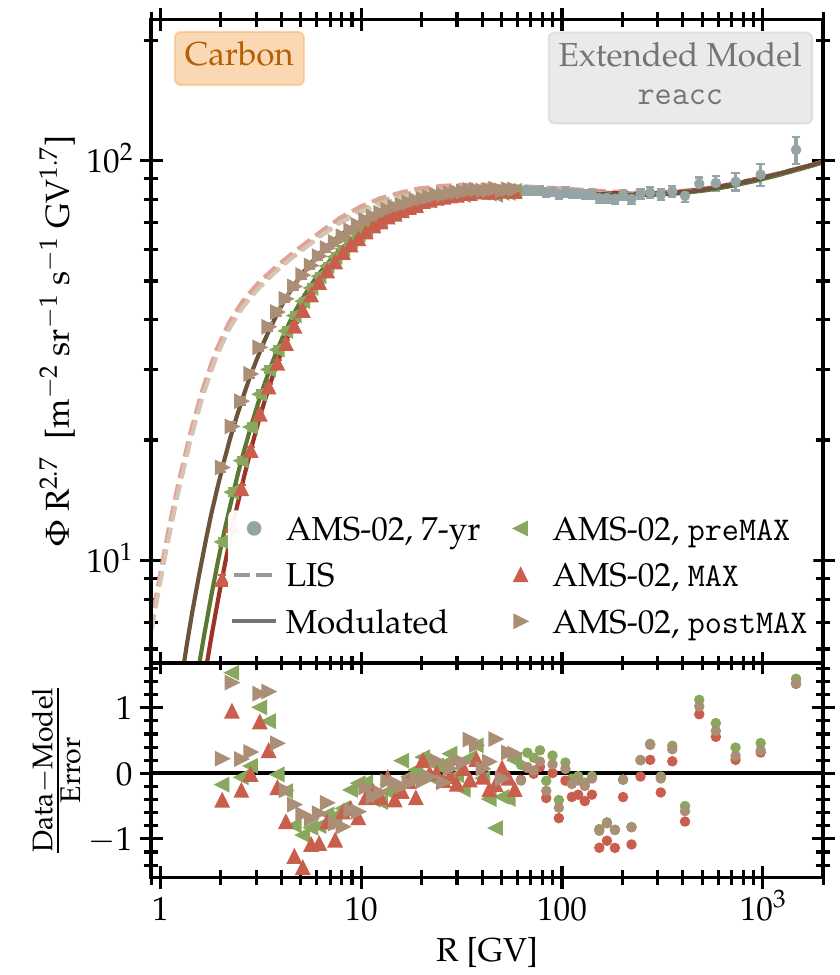}
    \end{minipage}
    \vfill\vspace{0.3cm}
    \begin{minipage}[t]{0.32\linewidth}
        \centering
        \includegraphics[width=1\linewidth]{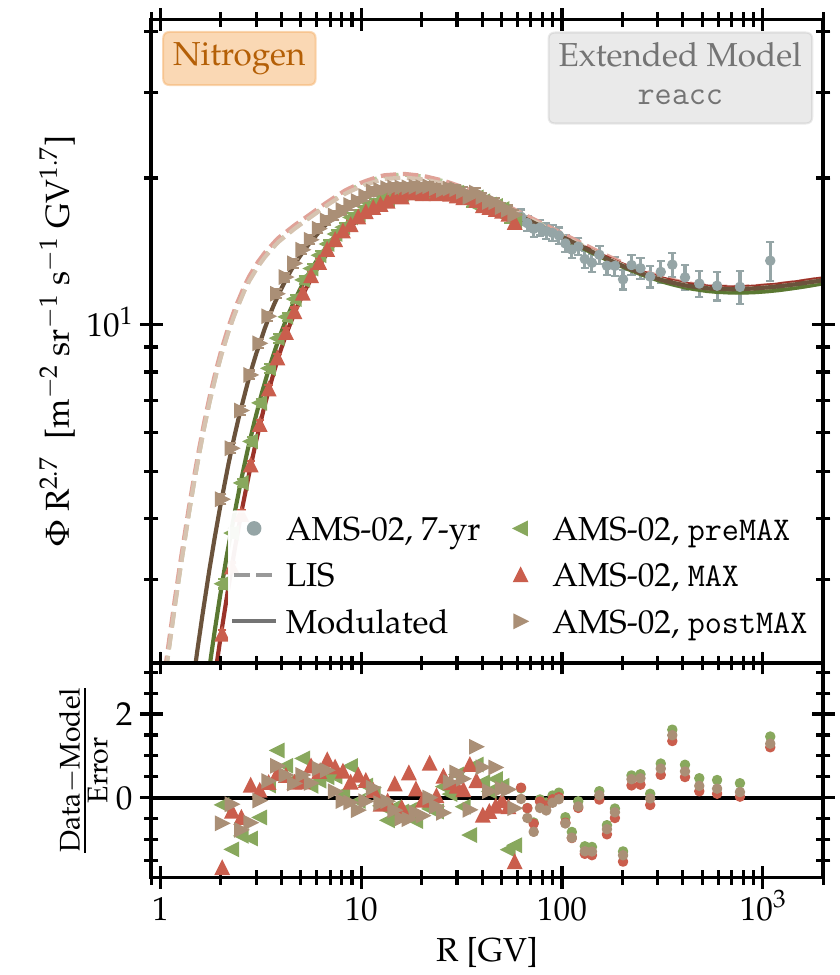}
    \end{minipage}
    \hfill
    \begin{minipage}[t]{0.32\linewidth}
        \centering
        \includegraphics[width=1\linewidth]{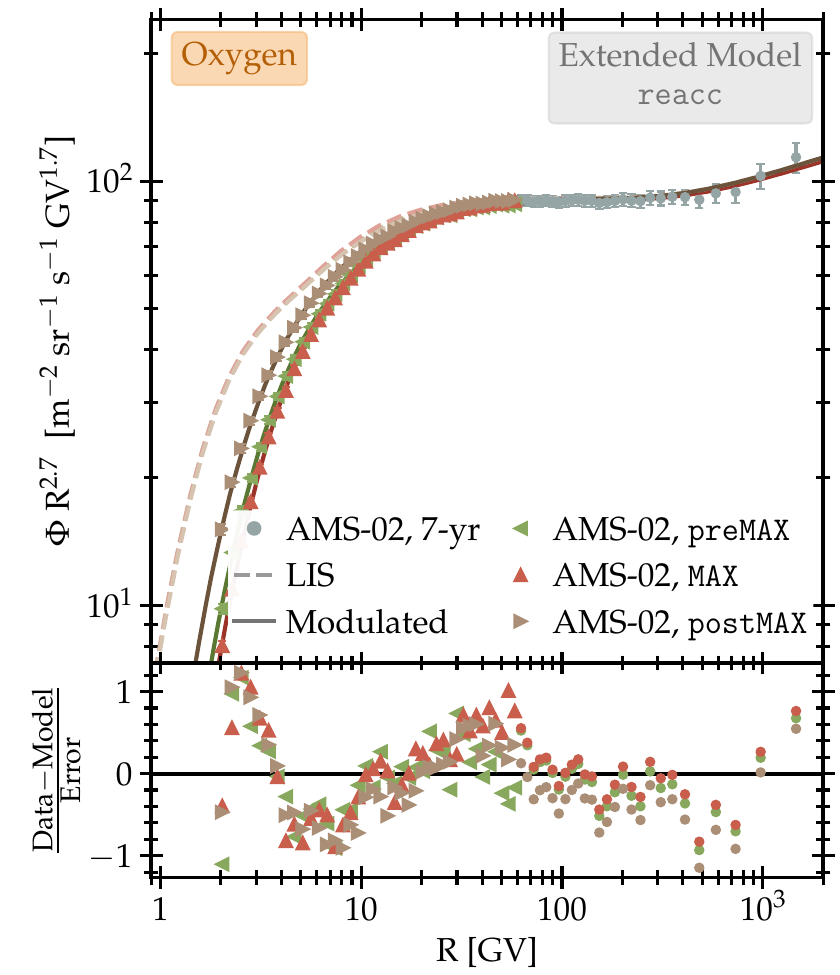}
    \end{minipage}
    \hfill
    \begin{minipage}[t]{0.32\linewidth}
        \centering
        \includegraphics[width=1\linewidth]{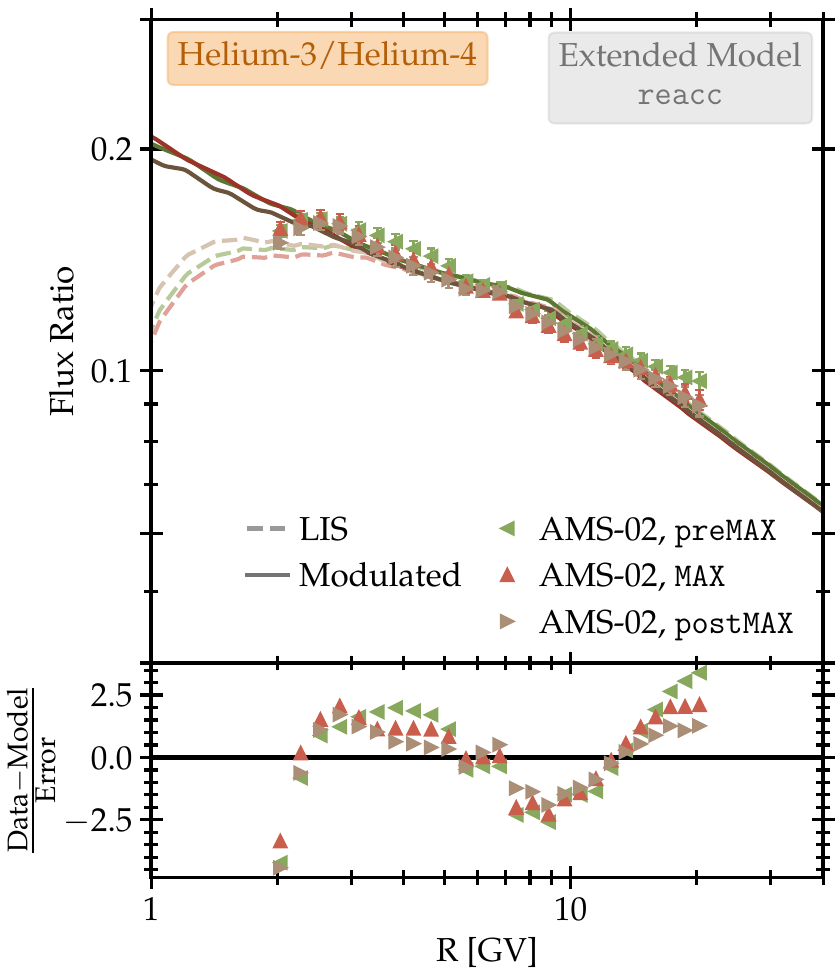}
    \end{minipage}
    \vfill\vspace{0.3cm}
    \begin{minipage}[t]{0.32\linewidth}
        \centering
        \includegraphics[width=1\linewidth]{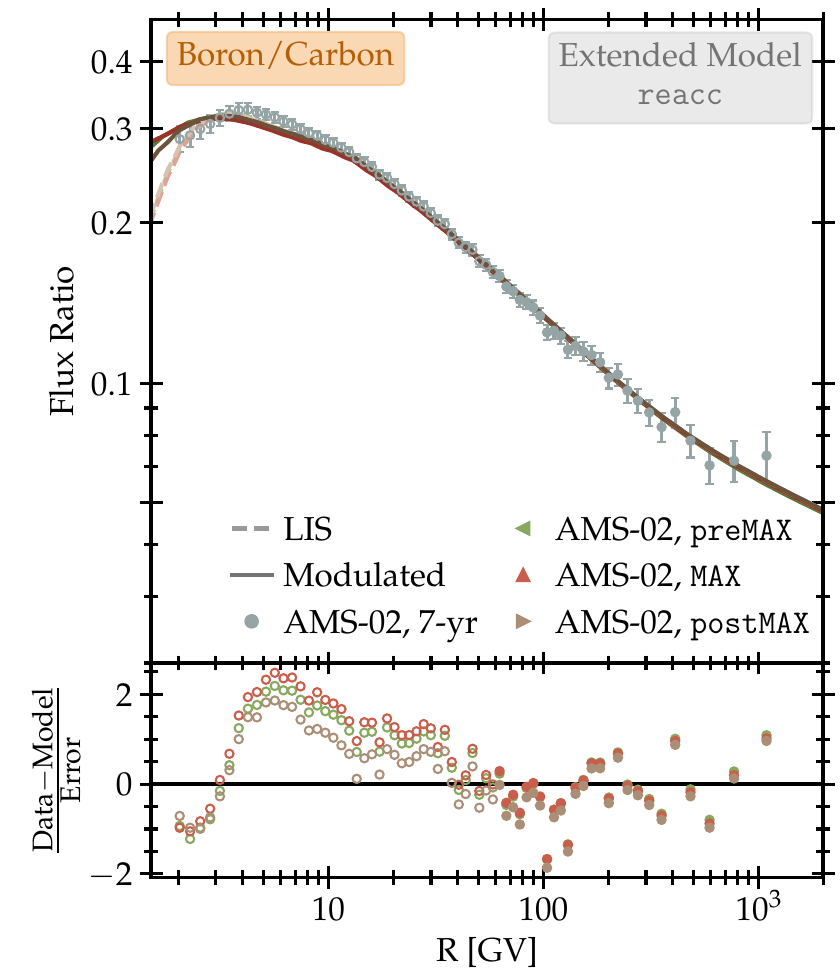}
    \end{minipage}
    \hfill
    \begin{minipage}[t]{0.32\linewidth}
        \centering
        \includegraphics[width=1\linewidth]{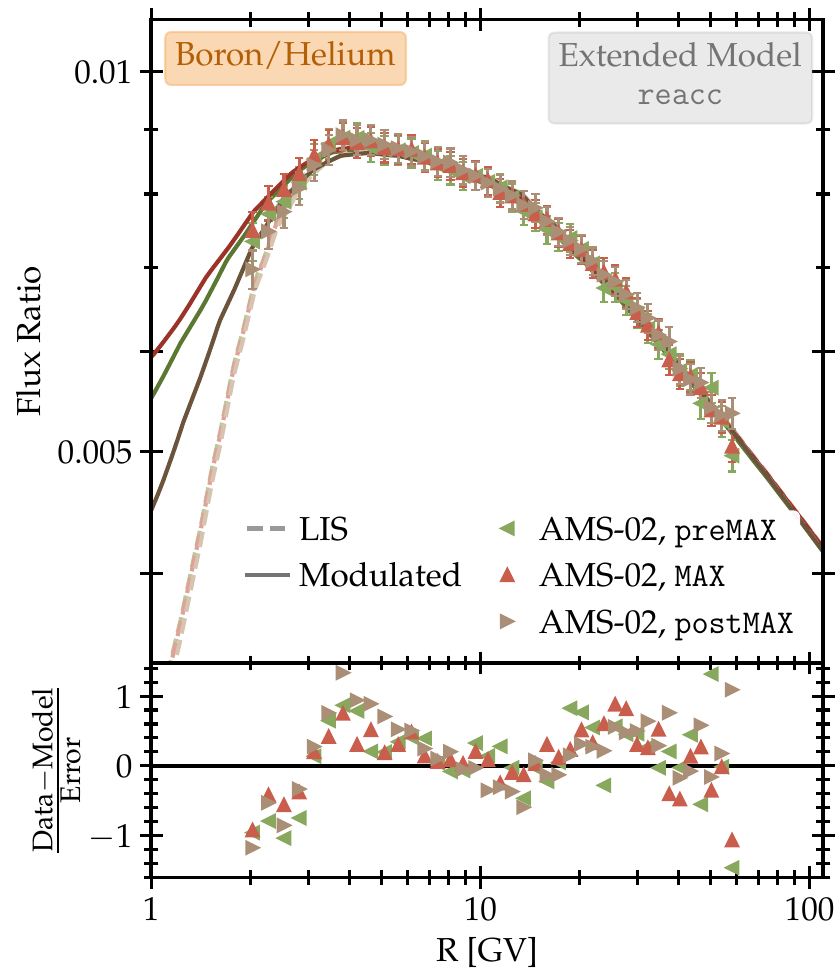}
    \end{minipage}
    \hfill
    \begin{minipage}[t]{0.32\linewidth}
        \centering
        \includegraphics[width=1\linewidth]{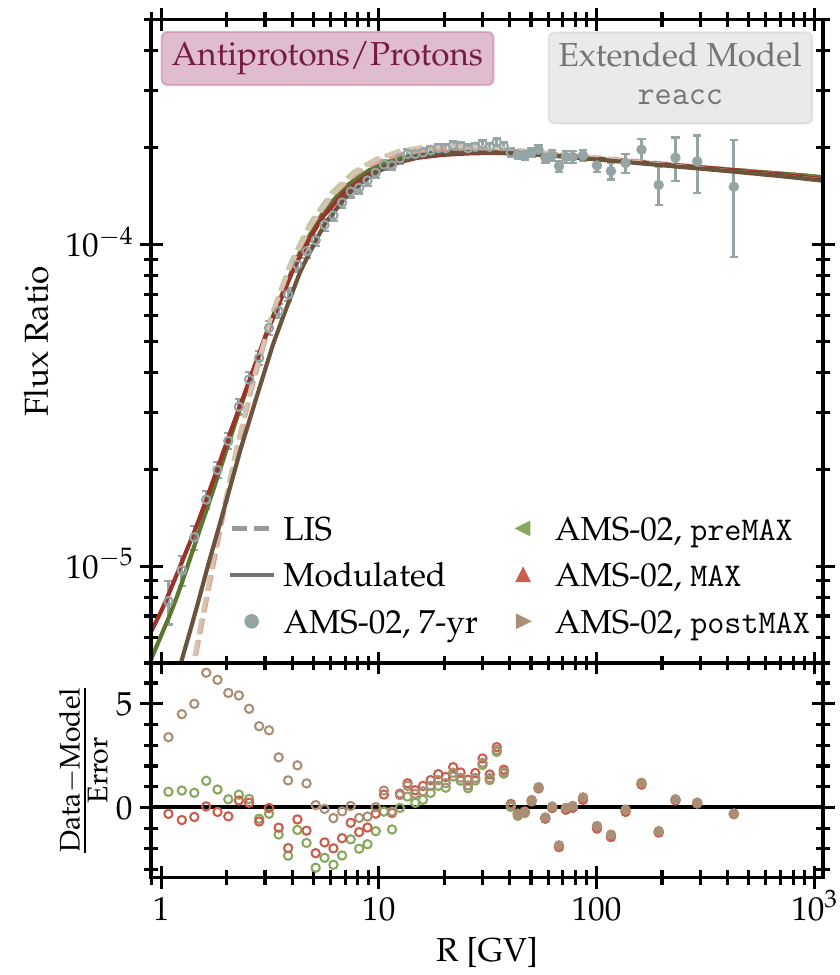}
    \end{minipage}
    \caption{Similar to Fig.~\ref{fig: LIS reacc}, the LIS and modulated fluxes of the fits to the 3 time-periods for the \reacc propagation and extended modulation model. Note that for the Boron/Carbon and antiproton/proton no time-dependent data is available.}
    \label{fig: LIS nuclei reacc Zhu}
\end{figure*}

Figures~\ref{fig: LIS nuclei conv FF} to~\ref{fig: LIS nuclei reacc Zhu} show the LIS and modulated fluxes for all cosmic-ray nuclei and ratios that are fitted in our analysis for the three time periods (see the main text for protons and antiprotons). Figs.~\ref{fig: LIS nuclei conv FF} and~\ref{fig: LIS nuclei conv Zhu} present the fluxes in the \conv propagation model with the force-field and extended solar modulation model, respectively, while Figs.~\ref{fig: LIS nuclei reacc FF} and~\ref{fig: LIS nuclei reacc Zhu} show the fluxes in the \reacc propagation model with the force-field and extended solar modulation model, respectively. The time-dependent AMS-02 data is complemented with the 7-year time-integrated data at higher rigidities (light grey circle markers). Note that in some cases, only time-dependent fluxes are available, as in the Helium-3-to-Helium-4 and Boron-to-Helium ratios, or only time-integrated data is available, as in the case of the Boron-to-Carbon and antiproton-to-proton ratios, where we only include the data above which no time-dependent data is available (see Section~\ref{sec: cosmic-ray data} for details). Data that is not included in the fit is indicated by empty markers. In general the quality of the fit is good in all cases, as indicated by the residuals which are at the level of $\sim 1-2\,\sigma$.

\section{Time-Independent Solar Modulation Potential}
\label{app: SM potential time-independent fit}
\setcounter{figure}{0} 
\setcounter{table}{0} 
\setcounter{equation}{0} 

\begin{figure*}[tbp]
    \centering
    \begin{minipage}[t]{0.485\textwidth}
        \centering
        \begin{minipage}[t]{0.7\linewidth}
            \centering
            \includegraphics[width=1\linewidth]{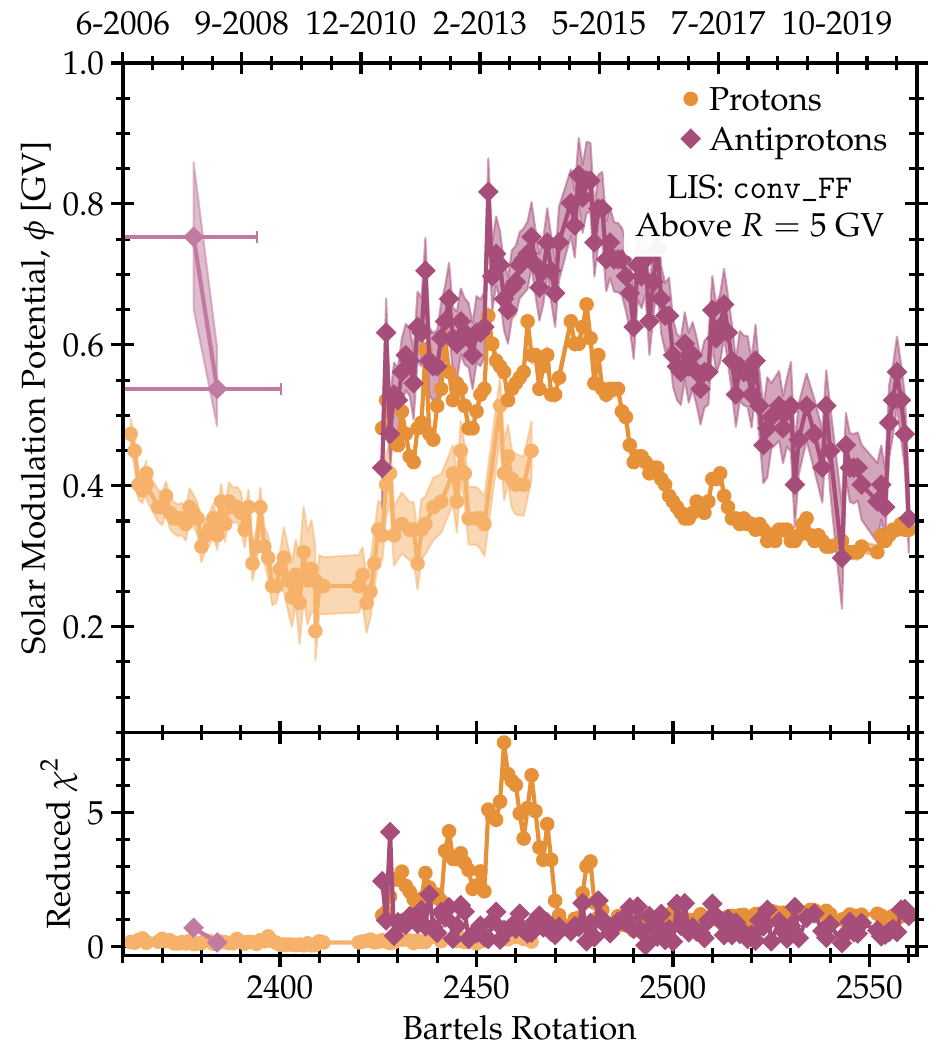}
        \end{minipage}
        \vfill\vspace{0.25cm}
        \begin{minipage}[t]{0.7\linewidth}
            \centering
            \includegraphics[width=1\linewidth]{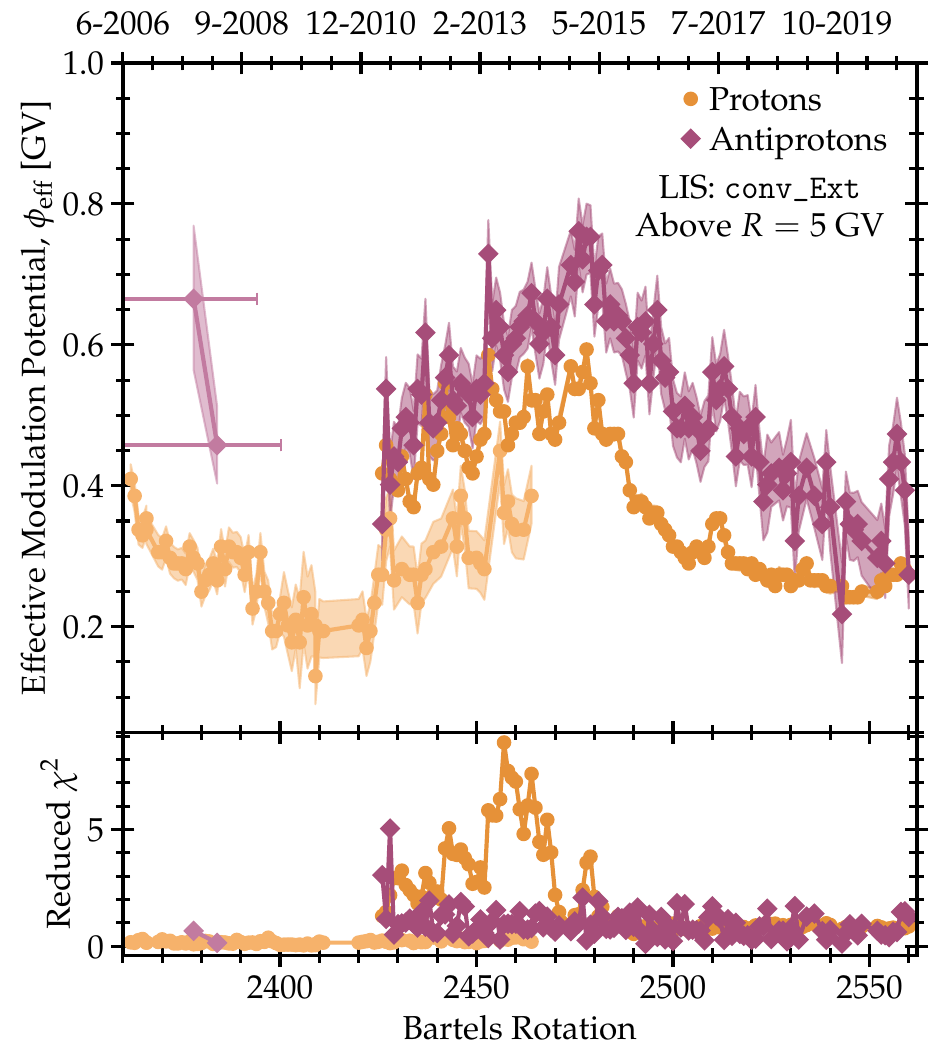}
        \end{minipage}
        \vfill\vspace{0.25cm}
        \begin{minipage}[t]{0.7\linewidth}
            \centering
            \includegraphics[width=1\linewidth]{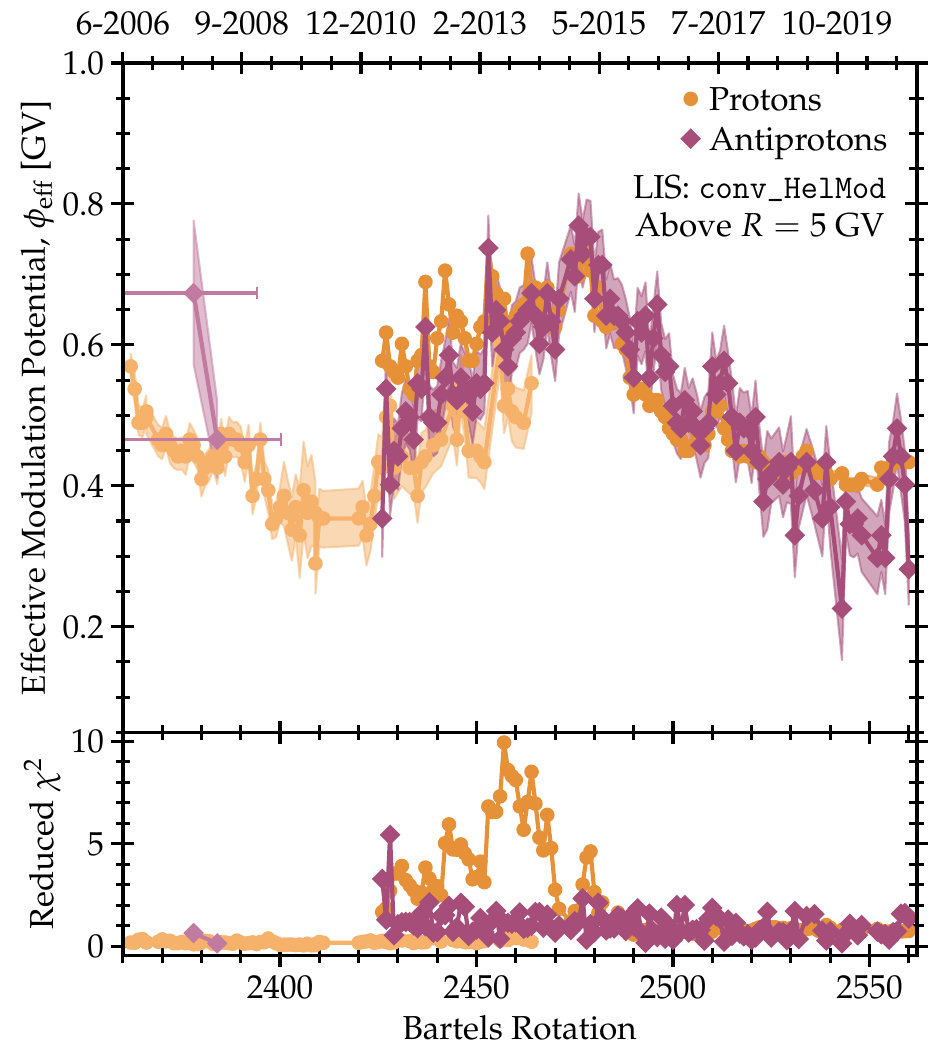}
        \end{minipage}
        \caption{Effective force-field solar modulation potential for each Bartels rotation, for protons (orange circle markers) and antiprotons (burgundy diamond markers). Results are shown for the \texttt{conv} propagation model, where the top panel uses the force-field (\texttt{FF}) 7-yr AMS-02 fit LIS, middle panel the extended model (\texttt{Ext}) LIS and bottom panel the \texttt{HelMod} LIS. Only rigidities above 5~GV are used to the derive the potential. Lighter shades correspond to points fitted to the PAMELA data, while darker points correspond to fits to AMS-02 data.}
        \label{fig: SM potential conv time-independent}
    \end{minipage}
    \hfill
    \begin{minipage}[t]{0.485\textwidth}
        \centering
        \begin{minipage}[t]{0.7\linewidth}
            \centering
            \includegraphics[width=1\linewidth]{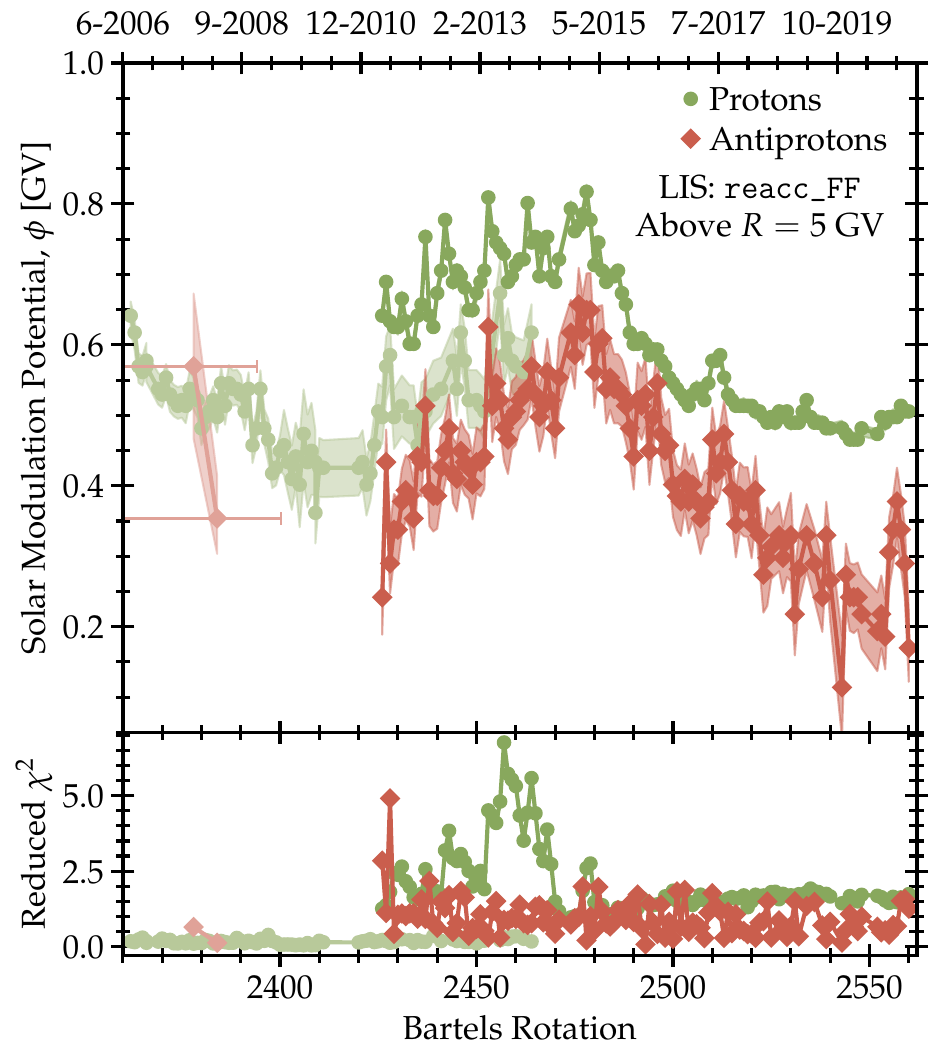}
        \end{minipage}
        \vfill\vspace{0.25cm}
        \begin{minipage}[t]{0.7\linewidth}
            \centering
            \includegraphics[width=1\linewidth]{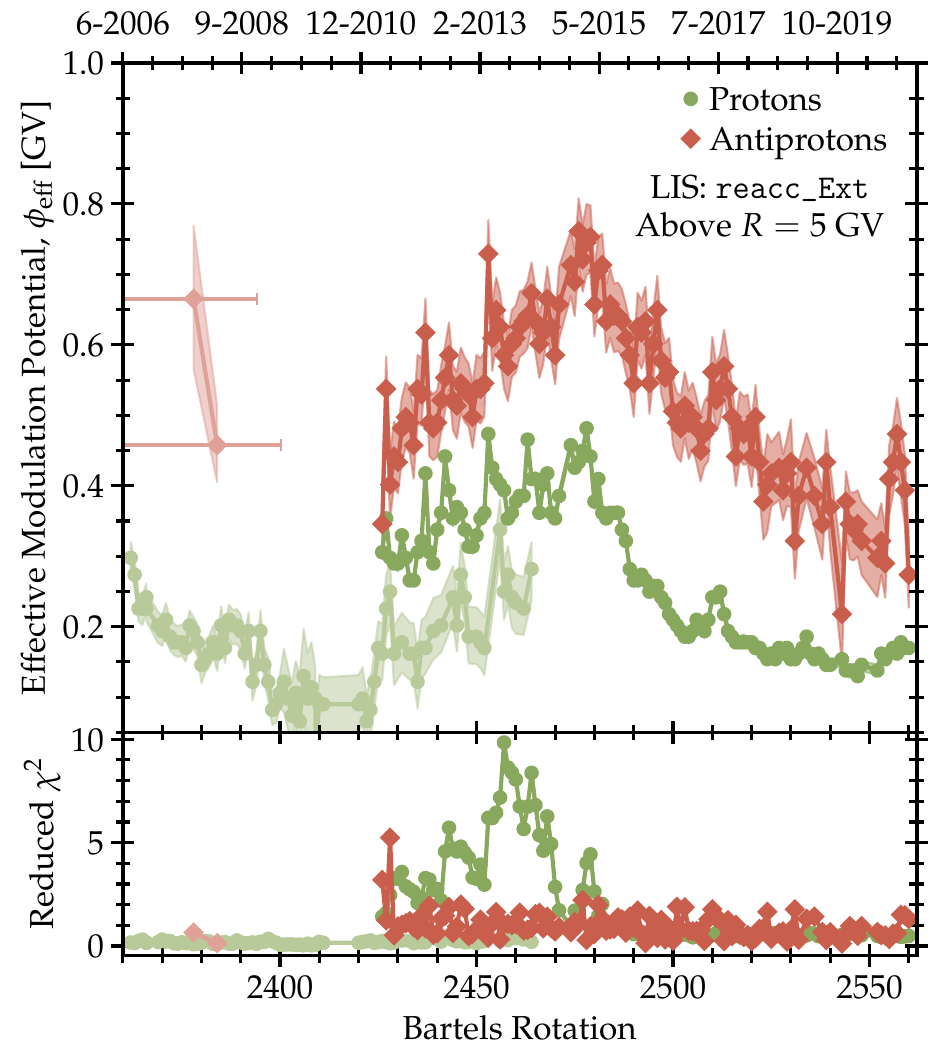}
        \end{minipage}
        \vfill\vspace{0.25cm}
        \begin{minipage}[t]{0.7\linewidth}
            \centering
            \includegraphics[width=1\linewidth]{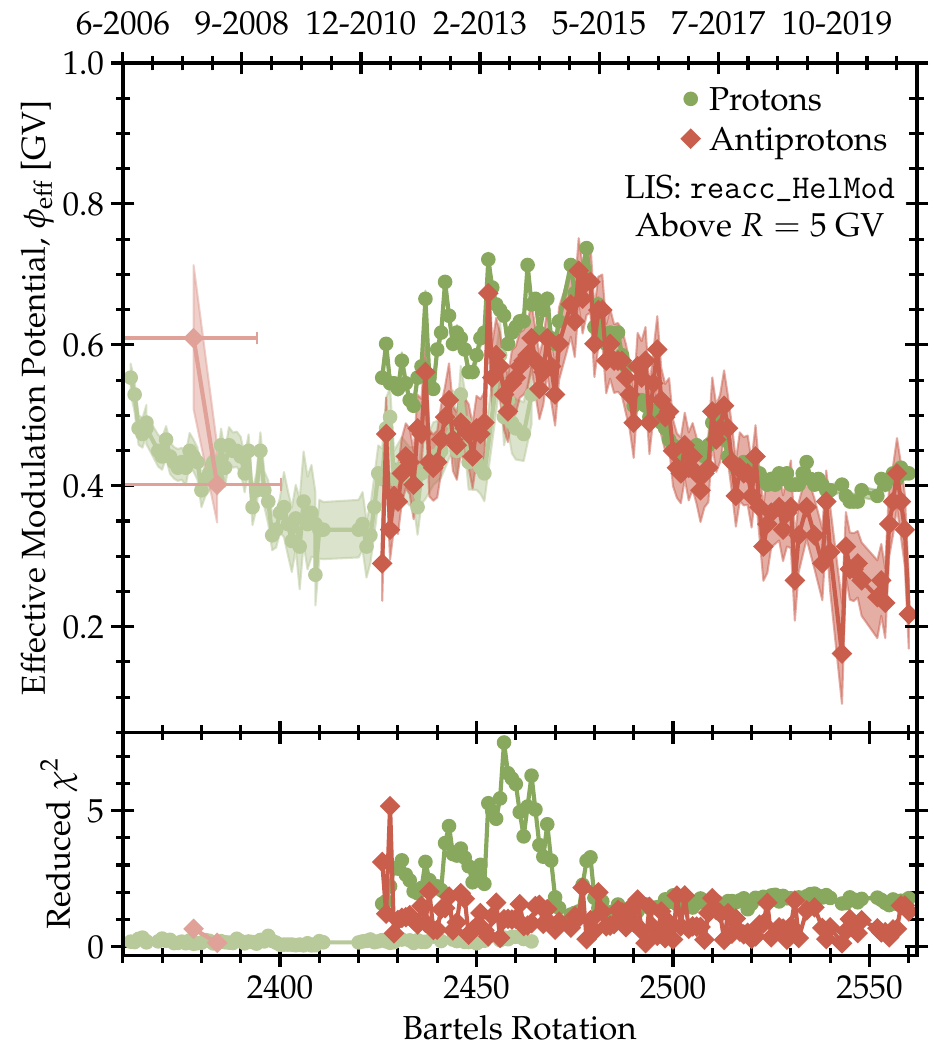}
        \end{minipage}
        \caption{Same as Fig.~\ref{fig: SM potential conv time-independent} but for the \reacc propagation model. Green circle markers correspond to protons and red diamond markers to antiprotons. Lighter shades represent the potentials based on PAMELA data.}
        \label{fig: SM potential reacc time-independent}
    \end{minipage}
\end{figure*}

Figures~\ref{fig: SM potential conv time-independent} and~\ref{fig: SM potential reacc time-independent} show the change in the effective force-field solar modulation potential for each Bartels rotation for protons and antiprotons. Each panel refers to a different LIS relative to the 7-yr AMS-02 dataset fit for the two propagation models (\conv and \reacc) and the three solar modulation models (force-field, extended force-field and \texttt{HelMod}). The shaded bands around the data points correspond to the $1\,\sigma$ uncertainty bands. In all plots, the reduced $\chi^2$ is typically the worst during the solar maximum, indicating that this period is especially hard to model. Beside the AMS-02 data we also use PAMELA data to cover the solar minimum at around 2008. In particular, we use the time-dependent proton data from~\cite{AdrianiEtAl2013PAMELAProtonTime}. Unfortunately, no time-dependent antiproton data from PAMELA is available. Instead, we use the two published time-integrated datasets, covering the period from July 2006 to December 2008~\cite{Adriani_2010}, and from July 2006 to December 2009~\cite{Adriani:2012paa}, respectively. Note that the LIS fluxes are only obtained from the fit to AMS-02 data. Nonetheless, we believe that a comparison of the time-dependent potentials to the PAMELA time period is still insightful.

Regarding the \conv propagation model results, shown in Fig.~\ref{fig: SM potential conv time-independent}, it can be seen that the force-field and extended modulation show a trend that disagrees with expectations (\textit{i.e.}, modulation is larger for protons in the \preMAX phase, similar in the \MAX phase, and larger for antiprotons in the \postMAX phase), contrary to what is observed for the time-dependent fits, discussed in the main text. This seems to confirm that using a single modulation across different solar phases, especially if the solar maximum is included, introduces biases in the fit and produce unreliable results. The \texttt{HelMod} modulation instead uses a different modulation appropriate for each time period, and indeed the resulting time-dependent potential (bottom panel of Fig.~\ref{fig: SM potential conv time-independent}) is more in agreement with expectations, although not perfect, especially in the \postMAX period. Regarding the PAMELA data, one can see an offset of the proton potential in the period in which the PAMELA and AMS-02 data overlap. This reflects a systematic offset originally present in the proton flux itself between the two datasets. There also exists a systematic offset in the two PAMELA antiproton datasets despite the fact that they cover almost the same time period. 

The results for the \reacc case, shown in Fig.~\ref{fig: SM potential reacc time-independent}, show the same trend observed for the time-dependent fits, with a general disagreement with expectations indicating some issue with the related LIS. The \texttt{HelMod} case, instead, gives results very similar to the \conv case. This is somehow expected, as the modulation model is fixed and thus seems to force the same LIS independently of the propagation model.

\section{Propagation Model parameters and full fit results}
\label{app: propagation models}
\setcounter{figure}{0} 
\setcounter{table}{0} 
\setcounter{equation}{0} 

The propagation model in our analysis is based on the propagation model developed in~\cite{Cuoco:2019kuu, Korsmeier:2021brc, Korsmeier:2021bkw, DiMauro:2023jgg}, where we refer the reader to for full details. In short, the model comes in two variations: the convection model (\conv) and the re-acceleration model (\reacc), where the latter allows for re-acceleration of cosmic rays during propagation (see also Sec.~\ref{sec: cosmic-ray models}). Our free parameters include:

\begin{itemize}
    \item Cosmic-ray injection spectra: in the \conv case the injection spectra are pure powerlaws with spectral index $\gamma_{1, p}$ for protons, $\gamma_{\rm 1, He}$ for Helium and $\gamma_{\rm 1, CNO}$ for heavier primary nuclei. For the fit, in practice, the differences with respect to the proton spectral index are used, \textit{i.e.}, $\gamma_{\rm 1, He-p}$ and $\gamma_{\rm 1, CNO-p}$. In the \reacc model, we also include a break in the injection spectrum with a smoothing factor, $R_{\rm inj}$ and $s_{\rm inj}$, respectively. Correspondingly, additional spectral indices below the break are given by $\gamma_{0, p}$, $\gamma_{\rm 0, He}$, $\gamma_{\rm 0, CNO}$.
    
    \item Diffusion: The diffusion coefficient is modelled as doubly broken power-law in the \conv model, with normalization $D_0$, breaks $R_{D,1}$ and $R_{D,2}$ and corresponding smoothing transition factors $s_{D,1}$ and $s_{D,2}$. The spectral indices at lower, intermediate and higher rigidities are $\delta_l$, $\delta$ and $\delta_h$, respectively. In the \reacc model the first diffusion break is not present, and is replaced instead by a break in the injection spectrum as mentioned above.
    
    \item The convection velocity $v_{0, c}$ in the \conv model, and the Alfv\'en velocity $v_A$ in the \reacc model, respectively.
    
    \item The abundances of the primary nuclei, $\text{Abd}_{^{12}\rm C}$, $\text{Abd}_{^{14}\rm N}$ and $\text{Abd}_{^{16}\rm O}$.
    
    \item Cross sections: The re-normalisation of the Helium-4 to Helium-3 cross section, together with a change of the slope, $A_{xs^4\rm He\rightarrow^3\rm He}$ and $\delta_{xs^4\rm He\rightarrow^3\rm He}$, respectively, and similarly for Carbon ($A_{xs\rightarrow \rm C}$, $\delta_{xs\rightarrow \rm C}$), Nitrogen ($A_{xs\rightarrow\rm N}$, $\delta_{xs\rightarrow\rm  N}$), and Boron ($\delta_{xs\rightarrow \rm B}$).

    \item Further normalization: The normalisation of the spectra for protons $A_p$, Helium $A_{\rm He}$, the Boron-to-Carbon ratio $A_{\rm B}$ and the antiproton-to-proton ratio $A_{\bar{p}}$.
    
    \item Parameters related to the specific solar modulation model, see Section~\ref{sec: solar modulation models}.
\end{itemize}

Tables~\ref{tab: conv FF} to~\ref{tab: HelMod} list all model parameters, priors, best-fit values and $\chi^2$ values for each of our fits. The results are also reported in graphical form in Figures~\ref{fig: parameter comparisons conv 1} and~\ref{fig: parameter comparisons reacc 1} for the \conv and \reacc models, respectively.

\clearpage

\begin{figure*}[tbp]
    \centering
    \includegraphics[page=1, width=1\linewidth]{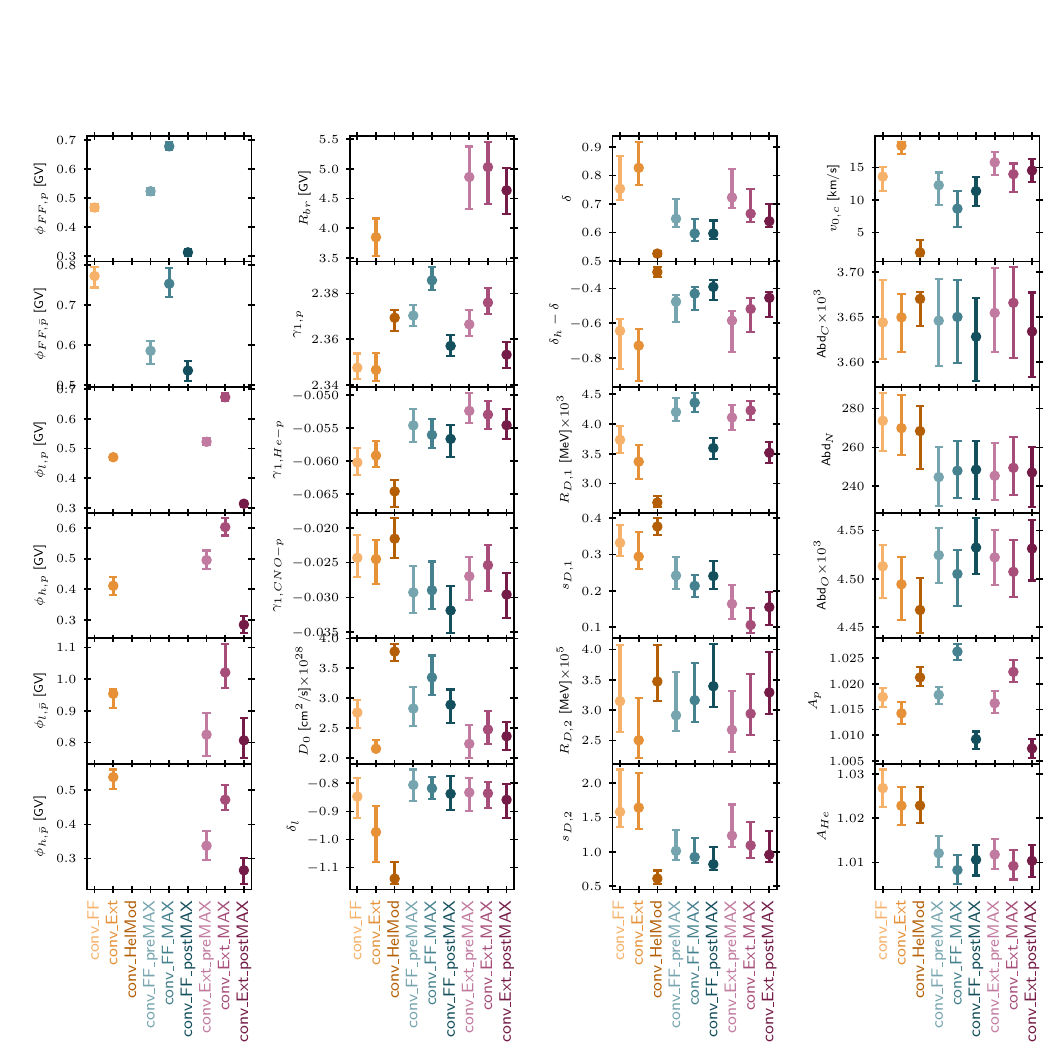}
    \caption{Best-fit values for the free parameters in the convection model, for the fits to the 7-yr AMS-02 data with solar modulation models for force-field (\texttt{conv\_FF}), extended force-field (\texttt{conv\_Ext}) and \texttt{HelMod} (\texttt{conv\_HelMod}), as well as for the 3 time-dependent force-field fits (\texttt{conv\_FF\_preMAX}, \texttt{conv\_FF\_MAX} and \texttt{conv\_FF\_postMAX}, respectively), and the 3 time-dependent extended force-field fits (\texttt{conv\_Ext\_preMAX}, \texttt{conv\_Ext\_MAX} and \texttt{conv\_Ext\_postMAX}, respectively). }
    \label{fig: parameter comparisons conv 1}
\end{figure*}
\begin{figure*}[tbp]
   \ContinuedFloat
    \centering
    \includegraphics[page=2, width=1\linewidth]{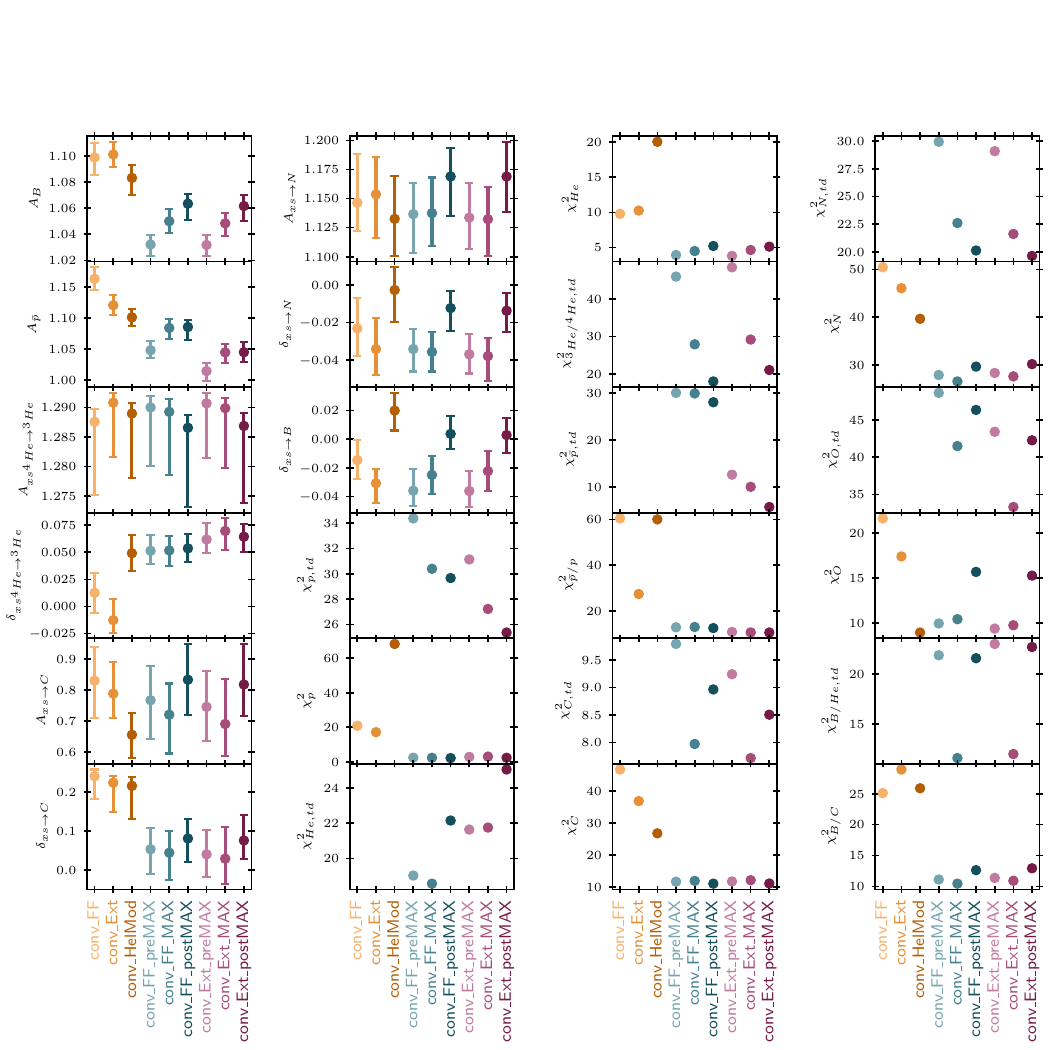}
    \caption{Same as Fig.~\ref{fig: parameter comparisons conv 1}. The $\chi^2$ are given for AMS-02 data, where the subscript $td$ indicates time-dependent data, unless otherwise indicated.}
    \label{fig: parameter comparisons conv 2}
\end{figure*}

\begin{figure*}[tbp]
    \centering
    \includegraphics[page=1, width=1\linewidth]{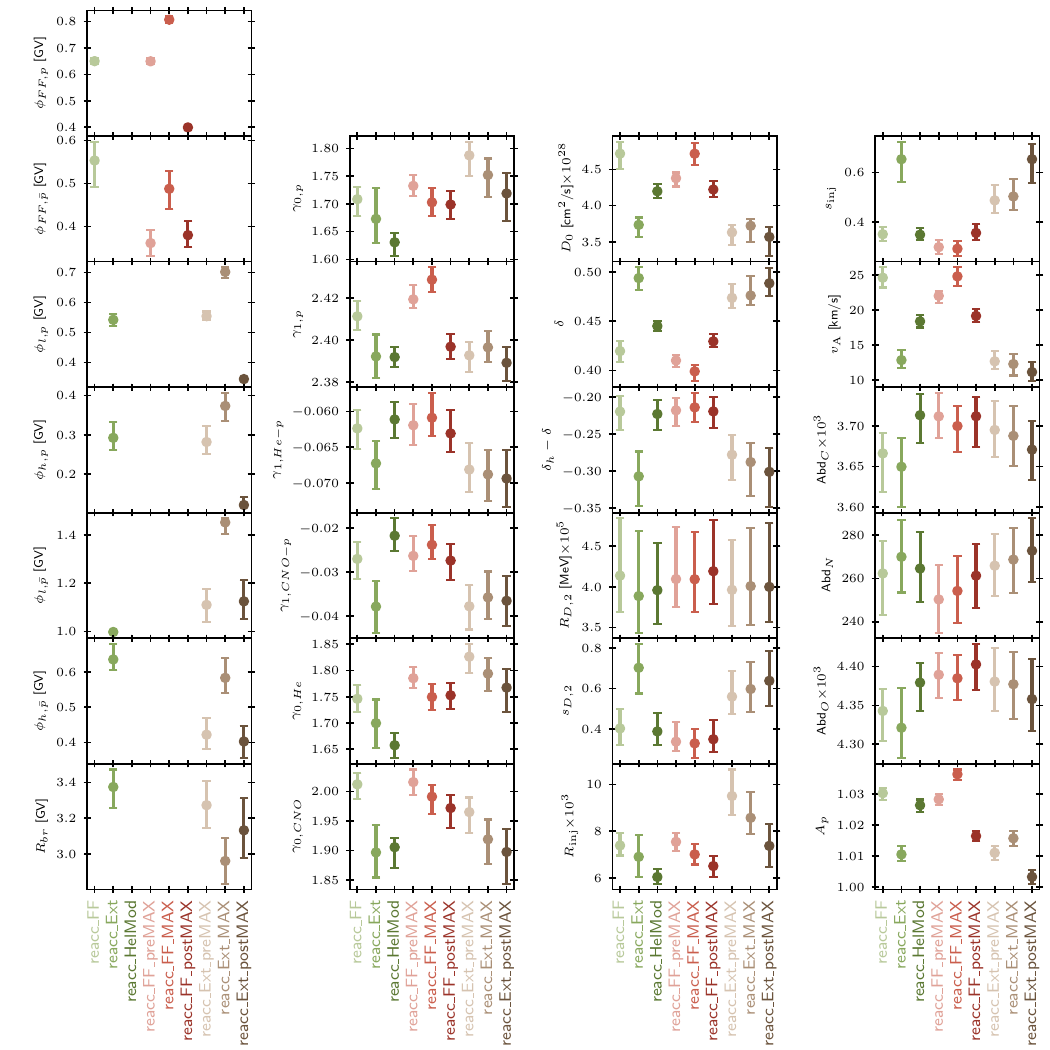}
    \caption{Best-fit values for the free parameters in the re-acceleration model, for the fits to the 7-yr AMS-02 data with solar modulation models for force-field (\texttt{reacc\_FF}), extended force-field (\texttt{reacc\_Ext}) and \texttt{HelMod} (\texttt{reacc\_HelMod}), as well as for the 3 time-dependent force-field fits (\texttt{reacc\_FF\_preMAX}, \texttt{reacc\_FF\_MAX} and \texttt{reacc\_FF\_postMAX}, respectively), and the 3 time-dependent extended force-field fits (\texttt{reacc\_Ext\_preMAX}, \texttt{reacc\_Ext\_MAX} and \texttt{reacc\_Ext\_postMAX}, respectively).}
    \label{fig: parameter comparisons reacc 1}
\end{figure*}
\begin{figure*}[tbp]
   \ContinuedFloat
    \centering
    \includegraphics[page=2, width=1\linewidth]{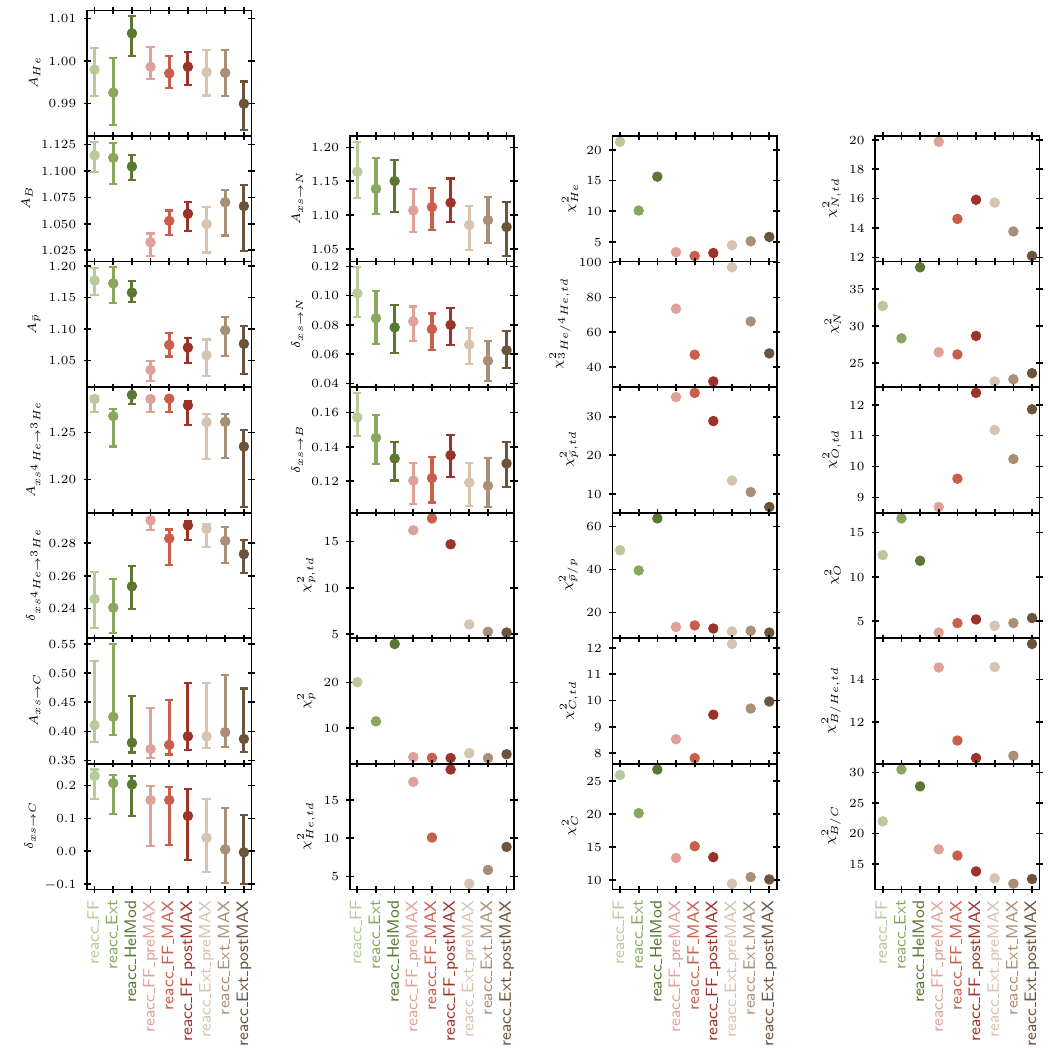}
    \caption{Same as Fig.~\ref{fig: parameter comparisons reacc 1}. The $\chi^2$ are given for AMS-02 data, where the subscript $td$ indicates time-dependent data, unless otherwise indicated. }
    \label{fig: parameter comparisons reacc 2}
\end{figure*}

\begin{table*}[tbp]
\centering
\caption{Summary of the best-fit parameters in the \texttt{conv} propagation model for the 3 time-dependent and the 7-yr time-integrated fit for the force-field solar modulation model (\texttt{FF}).}
\label{tab: conv FF}
\begin{tabular}{r|c||c|c|c|c}
Parameter & Prior & \texttt{conv\_FF\_preMAX} & \texttt{conv\_FF\_MAX} & \texttt{conv\_FF\_postMAX} & \texttt{conv\_FF} \\
\hline
$\phi_{FF, p}$ [GV] & $0.1 - 1.0 $ & $0.52_{-0.01}^{+0.01}$ & $0.68_{-0.01}^{+0.02}$ & $0.31_{-0.01}^{+0.01}$ & $0.47_{-0.01}^{+0.01}$ \\
$\phi_{FF, \bar{p}}$ [GV] & $0.1 - 1.0 $ & $0.59_{-0.03}^{+0.02}$ & $0.75_{-0.03}^{+0.04}$ & $0.54_{-0.03}^{+0.02}$ & $0.77_{-0.03}^{+0.02}$ \\
$\phi_{l, p}$ [GV] & $  -   $ &  $ - $  &  $ - $  &  $ - $  &  $ - $  \\
$\phi_{h, p}$ [GV] & $  -   $ &  $ - $  &  $ - $  &  $ - $  &  $ - $  \\
$\phi_{l, \bar{p}}$ [GV] & $  -   $ &  $ - $  &  $ - $  &  $ - $  &  $ - $  \\
$\phi_{h, \bar{p}}$ [GV] & $  -   $ &  $ - $  &  $ - $  &  $ - $  &  $ - $  \\
$R_{br}$ [GV] & $  -   $ &  $ - $  &  $ - $  &  $ - $  &  $ - $  \\
$\gamma_{1, p}$ & $2.3 - 2.42 $ & $2.370_{-0.005}^{+0.005}$ & $2.386_{-0.004}^{+0.006}$ & $2.357_{-0.005}^{+0.005}$ & $2.348_{-0.005}^{+0.006}$ \\
$\gamma_{1, He-p}$ & $-0.09 - -0.03 $ & $-0.055_{-0.002}^{+0.002}$ & $-0.056_{-0.002}^{+0.002}$ & $-0.057_{-0.003}^{+0.002}$ & $-0.060_{-0.002}^{+0.002}$ \\
$\gamma_{1, CNO-p}$ & $-0.05 - 0.0 $ & $-0.029_{-0.003}^{+0.004}$ & $-0.029_{-0.003}^{+0.004}$ & $-0.032_{-0.003}^{+0.003}$ & $-0.024_{-0.003}^{+0.003}$ \\
$D_0$ [cm$^2$/s] $\times 10^{28}$ & $1 - 6 $ & $2.8_{-0.3}^{+0.4}$ & $3.3_{-0.3}^{+0.4}$ & $2.9_{-0.3}^{+0.3}$ & $2.8_{-0.3}^{+0.2}$ \\
$\delta_l$ & $-1.2 - -0.4 $ & $-0.81_{-0.06}^{+0.06}$ & $-0.82_{-0.04}^{+0.04}$ & $-0.84_{-0.06}^{+0.06}$ & $-0.85_{-0.08}^{+0.06}$ \\
$\delta$ & $0.3 - 1.0 $ & $0.65_{-0.03}^{+0.07}$ & $0.60_{-0.03}^{+0.05}$ & $0.60_{-0.02}^{+0.05}$ & $0.75_{-0.04}^{+0.1}$ \\
$\delta_h - \delta$ & $-1.2 - -0.2 $ & $-0.48_{-0.1}^{+0.04}$ & $-0.43_{-0.09}^{+0.04}$ & $-0.39_{-0.07}^{+0.04}$ & $-0.64_{-0.2}^{+0.07}$ \\
$R_{D, 1}$ [MeV] $\times 10^{3}$ & $2.0 - 6 $ & $4.2_{-0.2}^{+0.2}$ & $4.4_{-0.2}^{+0.2}$ & $3.6_{-0.2}^{+0.2}$ & $3.7_{-0.2}^{+0.2}$ \\
$s_{D, 1}$ & $0.05 - 0.6 $ & $0.24_{-0.04}^{+0.05}$ & $0.21_{-0.03}^{+0.03}$ & $0.24_{-0.04}^{+0.04}$ & $0.33_{-0.04}^{+0.05}$ \\
$R_{D, 2}$ [MeV] $\times 10^{5}$ & $0.8 - 5 $ & $2.9_{-0.3}^{+0.7}$ & $3.2_{-0.4}^{+0.6}$ & $3.4_{-0.3}^{+0.7}$ & $3.1_{-0.5}^{+0.9}$ \\
$s_{D, 2}$ & $0.05 - 3.0 $ & $1.0_{-0.1}^{+0.3}$ & $0.93_{-0.09}^{+0.3}$ & $0.82_{-0.09}^{+0.3}$ & $1.6_{-0.2}^{+0.6}$ \\
$v_{0, c}$ [km/s] & $0.0 - 30 $ & $12_{-3}^{+2}$ & $8_{-3}^{+3}$ & $11_{-2}^{+2}$ & $13_{-2}^{+2}$ \\
Abd$_{C}$ $\times 10^{3}$ & $3.5 - 3.8 $ & $3.65_{-0.05}^{+0.05}$ & $3.65_{-0.05}^{+0.04}$ & $3.63_{-0.05}^{+0.04}$ & $3.64_{-0.04}^{+0.05}$ \\
Abd$_{N}$ & $200 - 350 $ & $244_{-15}^{+15}$ & $247_{-14}^{+15}$ & $248_{-15}^{+15}$ & $273_{-16}^{+14}$ \\
Abd$_{O}$ $\times 10^{3}$ & $4.35 - 4.65 $ & $4.52_{-0.03}^{+0.03}$ & $4.51_{-0.03}^{+0.02}$ & $4.53_{-0.03}^{+0.03}$ & $4.51_{-0.03}^{+0.02}$ \\
$A_{p}$ & $0.9 - 1.1 $ & $1.018_{-0.002}^{+0.001}$ & $1.026_{-0.002}^{+0.001}$ & $1.009_{-0.002}^{+0.001}$ & $1.017_{-0.002}^{+0.002}$ \\
$A_{He}$ & $0.9 - 1.1 $ & $1.012_{-0.003}^{+0.004}$ & $1.008_{-0.003}^{+0.003}$ & $1.011_{-0.004}^{+0.003}$ & $1.027_{-0.004}^{+0.004}$ \\
$A_{B}$ & $0.7 - 1.3 $ & $1.032_{-0.009}^{+0.008}$ & $1.050_{-0.009}^{+0.009}$ & $1.063_{-0.01}^{+0.007}$ & $1.10_{-0.01}^{+0.01}$ \\
$A_{\bar{p}}$ & $0.7 - 1.3 $ & $1.05_{-0.01}^{+0.02}$ & $1.08_{-0.02}^{+0.01}$ & $1.09_{-0.02}^{+0.01}$ & $1.16_{-0.02}^{+0.02}$ \\
$A_{xs ^{4}He \rightarrow ^{3}He}$ & $0.7 - 1.3 $ & $1.290_{-0.01}^{+0.002}$ & $1.289_{-0.01}^{+0.002}$ & $1.287_{-0.01}^{+0.002}$ & $1.288_{-0.01}^{+0.002}$ \\
$\delta_{xs ^{4}He \rightarrow ^{3}He}$ & $-0.3 - 0.3 $ & $0.05_{-0.01}^{+0.01}$ & $0.05_{-0.01}^{+0.01}$ & $0.05_{-0.01}^{+0.01}$ & $0.01_{-0.02}^{+0.02}$ \\
$A_{xs\rightarrow C}$ & $0.3 - 2.0 $ & $0.8_{-0.1}^{+0.1}$ & $0.7_{-0.1}^{+0.1}$ & $0.8_{-0.1}^{+0.1}$ & $0.8_{-0.1}^{+0.1}$ \\
$\delta_{xs\rightarrow C}$ & $-0.3 - 0.3 $ & $0.05_{-0.06}^{+0.05}$ & $0.04_{-0.07}^{+0.06}$ & $0.08_{-0.06}^{+0.05}$ & $0.24_{-0.06}^{+0.02}$ \\
$A_{xs\rightarrow N}$ & $0.7 - 1.3 $ & $1.14_{-0.03}^{+0.03}$ & $1.14_{-0.03}^{+0.03}$ & $1.17_{-0.03}^{+0.02}$ & $1.15_{-0.02}^{+0.04}$ \\
$\delta_{xs\rightarrow N}$ & $-0.3 - 0.3 $ & $-0.03_{-0.01}^{+0.01}$ & $-0.04_{-0.01}^{+0.01}$ & $-0.012_{-0.01}^{+0.009}$ & $-0.02_{-0.01}^{+0.02}$ \\
$\delta_{xs\rightarrow B}$ & $-0.3 - 0.3 $ & $-0.04_{-0.01}^{+0.02}$ & $-0.02_{-0.01}^{+0.01}$ & $0.00_{-0.01}^{+0.01}$ & $-0.01_{-0.01}^{+0.01}$
\end{tabular}
\end{table*}

\begin{table*}
\ContinuedFloat
\centering
\caption{Summary of the corresponding $\chi^2$ values. Where two numbers are given for the datapoints, the first number indicates the number of datapoints in the time-dependent fit (where only datapoints above a certain rigidity are included, where solar modulation effects become negligible, see main text), and the second number the number of datapoints in the 7-yr time-integrated fit.}
\begin{tabular}{r|c||c|c|c|c}
Parameter & \# Datapoints & \texttt{conv\_FF\_preMAX} & \texttt{conv\_FF\_MAX} & \texttt{conv\_FF\_postMAX} & \texttt{conv\_FF} \\
\hline
$\chi^2_{p, td}$ & $ 10 $ & $34.4$ & $30.4$ & $29.7$ &  $ - $  \\
$\chi^2_{p}$ & $ 27, 72 $ & $2.3$ & $2.3$ & $2.1$ & $20.9$ \\
$\chi^2_{p, CALET}$ & $ 23 $ & $6.7$ & $6.4$ & $6.7$ & $5.2$ \\
$\chi^2_{p, Voyager}$ & $ 6 $ & $9.1$ & $13.8$ & $7.9$ & $5.5$ \\
$\chi^2_{He, td}$ & $ 40 $ & $19.0$ & $18.6$ & $22.1$ &  $ - $  \\
$\chi^2_{He}$ & $ 28, 68 $ & $3.9$ & $4.5$ & $5.2$ & $9.8$ \\
$\chi^2_{He, DAMPE}$ & $ 21 $ & $7.1$ & $6.8$ & $6.9$ & $6.5$ \\
$\chi^2_{He, Voyager}$ & $ 7 $ & $25.2$ & $31.4$ & $25.3$ & $23.5$ \\
$\chi^2_{^{3}He/^{4}He, td}$ & $ 26 $ & $45.9$ & $27.9$ & $18.1$ &  $ - $  \\
$\chi^2_{\bar{p}, td}$ & $ 10 $ & $30.0$ & $29.8$ & $28.0$ &  $ - $  \\
$\chi^2_{\bar{p}/p}$ & $ 18, 58 $ & $12.7$ & $12.9$ & $12.4$ & $60.5$ \\
$\chi^2_{C, td}$ & $ 40 $ & $9.8$ & $8.0$ & $9.0$ &  $ - $  \\
$\chi^2_{C}$ & $ 28, 68 $ & $11.7$ & $12.0$ & $11.1$ & $46.7$ \\
$\chi^2_{N, td}$ & $ 40 $ & $29.9$ & $22.6$ & $20.1$ &  $ - $  \\
$\chi^2_{N}$ & $ 27, 66 $ & $28.0$ & $26.6$ & $29.7$ & $50.4$ \\
$\chi^2_{O, td}$ & $ 40 $ & $48.6$ & $41.5$ & $46.4$ &  $ - $  \\
$\chi^2_{O}$ & $ 28, 67 $ & $9.9$ & $10.4$ & $15.7$ & $21.7$ \\
$\chi^2_{B/He, td}$ & $ 41, - $ & $21.9$ & $11.6$ & $21.6$ &  $ - $  \\
$\chi^2_{B/C}$ & $ 27, 67 $ & $11.1$ & $10.5$ & $12.6$ & $25.1$ \\
$\chi^2_{B/C, DAMPE}$ & $ 13 $ & $14.7$ & $10.9$ & $19.2$ & $28.8$
\end{tabular}
\end{table*}

\begin{table*}[tbp]
\centering
\caption{Summary of the best-fit parameters in the \texttt{conv} propagation model for the 3 time-dependent and the 7-yr time-integrated fit for the extended solar modulation model (\texttt{Ext}).}
\label{tab: conv Zhu}
\begin{tabular}{r|c||c|c|c|c}
Parameter & Prior & \texttt{conv\_Ext\_preMAX} & \texttt{conv\_Ext\_MAX} & \texttt{conv\_Ext\_postMAX} & \texttt{conv\_Ext} \\
\hline
$\phi_{FF, p}$ [GV] & $  -   $ &  $ - $  &  $ - $  &  $ - $  &  $ - $  \\
$\phi_{FF, \bar{p}}$ [GV] & $  -   $ &  $ - $  &  $ - $  &  $ - $  &  $ - $  \\
$\phi_{l, p}$ [GV] & $0.1 - 1.5 $ & $0.52_{-0.01}^{+0.01}$ & $0.67_{-0.01}^{+0.01}$ & $0.31_{-0.01}^{+0.01}$ & $0.471_{-0.01}^{+0.009}$ \\
$\phi_{h, p}$ [GV] & $0.1 - 1.5 $ & $0.50_{-0.03}^{+0.03}$ & $0.60_{-0.03}^{+0.03}$ & $0.28_{-0.03}^{+0.03}$ & $0.41_{-0.03}^{+0.03}$ \\
$\phi_{l, \bar{p}}$ [GV] & $0.1 - 1.5 $ & $0.82_{-0.07}^{+0.07}$ & $1.02_{-0.05}^{+0.09}$ & $0.81_{-0.06}^{+0.07}$ & $0.95_{-0.05}^{+0.01}$ \\
$\phi_{h, \bar{p}}$ [GV] & $0.1 - 1.5 $ & $0.34_{-0.04}^{+0.04}$ & $0.47_{-0.03}^{+0.04}$ & $0.27_{-0.04}^{+0.04}$ & $0.54_{-0.03}^{+0.02}$ \\
$R_{br}$ [GV] & $1.0 - 6.0 $ & $4.9_{-0.5}^{+0.5}$ & $5.0_{-0.6}^{+0.4}$ & $4.6_{-0.4}^{+0.4}$ & $3.8_{-0.3}^{+0.3}$ \\
$\gamma_{1, p}$ & $2.3 - 2.42 $ & $2.367_{-0.005}^{+0.006}$ & $2.376_{-0.005}^{+0.006}$ & $2.353_{-0.006}^{+0.006}$ & $2.347_{-0.005}^{+0.007}$ \\
$\gamma_{1, He-p}$ & $-0.09 - -0.03 $ & $-0.052_{-0.002}^{+0.003}$ & $-0.053_{-0.002}^{+0.002}$ & $-0.055_{-0.002}^{+0.002}$ & $-0.059_{-0.002}^{+0.002}$ \\
$\gamma_{1, CNO-p}$ & $-0.05 - 0.0 $ & $-0.027_{-0.003}^{+0.003}$ & $-0.025_{-0.004}^{+0.003}$ & $-0.030_{-0.003}^{+0.003}$ & $-0.025_{-0.004}^{+0.003}$ \\
$D_0$ [cm$^2$/s] $\times 10^{28}$ & $1 - 6 $ & $2.2_{-0.2}^{+0.3}$ & $2.5_{-0.2}^{+0.3}$ & $2.4_{-0.2}^{+0.2}$ & $2.15_{-0.04}^{+0.1}$ \\
$\delta_l$ & $-1.2 - -0.4 $ & $-0.83_{-0.07}^{+0.05}$ & $-0.84_{-0.05}^{+0.04}$ & $-0.86_{-0.07}^{+0.06}$ & $-0.97_{-0.1}^{+0.09}$ \\
$\delta$ & $0.3 - 1.0 $ & $0.72_{-0.04}^{+0.1}$ & $0.67_{-0.03}^{+0.09}$ & $0.64_{-0.02}^{+0.06}$ & $0.83_{-0.06}^{+0.09}$ \\
$\delta_h - \delta$ & $-1.2 - -0.2 $ & $-0.58_{-0.2}^{+0.05}$ & $-0.52_{-0.1}^{+0.06}$ & $-0.45_{-0.1}^{+0.03}$ & $-0.73_{-0.2}^{+0.09}$ \\
$R_{D, 1}$ [MeV] $\times 10^{3}$ & $2.0 - 6 $ & $4.1_{-0.2}^{+0.2}$ & $4.2_{-0.2}^{+0.2}$ & $3.5_{-0.2}^{+0.2}$ & $3.4_{-0.3}^{+0.3}$ \\
$s_{D, 1}$ & $0.05 - 0.6 $ & $0.16_{-0.04}^{+0.05}$ & $0.11_{-0.02}^{+0.05}$ & $0.16_{-0.05}^{+0.04}$ & $0.29_{-0.03}^{+0.07}$ \\
$R_{D, 2}$ [MeV] $\times 10^{5}$ & $0.8 - 5 $ & $2.7_{-0.4}^{+0.6}$ & $2.9_{-0.4}^{+0.7}$ & $3.3_{-0.4}^{+0.7}$ & $2.5_{-0.3}^{+0.7}$ \\
$s_{D, 2}$ & $0.05 - 3.0 $ & $1.2_{-0.2}^{+0.5}$ & $1.1_{-0.2}^{+0.3}$ & $1.0_{-0.1}^{+0.3}$ & $1.6_{-0.3}^{+0.5}$ \\
$v_{0, c}$ [km/s] & $0.0 - 30 $ & $15_{-2}^{+2}$ & $13_{-3}^{+2}$ & $14_{-2}^{+2}$ & $18_{-1}^{+1}$ \\
Abd$_{C}$ $\times 10^{3}$ & $3.5 - 3.8 $ & $3.65_{-0.04}^{+0.05}$ & $3.67_{-0.06}^{+0.04}$ & $3.63_{-0.05}^{+0.04}$ & $3.65_{-0.04}^{+0.03}$ \\
Abd$_{N}$ & $200 - 350 $ & $245_{-12}^{+17}$ & $249_{-14}^{+16}$ & $247_{-18}^{+13}$ & $269_{-14}^{+17}$ \\
Abd$_{O}$ $\times 10^{3}$ & $4.35 - 4.65 $ & $4.52_{-0.03}^{+0.03}$ & $4.51_{-0.03}^{+0.03}$ & $4.53_{-0.03}^{+0.03}$ & $4.49_{-0.04}^{+0.03}$ \\
$A_{p}$ & $0.9 - 1.1 $ & $1.016_{-0.002}^{+0.002}$ & $1.022_{-0.002}^{+0.002}$ & $1.007_{-0.002}^{+0.002}$ & $1.014_{-0.002}^{+0.002}$ \\
$A_{He}$ & $0.9 - 1.1 $ & $1.012_{-0.003}^{+0.004}$ & $1.009_{-0.003}^{+0.004}$ & $1.010_{-0.004}^{+0.004}$ & $1.023_{-0.004}^{+0.004}$ \\
$A_{B}$ & $0.7 - 1.3 $ & $1.032_{-0.008}^{+0.008}$ & $1.048_{-0.01}^{+0.008}$ & $1.062_{-0.01}^{+0.008}$ & $1.10_{-0.01}^{+0.01}$ \\
$A_{\bar{p}}$ & $0.7 - 1.3 $ & $1.01_{-0.02}^{+0.01}$ & $1.04_{-0.02}^{+0.01}$ & $1.04_{-0.02}^{+0.02}$ & $1.12_{-0.02}^{+0.02}$ \\
$A_{xs ^{4}He \rightarrow ^{3}He}$ & $0.7 - 1.3 $ & $1.291_{-0.009}^{+0.002}$ & $1.290_{-0.01}^{+0.002}$ & $1.287_{-0.01}^{+0.002}$ & $1.291_{-0.009}^{+0.002}$ \\
$\delta_{xs ^{4}He \rightarrow ^{3}He}$ & $-0.3 - 0.3 $ & $0.06_{-0.01}^{+0.01}$ & $0.07_{-0.02}^{+0.01}$ & $0.06_{-0.01}^{+0.01}$ & $-0.01_{-0.01}^{+0.02}$ \\
$A_{xs\rightarrow C}$ & $0.3 - 2.0 $ & $0.7_{-0.1}^{+0.1}$ & $0.7_{-0.1}^{+0.1}$ & $0.8_{-0.1}^{+0.1}$ & $0.79_{-0.08}^{+0.1}$ \\
$\delta_{xs\rightarrow C}$ & $-0.3 - 0.3 $ & $0.04_{-0.06}^{+0.06}$ & $0.03_{-0.06}^{+0.08}$ & $0.08_{-0.05}^{+0.07}$ & $0.22_{-0.07}^{+0.02}$ \\
$A_{xs\rightarrow N}$ & $0.7 - 1.3 $ & $1.13_{-0.03}^{+0.03}$ & $1.13_{-0.03}^{+0.03}$ & $1.17_{-0.03}^{+0.03}$ & $1.15_{-0.04}^{+0.03}$ \\
$\delta_{xs\rightarrow N}$ & $-0.3 - 0.3 $ & $-0.04_{-0.01}^{+0.01}$ & $-0.04_{-0.01}^{+0.01}$ & $-0.014_{-0.01}^{+0.009}$ & $-0.03_{-0.01}^{+0.02}$ \\
$\delta_{xs\rightarrow B}$ & $-0.3 - 0.3 $ & $-0.04_{-0.01}^{+0.01}$ & $-0.02_{-0.01}^{+0.01}$ & $0.00_{-0.01}^{+0.01}$ & $-0.03_{-0.01}^{+0.01}$
\end{tabular}
\end{table*}

\begin{table*}[tbp]
\ContinuedFloat
\centering
\caption{Summary of the corresponding $\chi^2$ values.}
\begin{tabular}{r|c||c|c|c|c}
Parameter & \# Datapoints & \texttt{conv\_Ext\_preMAX} & \texttt{conv\_Ext\_MAX} & \texttt{conv\_Ext\_postMAX} & \texttt{conv\_Ext} \\
\hline
$\chi^2_{p, td}$ & $ 10 $ & $31.1$ & $27.2$ & $25.4$ &  $ - $  \\
$\chi^2_{p}$ & $ 27, 72 $ & $2.8$ & $3.0$ & $2.3$ & $17.1$ \\
$\chi^2_{p, CALET}$ & $ 23 $ & $8.0$ & $7.9$ & $8.1$ & $6.8$ \\
$\chi^2_{p, Voyager}$ & $ 6 $ & $7.6$ & $10.8$ & $6.6$ & $4.5$ \\
$\chi^2_{He, td}$ & $ 40 $ & $21.6$ & $21.7$ & $25.0$ &  $ - $  \\
$\chi^2_{He}$ & $ 28, 68 $ & $3.8$ & $4.6$ & $5.1$ & $10.3$ \\
$\chi^2_{He, DAMPE}$ & $ 21 $ & $6.8$ & $6.6$ & $6.7$ & $6.7$ \\
$\chi^2_{He, Voyager}$ & $ 7 $ & $22.8$ & $26.2$ & $22.1$ & $21.5$ \\
$\chi^2_{^{3}He/^{4}He, td}$ & $ 26 $ & $48.3$ & $29.2$ & $21.1$ &  $ - $  \\
$\chi^2_{\bar{p}, td}$ & $ 10 $ & $12.6$ & $10.0$ & $5.7$ &  $ - $  \\
$\chi^2_{\bar{p}/p}$ & $ 18, 58 $ & $10.7$ & $10.5$ & $10.4$ & $27.3$ \\
$\chi^2_{C, td}$ & $ 40 $ & $9.2$ & $7.7$ & $8.5$ &  $ - $  \\
$\chi^2_{C}$ & $ 28, 68 $ & $11.8$ & $12.2$ & $11.1$ & $36.8$ \\
$\chi^2_{N, td}$ & $ 40 $ & $29.1$ & $21.6$ & $19.6$ &  $ - $  \\
$\chi^2_{N}$ & $ 27, 66 $ & $28.4$ & $27.7$ & $30.2$ & $46.1$ \\
$\chi^2_{O, td}$ & $ 40 $ & $43.4$ & $33.3$ & $42.3$ &  $ - $  \\
$\chi^2_{O}$ & $ 28, 67 $ & $9.4$ & $9.7$ & $15.3$ & $17.4$ \\
$\chi^2_{B/He, td}$ & $ 41, - $ & $23.0$ & $12.0$ & $22.7$ &  $ - $  \\
$\chi^2_{B/C}$ & $ 27, 67 $ & $11.4$ & $10.9$ & $13.0$ & $28.9$ \\
$\chi^2_{B/C, DAMPE}$ & $ 13 $ & $15.5$ & $13.1$ & $19.8$ & $27.4$
\end{tabular}
\end{table*}

\begin{table*}[tbp]
\centering
\caption{Summary of the best-fit parameters in the \texttt{reacc} propagation model for the 3 time-dependent and the time-independent fit for the force-field solar modulation model (\texttt{FF}).}
\label{tab: reacc FF}
\begin{tabular}{r|c||c|c|c|c}
Parameter & Prior & \texttt{reacc\_FF\_preMAX} & \texttt{reacc\_FF\_MAX} & \texttt{reacc\_FF\_postMAX} & \texttt{reacc\_FF} \\
\hline
$\phi_{FF, p}$ [GV] & $0.1 - 1.0 $ & $0.65_{-0.01}^{+0.01}$ & $0.81_{-0.01}^{+0.01}$ & $0.40_{-0.01}^{+0.01}$ & $0.65_{-0.01}^{+0.01}$ \\
$\phi_{FF, \bar{p}}$ [GV] & $0.1 - 1.0 $ & $0.36_{-0.03}^{+0.03}$ & $0.49_{-0.05}^{+0.04}$ & $0.38_{-0.03}^{+0.03}$ & $0.55_{-0.06}^{+0.04}$ \\
$\phi_{l, p}$ [GV] & $  -   $ &  $ - $  &  $ - $  &  $ - $  &  $ - $  \\
$\phi_{h, p}$ [GV] & $  -   $ &  $ - $  &  $ - $  &  $ - $  &  $ - $  \\
$\phi_{l, \bar{p}}$ [GV] & $  -   $ &  $ - $  &  $ - $  &  $ - $  &  $ - $  \\
$\phi_{h, \bar{p}}$ [GV] & $  -   $ &  $ - $  &  $ - $  &  $ - $  &  $ - $  \\
$R_{br}$ [GV] & $  -   $ &  $ - $  &  $ - $  &  $ - $  &  $ - $  \\
$\gamma_{0, p}$ & $1.5 - 2.0 $ & $1.73_{-0.02}^{+0.02}$ & $1.70_{-0.02}^{+0.03}$ & $1.70_{-0.03}^{+0.02}$ & $1.71_{-0.03}^{+0.02}$ \\
$\gamma_{1, p}$ & $2.3 - 2.45 $ & $2.419_{-0.004}^{+0.007}$ & $2.429_{-0.006}^{+0.006}$ & $2.397_{-0.006}^{+0.006}$ & $2.411_{-0.007}^{+0.007}$ \\
$\gamma_{1, He-p}$ & $-0.09 - -0.03 $ & $-0.062_{-0.003}^{+0.003}$ & $-0.061_{-0.003}^{+0.004}$ & $-0.063_{-0.003}^{+0.003}$ & $-0.062_{-0.003}^{+0.003}$ \\
$\gamma_{1, CNO-p}$ & $-0.05 - 0.0 $ & $-0.026_{-0.003}^{+0.004}$ & $-0.024_{-0.003}^{+0.004}$ & $-0.027_{-0.004}^{+0.004}$ & $-0.027_{-0.004}^{+0.004}$ \\
$\gamma_{0, He}$ & $1.5 - 2.0 $ & $1.79_{-0.02}^{+0.02}$ & $1.75_{-0.02}^{+0.02}$ & $1.75_{-0.03}^{+0.02}$ & $1.75_{-0.03}^{+0.03}$ \\
$\gamma_{0, CNO}$ & $1.8 - 2.2 $ & $2.02_{-0.02}^{+0.02}$ & $1.99_{-0.03}^{+0.02}$ & $1.97_{-0.03}^{+0.02}$ & $2.01_{-0.03}^{+0.02}$ \\
$D_0$ [cm$^2$/s] $\times 10^{28}$ & $1 - 8 $ & $4.38_{-0.1}^{+0.08}$ & $4.7_{-0.2}^{+0.1}$ & $4.2_{-0.1}^{+0.1}$ & $4.7_{-0.2}^{+0.2}$ \\
$\delta$ & $0.2 - 0.8 $ & $0.410_{-0.006}^{+0.006}$ & $0.399_{-0.01}^{+0.007}$ & $0.429_{-0.006}^{+0.007}$ & $0.42_{-0.01}^{+0.01}$ \\
$\delta_h - \delta$ & $-1.0 - -0.1 $ & $-0.22_{-0.02}^{+0.02}$ & $-0.21_{-0.02}^{+0.02}$ & $-0.22_{-0.02}^{+0.02}$ & $-0.22_{-0.03}^{+0.02}$ \\
$R_{D, 2}$ [MeV] $\times 10^{5}$ & $1 - 7 $ & $4.1_{-0.3}^{+0.6}$ & $4.1_{-0.4}^{+0.6}$ & $4.2_{-0.4}^{+0.6}$ & $4.1_{-0.4}^{+0.7}$ \\
$s_{D, 2}$ & $0.05 - 1.0 $ & $0.34_{-0.05}^{+0.1}$ & $0.33_{-0.07}^{+0.07}$ & $0.35_{-0.06}^{+0.1}$ & $0.40_{-0.08}^{+0.1}$ \\
$R_{\rm inj}$ $\times 10^{3}$ & $3 - 15 $ & $7.5_{-0.4}^{+0.4}$ & $7.0_{-0.4}^{+0.4}$ & $6.5_{-0.5}^{+0.4}$ & $7.4_{-0.4}^{+0.5}$ \\
$s_{\rm inj}$ & $0.05 - 0.8 $ & $0.30_{-0.03}^{+0.03}$ & $0.29_{-0.03}^{+0.03}$ & $0.36_{-0.03}^{+0.04}$ & $0.35_{-0.03}^{+0.03}$ \\
$v_{\rm A}$ [km/s] & $0.0 - 35 $ & $22_{-1}^{+1}$ & $24_{-1}^{+1}$ & $19_{-1}^{+1}$ & $24_{-1}^{+1}$ \\
Abd$_{C}$ $\times 10^{3}$ & $3.5 - 3.8 $ & $3.71_{-0.03}^{+0.03}$ & $3.70_{-0.03}^{+0.02}$ & $3.71_{-0.04}^{+0.02}$ & $3.67_{-0.05}^{+0.03}$ \\
Abd$_{N}$ & $200 - 350 $ & $250_{-15}^{+16}$ & $254_{-15}^{+16}$ & $261_{-15}^{+14}$ & $262_{-19}^{+15}$ \\
Abd$_{O}$ $\times 10^{3}$ & $4.0 - 4.65 $ & $4.39_{-0.03}^{+0.03}$ & $4.38_{-0.03}^{+0.03}$ & $4.40_{-0.03}^{+0.03}$ & $4.34_{-0.04}^{+0.03}$ \\
$A_{p}$ & $0.9 - 1.1 $ & $1.028_{-0.002}^{+0.002}$ & $1.036_{-0.002}^{+0.002}$ & $1.016_{-0.002}^{+0.002}$ & $1.030_{-0.002}^{+0.002}$ \\
$A_{He}$ & $0.9 - 1.1 $ & $0.999_{-0.003}^{+0.005}$ & $0.997_{-0.003}^{+0.004}$ & $0.999_{-0.004}^{+0.004}$ & $0.998_{-0.006}^{+0.005}$ \\
$A_{B}$ & $0.7 - 1.3 $ & $1.033_{-0.01}^{+0.008}$ & $1.05_{-0.01}^{+0.01}$ & $1.06_{-0.02}^{+0.01}$ & $1.11_{-0.02}^{+0.01}$ \\
$A_{\bar{p}}$ & $0.7 - 1.3 $ & $1.04_{-0.02}^{+0.01}$ & $1.07_{-0.02}^{+0.02}$ & $1.07_{-0.02}^{+0.02}$ & $1.18_{-0.02}^{+0.02}$ \\
$A_{xs ^{4}He \rightarrow ^{3}He}$ & $0.7 - 1.3 $ & $1.286_{-0.01}^{+0.002}$ & $1.286_{-0.01}^{+0.003}$ & $1.279_{-0.02}^{+0.004}$ & $1.286_{-0.01}^{+0.003}$ \\
$\delta_{xs ^{4}He \rightarrow ^{3}He}$ & $-0.3 - 0.3 $ & $0.294_{-0.006}^{+0.001}$ & $0.283_{-0.02}^{+0.005}$ & $0.291_{-0.009}^{+0.002}$ & $0.25_{-0.02}^{+0.02}$ \\
$A_{xs\rightarrow C}$ & $0.3 - 2.0 $ & $0.37_{-0.02}^{+0.07}$ & $0.38_{-0.02}^{+0.08}$ & $0.39_{-0.02}^{+0.09}$ & $0.41_{-0.03}^{+0.1}$ \\
$\delta_{xs\rightarrow C}$ & $-0.3 - 0.3 $ & $0.16_{-0.1}^{+0.04}$ & $0.16_{-0.1}^{+0.04}$ & $0.11_{-0.1}^{+0.08}$ & $0.23_{-0.07}^{+0.02}$ \\
$A_{xs\rightarrow N}$ & $0.7 - 1.3 $ & $1.11_{-0.03}^{+0.03}$ & $1.11_{-0.03}^{+0.03}$ & $1.12_{-0.03}^{+0.04}$ & $1.16_{-0.04}^{+0.04}$ \\
$\delta_{xs\rightarrow N}$ & $-0.3 - 0.3 $ & $0.08_{-0.01}^{+0.01}$ & $0.08_{-0.01}^{+0.01}$ & $0.08_{-0.01}^{+0.01}$ & $0.10_{-0.02}^{+0.02}$ \\
$\delta_{xs\rightarrow B}$ & $-0.3 - 0.3 $ & $0.12_{-0.01}^{+0.01}$ & $0.12_{-0.01}^{+0.01}$ & $0.14_{-0.01}^{+0.01}$ & $0.16_{-0.01}^{+0.01}$
\end{tabular}
\end{table*}

\begin{table*}[tbp]
\ContinuedFloat
\centering
\caption{Summary of the corresponding $\chi^2$ values.}
\begin{tabular}{r|c||c|c|c|c}
Parameter & \# Datapoints & \texttt{reacc\_FF\_preMAX} & \texttt{reacc\_FF\_MAX} & \texttt{reacc\_FF\_postMAX} & \texttt{reacc\_FF} \\
\hline
$\chi^2_{p, td}$ & $ 10 $ & $16.2$ & $17.5$ & $14.7$ &  $ - $  \\
$\chi^2_{p}$ & $ 27, 72 $ & $3.7$ & $3.5$ & $3.5$ & $20.1$ \\
$\chi^2_{p, CALET}$ & $ 23 $ & $5.6$ & $6.0$ & $5.8$ & $5.8$ \\
$\chi^2_{p, Voyager}$ & $ 6 $ & $31.4$ & $37.3$ & $24.2$ & $32.3$ \\
$\chi^2_{He, td}$ & $ 40 $ & $17.3$ & $10.1$ & $18.9$ &  $ - $  \\
$\chi^2_{He}$ & $ 28, 68 $ & $3.3$ & $2.7$ & $3.2$ & $21.3$ \\
$\chi^2_{He, DAMPE}$ & $ 21 $ & $8.4$ & $8.1$ & $7.9$ & $7.7$ \\
$\chi^2_{He, Voyager}$ & $ 7 $ & $26.4$ & $34.5$ & $20.8$ & $29.8$ \\
$\chi^2_{^{3}He/^{4}He, td}$ & $ 26 $ & $73.5$ & $47.1$ & $31.6$ &  $ - $  \\
$\chi^2_{\bar{p}, td}$ & $ 10 $ & $35.1$ & $36.2$ & $28.9$ &  $ - $  \\
$\chi^2_{\bar{p}/p}$ & $ 18, 58 $ & $13.2$ & $13.9$ & $12.5$ & $48.9$ \\
$\chi^2_{C, td}$ & $ 40 $ & $8.5$ & $7.8$ & $9.5$ &  $ - $  \\
$\chi^2_{C}$ & $ 28, 68 $ & $13.3$ & $15.1$ & $13.5$ & $25.9$ \\
$\chi^2_{N, td}$ & $ 40 $ & $19.9$ & $14.6$ & $15.9$ &  $ - $  \\
$\chi^2_{N}$ & $ 27, 66 $ & $26.5$ & $26.2$ & $28.7$ & $32.7$ \\
$\chi^2_{O, td}$ & $ 40 $ & $8.7$ & $9.6$ & $12.4$ &  $ - $  \\
$\chi^2_{O}$ & $ 28, 67 $ & $3.7$ & $4.8$ & $5.2$ & $12.4$ \\
$\chi^2_{B/He, td}$ & $ 41, - $ & $14.6$ & $11.1$ & $10.3$ &  $ - $  \\
$\chi^2_{B/C}$ & $ 27, 67 $ & $17.4$ & $16.4$ & $13.8$ & $22.0$ \\
$\chi^2_{B/C, DAMPE}$ & $ 13 $ & $3.6$ & $2.7$ & $7.9$ & $6.7$
\end{tabular}
\end{table*}

\begin{table*}[tbp]
\centering
\caption{Summary of the best-fit parameters in the \texttt{reacc} propagation model for the 3 time-dependent and the 7-yr time-integrated fit for the extended solar modulation model (\texttt{Ext}).}
\label{tab: reacc Zhu}
\begin{tabular}{r|c||c|c|c|c}
Parameter & Prior & \texttt{reacc\_Ext\_preMAX} & \texttt{reacc\_Ext\_MAX} & \texttt{reacc\_Ext\_postMAX} & \texttt{reacc\_Ext} \\
\hline
$\phi_{FF, p}$ [GV] & $  -   $ &  $ - $  &  $ - $  &  $ - $  &  $ - $  \\
$\phi_{FF, \bar{p}}$ [GV] & $  -   $ &  $ - $  &  $ - $  &  $ - $  &  $ - $  \\
$\phi_{l, p}$ [GV] & $0.1 - 1.5 $ & $0.55_{-0.01}^{+0.02}$ & $0.70_{-0.02}^{+0.02}$ & $0.344_{-0.008}^{+0.007}$ & $0.54_{-0.02}^{+0.02}$ \\
$\phi_{h, p}$ [GV] & $0.1 - 1.5 $ & $0.28_{-0.03}^{+0.04}$ & $0.37_{-0.04}^{+0.03}$ & $0.120_{-0.005}^{+0.02}$ & $0.29_{-0.03}^{+0.04}$ \\
$\phi_{l, \bar{p}}$ [GV] & $0.1 - 1.5 $ & $1.11_{-0.07}^{+0.06}$ & $1.45_{-0.05}^{+0.02}$ & $1.12_{-0.07}^{+0.09}$ & $0.998_{-0.002}^{+0.002}$ \\
$\phi_{h, \bar{p}}$ [GV] & $0.1 - 1.5 $ & $0.42_{-0.04}^{+0.05}$ & $0.58_{-0.04}^{+0.05}$ & $0.40_{-0.05}^{+0.04}$ & $0.64_{-0.03}^{+0.04}$ \\
$R_{br}$ [GV] & $1.0 - 6.0 $ & $3.3_{-0.1}^{+0.1}$ & $3.0_{-0.1}^{+0.1}$ & $3.1_{-0.2}^{+0.2}$ & $3.4_{-0.1}^{+0.1}$ \\
$\gamma_{0, p}$ & $1.5 - 2.0 $ & $1.79_{-0.04}^{+0.02}$ & $1.75_{-0.04}^{+0.03}$ & $1.72_{-0.05}^{+0.04}$ & $1.67_{-0.04}^{+0.06}$ \\
$\gamma_{1, p}$ & $2.3 - 2.45 $ & $2.393_{-0.008}^{+0.006}$ & $2.396_{-0.007}^{+0.008}$ & $2.389_{-0.009}^{+0.007}$ & $2.39_{-0.01}^{+0.01}$ \\
$\gamma_{1, He-p}$ & $-0.09 - -0.03 $ & $-0.068_{-0.003}^{+0.004}$ & $-0.069_{-0.004}^{+0.003}$ & $-0.069_{-0.004}^{+0.004}$ & $-0.067_{-0.004}^{+0.003}$ \\
$\gamma_{1, CNO-p}$ & $-0.05 - 0.0 $ & $-0.038_{-0.005}^{+0.005}$ & $-0.036_{-0.005}^{+0.006}$ & $-0.036_{-0.006}^{+0.006}$ & $-0.038_{-0.006}^{+0.006}$ \\
$\gamma_{0, He}$ & $1.5 - 2.0 $ & $1.83_{-0.03}^{+0.02}$ & $1.79_{-0.03}^{+0.03}$ & $1.77_{-0.05}^{+0.04}$ & $1.70_{-0.05}^{+0.05}$ \\
$\gamma_{0, CNO}$ & $1.8 - 2.2 $ & $1.97_{-0.03}^{+0.02}$ & $1.92_{-0.04}^{+0.03}$ & $1.90_{-0.05}^{+0.04}$ & $1.90_{-0.04}^{+0.05}$ \\
$D_0$ [cm$^2$/s] $\times 10^{28}$ & $1 - 8 $ & $3.6_{-0.2}^{+0.1}$ & $3.7_{-0.2}^{+0.1}$ & $3.6_{-0.3}^{+0.1}$ & $3.7_{-0.2}^{+0.1}$ \\
$\delta$ & $0.2 - 0.8 $ & $0.47_{-0.01}^{+0.01}$ & $0.48_{-0.01}^{+0.02}$ & $0.49_{-0.01}^{+0.02}$ & $0.49_{-0.01}^{+0.01}$ \\
$\delta_h - \delta$ & $-1.0 - -0.1 $ & $-0.28_{-0.03}^{+0.03}$ & $-0.29_{-0.05}^{+0.03}$ & $-0.30_{-0.05}^{+0.03}$ & $-0.31_{-0.04}^{+0.03}$ \\
$R_{D, 2}$ [MeV] $\times 10^{5}$ & $1 - 7 $ & $4.0_{-0.4}^{+0.6}$ & $4.0_{-0.5}^{+0.7}$ & $4.0_{-0.4}^{+0.8}$ & $3.9_{-0.4}^{+0.8}$ \\
$s_{D, 2}$ & $0.05 - 1.0 $ & $0.56_{-0.09}^{+0.1}$ & $0.6_{-0.1}^{+0.1}$ & $0.6_{-0.1}^{+0.1}$ & $0.7_{-0.1}^{+0.1}$ \\
$R_{\rm inj}$ $\times 10^{3}$ & $3 - 15 $ & $9_{-1}^{+1}$ & $8_{-1}^{+1}$ & $7.4_{-0.9}^{+0.9}$ & $6.9_{-0.8}^{+0.9}$ \\
$s_{\rm inj}$ & $0.05 - 0.8 $ & $0.49_{-0.05}^{+0.06}$ & $0.50_{-0.05}^{+0.07}$ & $0.65_{-0.09}^{+0.06}$ & $0.65_{-0.09}^{+0.07}$ \\
$v_{\rm A}$ [km/s] & $0.0 - 35 $ & $12_{-1}^{+1}$ & $12_{-2}^{+1}$ & $11_{-1}^{+1}$ & $12_{-1}^{+1}$ \\
Abd$_{C}$ $\times 10^{3}$ & $3.5 - 3.8 $ & $3.69_{-0.03}^{+0.04}$ & $3.69_{-0.04}^{+0.04}$ & $3.67_{-0.04}^{+0.04}$ & $3.65_{-0.05}^{+0.04}$ \\
Abd$_{N}$ & $200 - 350 $ & $265_{-14}^{+15}$ & $268_{-15}^{+15}$ & $272_{-15}^{+15}$ & $269_{-16}^{+17}$ \\
Abd$_{O}$ $\times 10^{3}$ & $4.0 - 4.65 $ & $4.38_{-0.04}^{+0.04}$ & $4.38_{-0.04}^{+0.04}$ & $4.36_{-0.04}^{+0.05}$ & $4.32_{-0.04}^{+0.05}$ \\
$A_{p}$ & $0.9 - 1.1 $ & $1.011_{-0.002}^{+0.002}$ & $1.016_{-0.002}^{+0.002}$ & $1.003_{-0.002}^{+0.002}$ & $1.010_{-0.002}^{+0.003}$ \\
$A_{He}$ & $0.9 - 1.1 $ & $0.997_{-0.005}^{+0.005}$ & $0.997_{-0.006}^{+0.005}$ & $0.990_{-0.006}^{+0.005}$ & $0.993_{-0.008}^{+0.008}$ \\
$A_{B}$ & $0.7 - 1.3 $ & $1.05_{-0.03}^{+0.02}$ & $1.07_{-0.03}^{+0.01}$ & $1.07_{-0.04}^{+0.02}$ & $1.11_{-0.03}^{+0.01}$ \\
$A_{\bar{p}}$ & $0.7 - 1.3 $ & $1.06_{-0.03}^{+0.02}$ & $1.10_{-0.04}^{+0.02}$ & $1.08_{-0.05}^{+0.03}$ & $1.17_{-0.03}^{+0.03}$ \\
$A_{xs ^{4}He \rightarrow ^{3}He}$ & $0.7 - 1.3 $ & $1.261_{-0.04}^{+0.009}$ & $1.262_{-0.04}^{+0.008}$ & $1.24_{-0.06}^{+0.02}$ & $1.268_{-0.03}^{+0.007}$ \\
$\delta_{xs ^{4}He \rightarrow ^{3}He}$ & $-0.3 - 0.3 $ & $0.289_{-0.01}^{+0.003}$ & $0.281_{-0.01}^{+0.009}$ & $0.273_{-0.01}^{+0.009}$ & $0.24_{-0.02}^{+0.02}$ \\
$A_{xs\rightarrow C}$ & $0.3 - 2.0 $ & $0.39_{-0.02}^{+0.09}$ & $0.40_{-0.02}^{+0.1}$ & $0.39_{-0.02}^{+0.09}$ & $0.43_{-0.03}^{+0.1}$ \\
$\delta_{xs\rightarrow C}$ & $-0.3 - 0.3 $ & $0.0_{-0.1}^{+0.1}$ & $0.0_{-0.1}^{+0.1}$ & $-0.0_{-0.1}^{+0.1}$ & $0.21_{-0.09}^{+0.02}$ \\
$A_{xs\rightarrow N}$ & $0.7 - 1.3 $ & $1.09_{-0.04}^{+0.03}$ & $1.09_{-0.03}^{+0.03}$ & $1.08_{-0.04}^{+0.04}$ & $1.14_{-0.04}^{+0.05}$ \\
$\delta_{xs\rightarrow N}$ & $-0.3 - 0.3 $ & $0.07_{-0.01}^{+0.01}$ & $0.06_{-0.01}^{+0.01}$ & $0.06_{-0.01}^{+0.01}$ & $0.08_{-0.02}^{+0.02}$ \\
$\delta_{xs\rightarrow B}$ & $-0.3 - 0.3 $ & $0.12_{-0.01}^{+0.01}$ & $0.12_{-0.01}^{+0.02}$ & $0.13_{-0.01}^{+0.01}$ & $0.15_{-0.02}^{+0.01}$
\end{tabular}
\end{table*}

\begin{table*}[tbp]
\ContinuedFloat
\centering
\caption{Summary of the corresponding $\chi^2$ values.}
\begin{tabular}{r|c||c|c|c|c}
Parameter & \# Datapoints & \texttt{reacc\_Ext\_preMAX} & \texttt{reacc\_Ext\_MAX} & \texttt{reacc\_Ext\_postMAX} & \texttt{reacc\_Ext} \\
\hline
$\chi^2_{p, td}$ & $ 10 $ & $6.0$ & $5.2$ & $5.2$ &  $ - $  \\
$\chi^2_{p}$ & $ 27, 72 $ & $4.5$ & $3.4$ & $4.3$ & $11.5$ \\
$\chi^2_{p, CALET}$ & $ 23 $ & $4.7$ & $4.7$ & $4.7$ & $4.8$ \\
$\chi^2_{p, Voyager}$ & $ 6 $ & $9.4$ & $9.2$ & $9.3$ & $11.5$ \\
$\chi^2_{He, td}$ & $ 40 $ & $4.1$ & $5.8$ & $8.9$ &  $ - $  \\
$\chi^2_{He}$ & $ 28, 68 $ & $4.4$ & $5.1$ & $5.8$ & $10.1$ \\
$\chi^2_{He, DAMPE}$ & $ 21 $ & $7.8$ & $7.4$ & $7.5$ & $7.0$ \\
$\chi^2_{He, Voyager}$ & $ 7 $ & $3.1$ & $3.7$ & $3.1$ & $5.3$ \\
$\chi^2_{^{3}He/^{4}He, td}$ & $ 26 $ & $97.3$ & $66.2$ & $47.8$ &  $ - $  \\
$\chi^2_{\bar{p}, td}$ & $ 10 $ & $13.4$ & $10.4$ & $6.6$ &  $ - $  \\
$\chi^2_{\bar{p}/p}$ & $ 18, 58 $ & $11.1$ & $11.4$ & $10.6$ & $39.6$ \\
$\chi^2_{C, td}$ & $ 40 $ & $12.2$ & $9.7$ & $10.0$ &  $ - $  \\
$\chi^2_{C}$ & $ 28, 68 $ & $9.4$ & $10.4$ & $10.1$ & $20.1$ \\
$\chi^2_{N, td}$ & $ 40 $ & $15.7$ & $13.8$ & $12.1$ &  $ - $  \\
$\chi^2_{N}$ & $ 27, 66 $ & $22.5$ & $22.8$ & $23.6$ & $28.4$ \\
$\chi^2_{O, td}$ & $ 40 $ & $11.2$ & $10.2$ & $11.9$ &  $ - $  \\
$\chi^2_{O}$ & $ 28, 67 $ & $4.5$ & $4.8$ & $5.4$ & $16.6$ \\
$\chi^2_{B/He, td}$ & $ 41, - $ & $14.6$ & $10.4$ & $15.7$ &  $ - $  \\
$\chi^2_{B/C}$ & $ 27, 67 $ & $12.6$ & $11.8$ & $12.5$ & $30.5$ \\
$\chi^2_{B/C, DAMPE}$ & $ 13 $ & $9.5$ & $9.2$ & $11.0$ & $11.9$
\end{tabular}
\end{table*}

\begin{table*}[tbp]
\centering
\caption{Summary of the best-fit parameters in the \texttt{conv} and \texttt{reacc} propagation models for the 7-yr time-integrated \texttt{HelMod} solar modulation fits.}
\label{tab: HelMod}
\begin{tabular}{r|c||c|c|c|c}
Parameter & Prior & \texttt{conv\_HelMod} & \texttt{reacc\_HelMod} \\
\hline
$\gamma_{0, p}$ & $1.5 - 2.0 $ & $2.369_{-0.006}^{+0.004}$ & $1.63_{-0.02}^{+0.02}$ \\
$\gamma_{1, p}$ & $2.3 - 2.4 $ & $2.369_{-0.006}^{+0.004}$ & $2.392_{-0.005}^{+0.005}$ \\
$\gamma_{1, He-p}$ & $-0.09 - -0.03 $ & $-0.065_{-0.002}^{+0.002}$ & $-0.061_{-0.003}^{+0.002}$ \\
$\gamma_{1, CNO-p}$ & $-0.05 - 0.0 $ & $-0.022_{-0.003}^{+0.003}$ & $-0.022_{-0.003}^{+0.004}$ \\
$\gamma_{0, He}$ & $1.5 - 2.0 $ & $2.305_{-0.004}^{+0.005}$ & $1.66_{-0.02}^{+0.02}$ \\
$\gamma_{0, CNO}$ & $1.8 - 2.2 $ & $2.348_{-0.004}^{+0.005}$ & $1.91_{-0.03}^{+0.02}$ \\
$R_{\rm inj}$ $\times 10^{3}$ & $3 - 15 $ &  $ - $  & $6.0_{-0.3}^{+0.4}$ \\
$s_{\rm inj}$ & $0.05 - 0.8 $ &  $ - $  & $0.35_{-0.02}^{+0.03}$ \\
$D_0$ [cm$^2$/s] $\times 10^{28}$ & $2 - 6 $ & $3.8_{-0.2}^{+0.1}$ & $4.20_{-0.09}^{+0.1}$ \\
$\delta_l$ & $-1.2 - 0.0 $ & $-1.14_{-0.02}^{+0.06}$ &  $ - $  \\
$\delta$ & $0.2 - 1.0 $ & $0.526_{-0.008}^{+0.01}$ & $0.445_{-0.005}^{+0.005}$ \\
$\delta_h - \delta$ & $-1.2 - 0.0 $ & $-0.30_{-0.03}^{+0.03}$ & $-0.22_{-0.02}^{+0.02}$ \\
$R_{D, 1}$ [MeV] $\times 10^{3}$ & $1.5 - 8 $ & $2.68_{-0.07}^{+0.1}$ &  $ - $  \\
$s_{D, 1}$ & $0.05 - 0.6 $ & $0.38_{-0.02}^{+0.02}$ &  $ - $  \\
$R_{D, 2}$ [MeV] $\times 10^{5}$ & $0.8 - 5 $ & $3.5_{-0.3}^{+0.6}$ & $4.0_{-0.4}^{+0.6}$ \\
$s_{D, 2}$ & $0.05 - 3.0 $ & $0.61_{-0.07}^{+0.1}$ & $0.39_{-0.07}^{+0.09}$ \\
$v_{0, c}$ [km/s] & $0.0 - 30 $ & $1_{-1}^{+2}$ &  $ - $  \\
$v_{\rm A}$ [km/s] & $10.0 - 40 $ &  $ - $  & $18.4_{-0.9}^{+0.9}$ \\
Abd$_{C}$ $\times 10^{3}$ & $3.52 - 3.7 $ & $3.670_{-0.03}^{+0.008}$ & $3.71_{-0.03}^{+0.03}$ \\
Abd$_{N}$ & $220 - 320 $ & $268_{-20}^{+13}$ & $264_{-15}^{+17}$ \\
Abd$_{O}$ $\times 10^{3}$ & $4.4 - 4.6 $ & $4.47_{-0.02}^{+0.03}$ & $4.38_{-0.04}^{+0.03}$ \\
$A_{p}$ & $0.9 - 1.1 $ & $1.021_{-0.002}^{+0.002}$ & $1.026_{-0.002}^{+0.002}$ \\
$A_{He}$ & $0.9 - 1.1 $ & $1.023_{-0.004}^{+0.004}$ & $1.006_{-0.005}^{+0.004}$ \\
$A_{B}$ & $0.7 - 1.3 $ & $1.08_{-0.01}^{+0.01}$ & $1.10_{-0.01}^{+0.01}$ \\
$A_{\bar{p}}$ & $0.7 - 1.3 $ & $1.10_{-0.01}^{+0.01}$ & $1.16_{-0.02}^{+0.02}$ \\
$A_{xs ^{4}He \rightarrow ^{3}He}$ & $0.7 - 1.3 $ & $1.289_{-0.01}^{+0.002}$ & $1.290_{-0.01}^{+0.002}$ \\
$\delta_{xs ^{4}He \rightarrow ^{3}He}$ & $-0.3 - 0.3 $ & $0.05_{-0.02}^{+0.02}$ & $0.25_{-0.01}^{+0.01}$ \\
$A_{xs\rightarrow C}$ & $0.3 - 2.0 $ & $0.66_{-0.08}^{+0.07}$ & $0.38_{-0.02}^{+0.08}$ \\
$\delta_{xs\rightarrow C}$ & $-0.3 - 0.3 $ & $0.22_{-0.08}^{+0.02}$ & $0.20_{-0.1}^{+0.02}$ \\
$A_{xs\rightarrow N}$ & $0.7 - 1.3 $ & $1.13_{-0.03}^{+0.04}$ & $1.15_{-0.05}^{+0.03}$ \\
$\delta_{xs\rightarrow N}$ & $-0.3 - 0.3 $ & $-0.00_{-0.02}^{+0.01}$ & $0.08_{-0.02}^{+0.02}$ \\
$\delta_{xs\rightarrow B}$ & $-0.3 - 0.3 $ & $0.02_{-0.01}^{+0.01}$ & $0.13_{-0.01}^{+0.01}$
\end{tabular}
\end{table*}

\begin{table*}[tbp]
\ContinuedFloat
\centering
\caption{Summary of the corresponding $\chi^2$ values.}
\begin{tabular}{r|c||c|c|c|c}
Parameter & \# Datapoints & \texttt{conv\_HelMod} & \texttt{reacc\_HelMod} \\
\hline
$\chi^2_{p}$ & $ 72 $ & $68.3$ & $28.5$ \\
$\chi^2_{p, CALET}$ & $ 23 $ & $6.8$ & $6.7$ \\
$\chi^2_{p, Voyager}$ & $ 6 $ & $9.8$ & $37.4$ \\
$\chi^2_{He}$ & $ 68 $ & $20.1$ & $15.6$ \\
$\chi^2_{He, DAMPE}$ & $ 21 $ & $8.2$ & $8.0$ \\
$\chi^2_{He, Voyager}$ & $ 7 $ & $99.7$ & $14.0$ \\
$\chi^2_{\bar{p}/p}$ & $ 58 $ & $60.0$ & $63.8$ \\
$\chi^2_{C}$ & $ 68 $ & $26.8$ & $26.7$ \\
$\chi^2_{N}$ & $ 66 $ & $39.7$ & $38.0$ \\
$\chi^2_{O}$ & $ 67 $ & $8.9$ & $11.8$ \\
$\chi^2_{B/C}$ & $ 67 $ & $25.9$ & $27.7$ \\
$\chi^2_{B/C, DAMPE}$ & $ 13 $ & $17.4$ & $11.3$
\end{tabular}
\end{table*}

\end{document}